\documentclass[a4paper]{sapthesis}

\usepackage{microtype}
\usepackage[english]{babel}
\usepackage[utf8]{inputenc}
\usepackage{imakeidx}
\usepackage{subfigure}
\usepackage[acronym]{glossaries}
\usepackage{graphicx}
\usepackage{array}
\usepackage{ragged2e}
\usepackage[table]{xcolor}
\usepackage{hyperref}
\usepackage{blindtext}
\usepackage{comment}
\usepackage[linguistics]{forest} 

\usepackage{xcolor}
\usepackage{amsmath, amssymb}
\usepackage{lmodern}
\usepackage{adjustbox}
\usepackage{makecell}
\usepackage[font=small,labelfont=normalfont,labelsep=quad]{caption}

\usepackage{feynmp-auto}

\usepackage{float} 

\usepackage{slashed} 

\usepackage{enumerate} 

\usepackage{cite}

\newcommand*{\belowrulesepcolor}[1]{%
  \noalign{%
  \kern-\belowrulesep \begingroup \color{#1}%
  \hrule height\belowrulesep \endgroup }%
}
\newcommand*{\aboverulesepcolor}[1]{%
  \noalign{%
  \begingroup \color{#1}%
  \hrule height\aboverulesep \endgroup \kern-\aboverulesep }%
}

\usepackage{xcolor}
\definecolor{emerald}{RGB}{0,153,102}

\newcommand{\mg}[1]{}
\newcommand{\fm}[1]{}
\newcommand{\FM}[1]{}
\newcommand{\lv}[1]{}
\newcommand{\gl}[1]{}

\renewcommand{\arraystretch}{1.5}  

\hypersetup{pdftitle={Light dark sector via thermal decays of Dark Matter: the case of a 17 MeV particle coupled to electrons}, pdfauthor={Marco Graziani}}
\title{Light dark sector via thermal decays of Dark Matter: the case of a 17 MeV particle coupled to electrons}
\author{Marco Graziani}
\IDnumber{1962894}
\course{Laurea Magistrale in Fisica}
\courseorganizer{Facoltà di Scienze matematiche, fisiche e naturali}
\AcademicYear{2024/2025}

\advisor{Dott. Ludovico Vittorio}
\customadvisorlabel{Relatore Interno}
\coadvisor{Dott. Giacomo Landini}
\customcoadvisorlabel{Relatore Esterno}
\authoremail{graziani.1962894@studenti.uniroma1.it}
\copyyear{2026}
\thesistype{}
\examdate{26 Gennaio 2026}
\examiner{Prof. Enzo Marinari} 
\examiner{Prof. Roberto Contino} 
\examiner{Prof. Anna Maria Siani}
\examiner{Prof. Sandro De Cecco}
\examiner{Prof. Luca Graziani}
\examiner{Prof. Valeria Cimini}
\examiner{Prof. Luca Angelani}

\makeindex

\begin{document}
\frontmatter
\maketitle
\dedication{}
\newpage

\begin{center}
\textbf{\large Abstract}
\end{center}

\noindent

Recent experimental observations, most notably those reported by the ATOMKI and Positron Annihilation into Dark Matter Experiment (PADME) collaborations, have hinted anomalies that may indicate the presence of a new resonance with a mass around $17\,\text{MeV}$, potentially interacting with both nucleons and electrons. Since 2020, ATOMKI has observed this resonance in nuclear transitions from excited to ground states in ${}^{8}\mathrm{Be}$, ${}^{4}\mathrm{He}$, and ${}^{12}\mathrm{C}$. More recently, in 2025, PADME, operating at the Laboratori Nazionali di Frascati, has also hinted a similar excess, in this case in the $e^{+}e^{-}$ final-state events originating from positron annihilation on  fixed-target atomic electrons of Carbonium. This concordance strengthens the case for a common underlying origin, potentially involving a new boson, conventionally referred to as $X_{17}$.

Despite these intriguing developments, the global experimental landscape remains highly dynamic, particularly in light of recent MEG~II constraints, and a definitive confirmation or exclusion of the $X_{17}$ hypothesis is still lacking.

Within this evolving and exciting context, this thesis investigates whether a hypothetical $17\,\text{MeV}$ particle, coupled to electrons as suggested by the PADME observations, could function as a mediator between the Standard Model and previously unexplored hidden sectors. Such a mediator could, in principle, offer a novel pathway toward addressing one of the principal outstanding inconsistencies of the Standard Model: the nature and origin of dark matter.

\tableofcontents


\mainmatter
\chapter{Introduction and Motivation}
\label{cap:MotAndResObj}

Recently, a series of experiments, most notably those carried out by ATOMKI \cite{Krasznahorkay:2015iga,Krasznahorkay:2021joi,Krasznahorkay:2024universe} and PADME \cite{Bossi:2025ptv}, have reported tensions that might hint at a new resonance with a mass around $17\,\text{MeV}$,  possibly interacting with nucleons and electrons. At present, however, the experimental situation remains highly dynamic, also in light of the recent MEG~II result \cite{MEGII:2024urz}, and a definitive confirmation or exclusion of the $X_{17}$ hypothesis is therefore still lacking.

Within this evolving and exciting context, we aim to address in this thesis the following question: \emph{could a hypothetical $17\,{\rm{ MeV}}$ particle, coupled to electrons as suggested by the PADME observations, serve as a mediator describing interactions between the Standard Model and new hidden sectors?}

\medskip
In this chapter we will introduce the background of this thesis, reviewing the particle content of the Standard Model, summarizing the experimental evidence supporting the existence of dark matter, and examining the current status of the $17\,\text{MeV}$ anomalies in light of both existing data and forthcoming experimental efforts. This overview chapter summarizes the context and motivates the analyses presented in the rest of the thesis.

\section{The Standard Model of Particle Physics}
\label{sec:SM}

The Standard Model (SM) of particle physics provides a remarkably successful
description of the fundamental constituents of matter and of their interactions.
It is a quantum field theory based on the gauge symmetry group
\begin{equation}
\mathcal{G_{SM}}\equiv SU(3)_C \times SU(2)_L \times U(1)_Y \, .
\label{eq:Gsm}
\end{equation}
We now briefly review the main features of the SM that will be used as the theoretical basis for the developments discussed in this thesis.

The SM Lagrangian can be decomposed into three main sectors: the fermionic sector, the gauge sector, and the Higgs sector.

The first SM sector consists of fermions, arranged in three
replicated generations.
Each generation contains two quarks and two leptons, differing only by their
masses and mixing patterns.
Quarks carry color charge and therefore participate in strong interactions,
while leptons are color singlets. The SM fermions are chiral, meaning that left–handed fields transform as doublets under $SU(2)_L$, while right–handed fields are singlets.
For one generation, the fermionic content can be written as
\begin{equation}
Q_L = \begin{pmatrix} u_L \\ d_L \end{pmatrix}, \quad
L_L = \begin{pmatrix} \nu_L \\ e_L \end{pmatrix}, \quad
u_R,\ d_R,\ e_R \, ,
\end{equation}
where $Q_L$ and $L_L$ denote the quark and lepton weak doublets, respectively. In the SM, right–handed neutrinos are absent and neutrinos are exactly massless.

The gauge representations and hypercharges of the SM fermion fields are summarized in tab.~\ref{tab:FerQN}. 

\begin{table}[h]
    \centering
    \renewcommand{\arraystretch}{1.3}
    \begin{tabular}{c|c c c}
        \hline
        & $SU(3)_C$ & $SU(2)_L$ & $U(1)_Y$ \\
        \hline
        \hline
        $Q_L$   & $\mathbf{3}$ & $\mathbf{2}$ & $+1/6$ \\
        $u_R$   & $\mathbf{3}$ & $\mathbf{1}$ & $+2/3$ \\
        $d_R$   & $\mathbf{3}$ & $\mathbf{1}$ & $-1/3$ \\
        \hline
        \hline
        $L_L$   & $\mathbf{1}$ & $\mathbf{2}$ & $-1/2$ \\
        $\ell_R$& $\mathbf{1}$ & $\mathbf{1}$ & $-1$   \\
        \hline
    \end{tabular}
    \caption{Gauge quantum numbers of one fermion generation in the SM under 
    $SU(3)_C\times SU(2)_L\times U(1)_Y$.}
    \label{tab:FerQN}
\end{table}

The interactions among SM particles are mediated by gauge bosons associated
with the gauge symmetries.
The $SU(3)_C$ group describes quantum chromodynamics (QCD) and is mediated by
eight massless gluons.
The electroweak sector is governed by $SU(2)_L \times U(1)_Y$ and involves four
gauge fields, which after electroweak symmetry breaking (EWSB) give rise to the massive $W^\pm$ and $Z$ bosons, as well as the massless photon.

A central ingredient of the SM is the Higgs mechanism, responsible for EWSB.
The Higgs sector contains a single complex scalar doublet $H\sim(\mathbf{1},\mathbf{2},1/2)$, with potential
\begin{equation}
V(H) = -\mu^2 |H|^2 + \lambda |H|^4 \, .
\end{equation}
When $\mu^2 > 0$, the Higgs field acquires a non–vanishing vacuum expectation value (vev),
defined as
\begin{equation}
\langle H \rangle = \frac{1}{\sqrt{2}}
\begin{pmatrix}
0 \\ v_h
\end{pmatrix},
\qquad
v_h \equiv \sqrt{\frac{\mu^2}{\lambda}} \simeq 246~\mathrm{GeV},
\end{equation}
which triggers the spontaneous breaking of the electroweak gauge symmetry
$SU(2)_L \times U(1)_Y$ down to the electromagnetic subgroup $U(1)_{\rm em}$.
As a consequence, the $W^\pm$ and $Z$ bosons acquire masses, while the photon
remains massless.
The Higgs mechanism also generates fermion masses through Yukawa interactions
of the form
\begin{equation}
\mathcal{L}_Y = - y\, \bar{\psi}_{L} H \psi_{R} + \mathrm{h.c.}\;,
\end{equation}
where $y$ are dimensionless Yukawa couplings.
After EWSB, fermion masses are given by $m= y \,v_h/\sqrt{2}$.

The SM successfully describes a vast range of phenomena,
nevertheless, it is widely regarded as an incomplete theory.
It does not account for several well–established observations, including the
existence of Dark Matter (DM), the small but nonzero neutrino masses, the matter–antimatter asymmetry of the Universe, and the origin of gravity.
In particular, the absence of viable DM candidates within the SM strongly points to new physics, possibly indicating the existence of an additional sector, commonly referred to as a \emph{dark sector}.

This observation forms the conceptual basis for the models explored in this thesis, where additional degrees of freedom are introduced while preserving the SM gauge structure at low energies.

\section{Dark Matter}
\label{sec:DM}

Observations of astrophysical systems over a wide range of length scales, from individual galaxies to the largest structures in the universe, indicate the presence of a non–luminous matter component, commonly referred to as DM.
In this section we briefly summarize the main observational arguments that historically led to the formulation of the  DM hypothesis.
We will then review the general properties that DM must satisfy in order to be
consistent with cosmological and astrophysical constraints, setting the basis for the specific framework adopted in this thesis.

\subsection{Evidences of Dark Matter}
\begin{figure}[t]
    \centering
    \begin{minipage}{0.46\linewidth}
        \centering
        \includegraphics[width=0.95\linewidth]{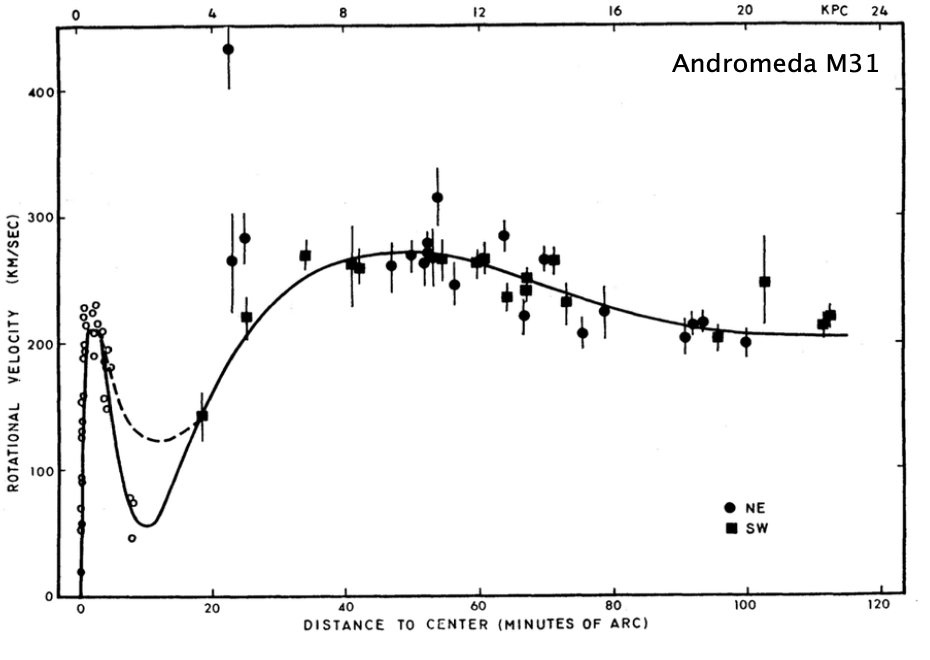}
    \end{minipage}
    \hfill
    \begin{minipage}{0.51\linewidth}
        \centering
        \includegraphics[width=0.95\linewidth]{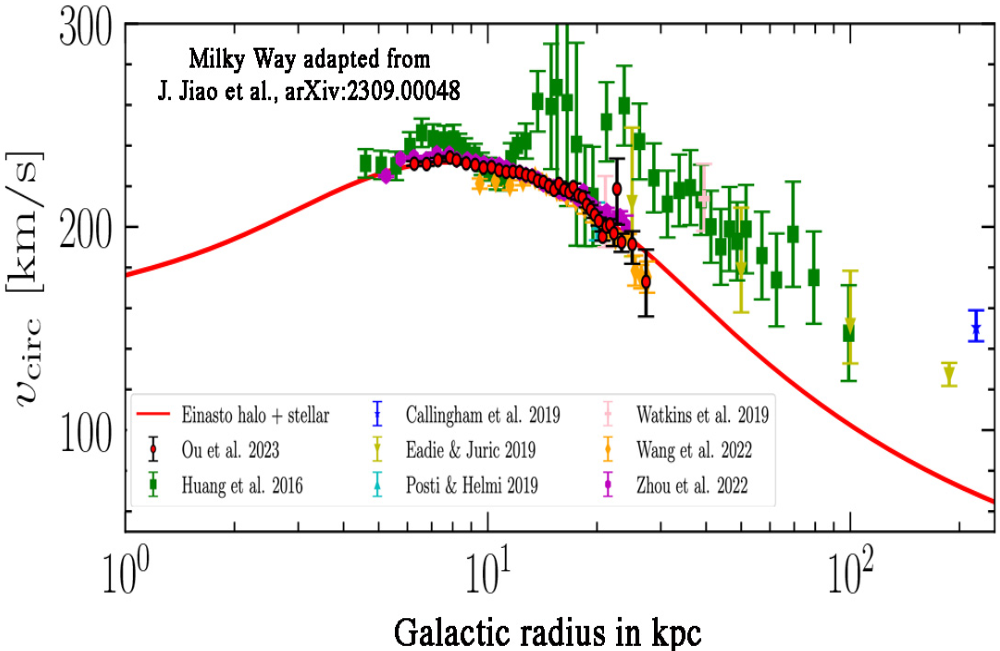}
    \end{minipage}
    \caption{Rotation curves of spiral galaxies taken from \cite{Cirelli:2024ssz}.
    On the left, the original rotation curve of Andromeda by Rubin and Ford (1970).
    On the right, a recent rotation curve for the Milky Way adapted from \cite{refId0}. Both are plotted in function of the distance with respect to the center of the galaxy}
    \label{fig:RotCur}
\end{figure}

Spiral galaxies, including the Milky Way, exhibit rotational motion around their symmetry axis.
The corresponding tangential velocity profile can be experimentally reconstructed through Doppler measurements of spectral lines emitted by stars and gas clouds. An historical and a recent example are plotted in fig. \ref{fig:RotCur}.

The tangential velocity ($v_{tan}$) of a test particle $m$ (e.g., a star) at a distance $r$ from the galactic center can be obtained from Newtonian dynamics.
Equating the centripetal force to the gravitational attraction generated by the enclosed mass ($M(r)$), one finds
\begin{equation}
    m\,\frac{v^2_{\mathrm{tan}}(r)}{r}
    =
    \frac{G\,m\,M(r)}{r^2}
    \qquad \Rightarrow \qquad
    v_{\mathrm{tan}}(r)
    =
    \sqrt{\frac{G\,M(r)}{r}} \, ,
    \label{eq:vtan}
\end{equation}
where $G$ denotes Newton’s gravitational constant. In spiral galaxies, the luminous matter is predominantly concentrated in the central bulge and in the disk.
As a consequence, beyond a characteristic radius of order several tens of kiloparsecs, the enclosed mass is expected to approach a constant value, $M(r)\to M$.
In this regime, eq. \eqref{eq:vtan} predicts a Keplerian decrease of the tangential velocity, $v_{\mathrm{tan}}(r)\propto r^{-1/2}$.

Observationally, this behavior is not seen, as shown in fig. \ref{fig:RotCur}.
Measurements suggest that the tangential velocity remains approximately constant out to the largest probed radii, leading to so-called flat rotation curves.

To maintain a flat rotation curve, additional gravitating matter must be present at large radii.
This motivates the hypothesis that galaxies are embedded in extended DM halos.
From eq. \eqref{eq:vtan}, it follows that a flat rotation curve requires the enclosed mass to grow linearly with radius, $M(r)\propto r$.
This behavior can be achieved if the halo density profile scales as $\rho(r)\propto \frac{1}{r^2}$, since
\begin{equation}
    M(r)
    =
    4\pi \int_0^r dr'\, r'^2\,\rho(r')
    \propto 4\pi r
    \qquad \Rightarrow \qquad
    v_{\mathrm{tan}}(r)
    =
    \sqrt{\frac{G\,M(r)}{r}}
    \simeq \mathrm{const}.
\end{equation}
At sufficiently large distances from the galactic center, the DM halo is expected to become less dense, and the rotation curve eventually starts to decline.
This qualitative behavior is supported by observations of large samples of spiral galaxies, as discussed in detail in ref.~\cite{Cirelli:2024ssz}.

Galaxy clusters are the largest gravitationally bound systems in the Universe, and their
dynamics provide an important probe of the total matter content on large scales.

One of the earliest and most influential indications of missing mass in galaxy clusters
was obtained by Fritz Zwicky in 1933 through his study of the Coma Cluster.
By measuring the radial velocities of cluster member galaxies, Zwicky observed velocity
dispersions of order $v \sim 2000~\mathrm{km/s}$, significantly larger than what could be
accounted for by the luminous matter alone.
While similar discrepancies had already been noted by Hubble and Humason, Zwicky was the
first to interpret them quantitatively by applying the virial theorem.

For a system of $N$ galaxies of mass $m$ bound by gravity, the virial theorem implies
\begin{equation}
\langle K \rangle = -\frac{1}{2}\langle V \rangle
\qquad \Longrightarrow \qquad
N\,\frac{m v^2}{2} = \frac{N^2 G m^2}{4R}
\Longrightarrow 
M \equiv Nm = \frac{2 R v^2}{G},
\end{equation}
where $R$ denotes the characteristic size of the cluster.
Inserting the observed values for the Coma Cluster, Zwicky found that the inferred total
mass exceeds the mass associated with luminous matter by more than an order of magnitude.

The observations discussed above represent only the earliest historical evidence
that motivated the introduction of DM.
Over the years, a wide range of independent and complementary observations has
further strengthened this hypothesis, including gravitational lensing,
cosmic microwave background anisotropies, large–scale structure formation,
and precision cosmology.
A detailed discussion of these additional probes goes beyond the scope of this
thesis and can be found in Ref.\,\cite{Cirelli:2024ssz}.

Having established the empirical motivation for the existence of DM, we now turn
to a discussion of its general properties, which will guide the construction and
analysis of the particle–physics models considered in this thesis.

\subsection{General Properties and Theoretical models of DM}
\label{sec:DMGP}
A first and fundamental requirement concerns the present-day DM density ($\rho_{DM}$).
In cosmology, the quantity used to account for $\rho_{DM}$ is the parameters combination $\Omega_{\rm DM} = \rho_{\rm DM}/\rho_{\rm cr}$, where $\rho_{cr} =3H_0^2/8\pi G$ is the critical density of the universe. Defining the parameter $h$ as $H_0 = h \times 100 \,\text{km}\,\text{s}^{-1}\,\text{Mpc}^{-1}$, the explicit dependence of the Hubble constant $H_0$ may be removed, hence hereafter we will consider directly the quantity $\Omega_{\rm DM}\,h^2$. This observable is known with extreme precision through the measurement of the peaks of the cosmic microwave background (CMB) by the Planck Collaboration \cite{Planck:2018vyg}
\begin{equation}
    \Omega_{DM} h^2=0.1200 \pm 0.0012\,.
\end{equation}

At present, the mass of DM is essentially unconstrained.
Both theoretical arguments and observational considerations allow it to span an
enormous range, covering roughly ninety orders of magnitude. Constraints however can nonetheless be derived by considering generic astrophysical arguments.

A lower bound on the DM mass follows from the requirement that its de Broglie wavelength fits inside dwarf galaxies, leading to (see \cite{Zimmermann:2024xvd})
\begin{equation}
    m_{DM} \gtrsim 10^{-21}\,\mathrm{eV}.
\end{equation}
In principle a particle cannot be heavier than $M_{Pl}$, otherwise it would be a black hole. However, Dark matter could be composed of composite compact objects. 
In this case, the relevant requirement is that the constituents of such objects are not themselves black holes. 
A simple heuristic estimate reported in \cite{Cirelli:2024ssz} is that their mass should not exceed that of the smallest dwarf galaxies
\begin{equation}
    m \lesssim 10^{4}\,M_\odot \simeq 10^{34}\,\mathrm{kg}\,,
\end{equation}
and this gives an upper bound on DM mass.

\begin{figure}[t]
    \centering
    \vspace*{0.25cm}
    \includegraphics[width=0.9\linewidth]{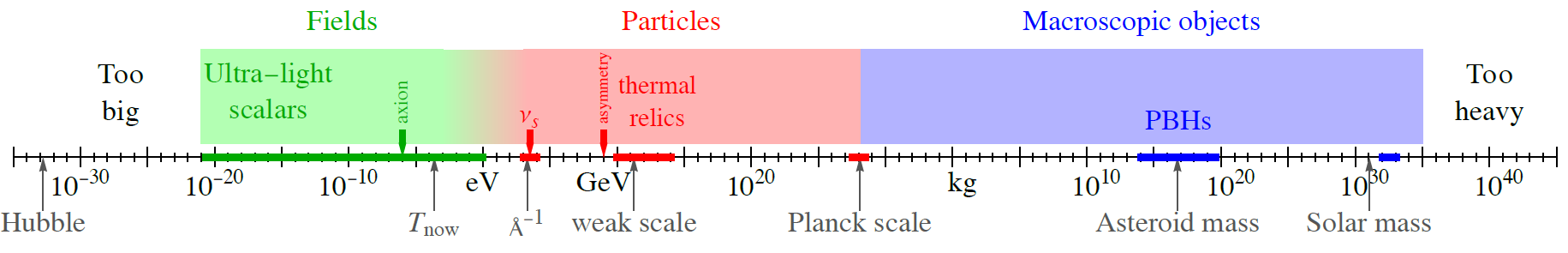}
    \caption{Possible mass range for  DM candidates. In this work, we focus specifically on the MeV region.} 
    \label{fig:DMMasRan}
\end{figure}

Depending on its mass, DM can therefore exhibit qualitatively different physical
behaviors, motivating the classification of DM candidates into three distinct regimes,
as illustrated in fig. \ref{fig:DMMasRan}: fields, particles and macroscopic objects.

\begin{enumerate}
    \item \textbf{Fields}: if the  DM mass is $M_{DM} \ll \text{eV}$, it will behave like a classical field.\\
    This is analogous to the electromagnetic field: for such light  DM, the number of particles required to account for the observed gravitational effects is enormous, so—as in the case of many photons—it is more convenient to describe it as a classical field.
    \item \textbf{Particles}: in this case, we are in the range $\text{eV} < m_{DM} < M_{\text{Pl}}$. The upper limit is motivated by the fact that, for masses larger than the Planck scale, an object would collapse and should be regarded as a black hole rather than as an elementary particle.

    \item \textbf{Macroscopic objects}: in this regime the DM is not described in terms of
    elementary particles, but rather as composite objects with masses exceeding the
    Planck scale, $m_{\rm DM} > M_{\rm Pl}$.
    Such objects are intrinsically macroscopic and behave as classical gravitating bodies.
    Examples that are often discussed in the literature include Massive Astrophysical
    Compact Halo Objects (MACHOs) and primordial black holes, among other possibilities.
\end{enumerate}

In this thesis we focus on the second hypothesis, namely that DM is composed of elementary particles.
Under this assumption, a viable DM particle candidate must satisfy a number of general requirements, which can be summarized as follows:

\begin{itemize}
    \item Particle DM must be \emph{cold}, meaning that it is non–relativistic at the time of matter–radiation equality.
    This condition is required in order for DM to efficiently cluster and drive the formation of large–scale structures.
    
    \item DM must be electrically neutral, or carry at most an extremely small
    electric charge.
    Strong bounds from cosmology, astrophysics and laboratory experiments constrain any possible electromagnetic interaction to be many orders of magnitude weaker than ordinary SM interactions.

    \item The interaction strength of DM with SM particles must be sufficiently suppressed.
    In particular, experimental bounds imply that the elastic scattering cross section of DM on ordinary matter is well below a typical weak–scale cross section over a wide range of masses.
    This requirement severely restricts the allowed range of couplings between DM and SM fields.
    
    \item DM self–interactions must also be limited.
    Large self–interaction cross sections would modify the dynamical behavior of DM in galaxies and clusters, in conflict with observations that indicate that DM behaves, to a good approximation, as a collisionless component on astrophysical scales.

    \item Finally, DM must be stable on cosmological timescales, or at least it's lifetime ($\tau_{DM}$) must be greater than the age of the universe $T_U \approx13.8\,\text{Gyr}$.
    If DM were unstable, its decay products would leave observable imprints in cosmological and astrophysical data, which are not seen.
    This requirement is often enforced by imposing an exact or approximate symmetry that forbids DM decay.

\end{itemize}

It is worth commenting on the allowed mass range for fermionic DM.
Unlike bosonic candidates, fermionic DM is subject to the Pauli exclusion principle, which imposes the stronger, well–known Gunn-Tremaine bound \cite{PhysRevLett.42.407,10.1093/mnras/staa3640}.
For the Milky Way halo, this implies a lower limit of order
\begin{equation}
m_{\rm DM} \gtrsim \mathcal{O}(0.1\text{-}1)\,\mathrm{keV}.
\end{equation}
For the purposes of this thesis, and in order to keep the phenomenological analysis as simple and transparent as possible, we will restrict ourselves to DM masses in the range from the MeV to the GeV scale.
This choice is primarily motivated by the experimental anomalies and detection strategies discussed in the following sections.

\subsection{Dark Matter Constraints and Hints of New  Physics Resonances}
The above general considerations define the theoretical framework within which
particle DM models are constructed and will guide the model-building
strategy and the phenomenological analyses developed in the following chapters.
In particular, they motivate the study of minimal and predictive extensions of
the SM that can provide a {\it viable DM candidate} while remaining consistent with the wide range of existing experimental constraints.

\begin{figure}[t]
    \centering
    \includegraphics[width=0.55\linewidth]{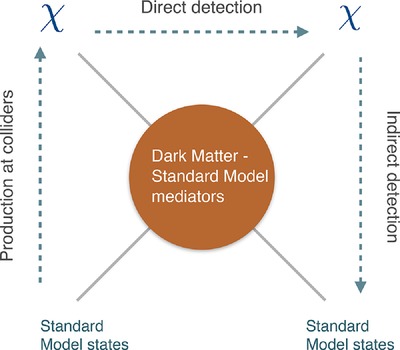}
    \caption{Schematic representation of the three main experimental strategies to probe particle  DM: indirect detection, direct detection, and collider searches. They correspond respectively to SM final states from DM, DM–SM scattering, and DM production from SM collisions.The symbol $\chi$ denotes the dark–matter particle. Image taken from \cite{https://doi.org/10.1002/andp.201500114}.}
    \label{fig:DMtriad}
\end{figure}

Processes involving DM and SM particles can be grouped into three broad phenomenological categories, according to the experimental strategy that is sensitive to them. Indirect-detection (ID) experiments search for SM particles produced in DM annihilations or decays in astrophysical environments. Direct-detection experiments instead aim at observing the effect of DM particles coming from the galactic halo scattering off SM targets. Finally, collider searches look for the production of  DM in high-energy collisions of SM particles, typically through missing-energy signatures.
These three classes of processes, and their connection through the DM–SM mediator, are schematically summarized in fig. \ref{fig:DMtriad}.

In this broader context, it is also useful to consider experimental hints
that may point at physics beyond the SM and that may be properly connected to DM physics.
Among these, particular attention has been drawn to the so-called $X_{17}$ anomaly,
which if confirmed is a new physics particles with light mass.
In the next section, we summarize the current experimental status of this $X_{17}$ anomaly,
and discuss its possible theoretical implications within the framework outlined above.

\section{Hints of a Hypothetical Resonance at 17 MeV, $X_{17}$ Boson}
\label{sec:X17}

Over the last decade, a series of intriguing anomalies observed by the ATOMKI collaboration in nuclear transition experiments has sparked considerable interest in the possibility of new physics at the MeV scale.
The anomalies were first reported in processes associated with excited nuclear states of ${}^{8}\mathrm{Be}$, and were later reinforced by similar observations in transitions involving ${}^{4}\mathrm{He}$ and ${}^{12}\mathrm{C}$.

In this section we will first review these experimental results and discuss their possible interpretation in terms of a new light neutral boson, commonly referred to as $X_{17}$.
We will then revisit how nuclear transition properties can be used to assess which combinations of spin and parity are more compatible with the available data.
Finally, we will then turn to a putative 17 MeV resonance hinted by PADME experiment  at 1.7$\sigma$ via  $e^+e^-$ annihilation.

\subsection{Nuclear Transitions and Spin-parity Selection Rules}

Since 2020, anomalous features have been reported in the angular distributions of electron-positron pairs emitted in nuclear transitions by the ATOMKI collaboration.
These anomalies were first observed in internal pair creation (IPC) processes associated with excited states of ${}^8\mathrm{Be}$ \cite{Krasznahorkay:2015iga}.
Subsequently, similar deviations from SM expectations were reported in nuclear transitions of ${}^4\mathrm{He}$\cite{Krasznahorkay:2021joi} and ${}^{12}\mathrm{C}$\cite{Krasznahorkay:2024universe}.
In all cases, the anomaly manifests itself as an excess of $e^+e^-$ pairs at large opening angles between the electron and the positron.
Such a behavior cannot be accounted for by conventional IPC alone and suggests the presence of an intermediate particle emitted in the nuclear transition and decaying promptly into an electron-positron pair.
The data can be consistently interpreted by postulating the emission of a new, light, electrically neutral boson with a mass of about $17\,\mathrm{MeV}$, commonly referred to as $X_{17}$.
In this interpretation, the nuclear transition proceeds through the emission of the $X_{17}$, followed by the decay $X_{17}\to e^+e^-$, leading to a characteristic angular correlation that matches the observed excess.

Given the set of nuclear anomalies reported by the ATOMKI collaboration, a natural question concerns the spin and parity quantum numbers of the hypothetical $X_{17}$ boson.
Once the mass scale is fixed by kinematics, the consistency of the interpretation hinges on whether the observed nuclear transitions are dynamically allowed for a given spin-parity assignment of $X_{17}$.
For each nuclear transition, angular momentum and parity conservation impose selection rules.
These constraints determine whether the emission of an intermediate boson with given spin $J$ and parity $\pi$ is allowed, and, if so, which values of the orbital angular momentum $L$ between the final nuclear state and the emitted boson are permitted.
In this way, each anomaly can be mapped onto a set of allowed or forbidden spin-parity assignments for $X_{17}$.

The results of this analysis are summarized in table \ref{tab:spin_parity}, where we report, for each nuclear transition involved in the ATOMKI anomalies, the allowed values of the orbital angular momentum $L$ associated with the emission of an $X_{17}$ boson with definite spin and parity.
Entries marked as "/" correspond to forbidden transitions.
\begin{table}[t]
    \centering
    \begin{tabular}{l|cccc}
        \hline
        Process & $S^\pi = 1^{-}$ & $S^\pi = 1^{+}$ & $S^\pi = 0^{-}$ & $S^\pi = 0^{+}$ \\
        \hline
        \hline
        ${}^{8}\mathrm{Be}(18.15) \to {}^{8}\mathrm{Be}$ & 1     & $0,\,2$ & 1   & / \\
        ${}^{8}\mathrm{Be}(17.64) \to {}^{8}\mathrm{Be}$ & 1     & $0,\,2$ & 1   & / \\
        \hline
        \hline
        ${}^{4}\mathrm{He}(21.01) \to {}^{4}\mathrm{He}$ & /     & 1       & 0   & / \\
        ${}^{4}\mathrm{He}(20.21) \to {}^{4}\mathrm{He}$ & 1     & /       & /   & 0 \\
        \hline
        \hline
        ${}^{12}\mathrm{C}(17.23) \to {}^{12}\mathrm{C}$ & $0,\,2$ & 1     & /   & 1 \\
        \hline
    \end{tabular}
    \caption{Allowed values of the orbital angular momentum $L$ of the emitted $X$ boson
    for different choices of its spin-parity $S^\pi$. The symbol "/" corresponds to forbidden transitions, excluded by angular momentum and parity conservation}. 
    \label{tab:spin_parity}
\end{table}
From the table, two robust conclusions can be drawn.
First, the beryllium transitions exclude a CP-even scalar interpretation of $X_{17}$, as no value of the orbital angular momentum satisfies the required selection rules.
Second, the carbon transition rules out a pseudoscalar assignment, since the corresponding emission is forbidden by parity conservation.
As a consequence, not all spin-parity hypotheses are compatible with the full set of observed anomalies.
These considerations significantly restrict the viable quantum numbers of $X_{17}$ and play a central role in guiding the theoretical interpretation of the ATOMKI results.

At this point it is useful to add a brief remark motivated by recent experimental activity.
The MEG-II collaboration has investigated the same excited ${}^{8}\mathrm{Be}$ transitions that, according to table \ref{tab:spin_parity}, would exclude a CP-even scalar interpretation of $X_{17}$.
No statistically significant signal was observed in this channel.
This result has prompted a renewed discussion in the literature, and in particular a re-assessment of whether a scalar $X$ may still be viable once theoretical uncertainties and modelling assumptions entering the nuclear--transition analysis are taken into account, see ref. \cite{Barducci:2025hpg}.

\subsection{A 17 MeV Particle Coupled to electrons and PADME hints}

An important ingredient of the phenomenological analysis developed in this thesis is the coupling of the $X_{17}$ boson to electrons ($g_e$), appearing in a low-energy Lagrangian as
\begin{equation}
    \mathcal{L}\,\supset \,g_e \,X\, \bar{e}\,e\,.
\end{equation}
The ATOMKI experiment, however, does not provide a direct measurement of $g_e$.
The observable signal at ATOMKI arises from nuclear transitions of the form $p + \text{Target} \to N^\ast \to N + e^+ e^-$, where $N={}^{8}\mathrm{Be},{}^{4}\mathrm{He},{}^{12}\mathrm{C}$.
If the $e^+e^-$ pair originates from the decay of an on-shell intermediate boson $X_{17}$, the measured signal rate is proportional to
\begin{equation}
\text{Signal\,Rate}\;\propto\;\sigma(p + \text{Target} \to N^\ast)\,\mathrm{BR}(X_{17}\to e^+e^-)\,.   
\end{equation}

As a consequence, the experimental sensitivity is only to the branching ratio of the decay into electron-positron pairs, and not to the electron coupling itself.
Under the assumption that $X_{17}$ decays exclusively into $e^+e^-$, one has $\mathrm{BR}(X_{17}\to e^+e^-)\simeq 1$, and the explicit dependence on $g_e$ is lost.

An independent probe of the $X_{17}$ hypothesis is provided by the PADME experiment.
PADME is a fixed-target experiment in which a positron beam impinges on a thin diamond target.
In this setup, new light bosons can be searched for through their direct production in purely leptonic processes.
Unlike nuclear experiments, this approach does not rely on nuclear transitions and provides direct sensitivity to the coupling of $X_{17}$ to electrons.

By varying the energy of the incident positron beam, PADME has scanned the center-of-mass energy range $14 \lesssim \sqrt{s} \lesssim 23~\mathrm{MeV}$.
This interval includes the region around $\sqrt{s}\simeq 17~\mathrm{MeV}$, where the production of an $X_{17}$ boson is expected.
Within this energy window, PADME has reported a mild excess in the $e^+e^-$ channel corresponding to an invariant mass around $m_X \simeq 16.90~\mathrm{MeV}$.
While the statistical significance of this fluctuation is limited, it is compatible with the mass scale suggested by the ATOMKI anomalies.
Interpreting this excess under the assumption that $X$ is a vector mediator, PADME performed a fit to the data and extracted a preferred value for the electron coupling,
$g_{e} \simeq 5.6 \times 10^{-4}$ .
This result provides an additional, purely leptonic indication that is consistent with the $X_{17}$ hypothesis and complements the constraints derived from nuclear transitions.

\medskip
This overview sets the stage for the theoretical framework developed in the following chapter, where the $X_{17}$ boson will be embedded into a consistent particle-physics model and its implications for  DM phenomenology will be explored. 
We also highlight that in cap \ref{cap:DMExpSig} we will place our results in a broader context by reviewing the comprehensive analysis of ref. \cite{DiLuzio:2025ojt} and by comparing them with a wide set of laboratory constraints on light bosons coupled to electrons.

Before moving to the next chapter, it is worth noting that possible connections between $X_{17}$ and the dark sector have already been explored in the literature. For instance, in Refs. \cite{Alves:2023ree,Barman:2021yaz} $X_{17}$ is interpreted as a dark photon and its role as a portal between DM and the SM sector is analyzed in detail. The work in this thesis, however, is highly different from this set-up for two main reasons: $i)$ we will explicitly consider $X_{17}$ as a scalar boson; $ii)$ we will not forbid direct interactions among SM, DM and the $X_{17}$. The second item implies, as we will see, that particular care will be devoted to ensuring the requirement of stability of the DM in our framework.

\chapter{Theoretical Framework Of Our Model}
\label{cap:TheFra}

Here begins the original contribution of this thesis.  In this chapter we will introduce the theoretical setup that will be used throughout the rest of the thesis.  
Starting from the SM field content and gauge symmetries, we will explore the possibility that $X_{17}$-boson may mediate interactions between SM and DM particles.
We define the relevant interaction Lagrangian, discuss the assumptions underlying the effective description, and identify the couplings that will control the phenomenology studied in the following chapters.  

The dark sector is assumed to be minimal and consists of a real scalar field
$X$ and a fermionic vector-like DM particle $\chi$.  While
our original motivation arises from anomalies observed at an energy of
approximately $17\,\mathrm{MeV}$,  we will present results for a generic
mediator $X$ with mass $m_X$, allowing our analysis to extend beyond the specific
$X_{17}$ case.

In addition, we extend the SM leptonic sector by introducing a right–handed Majorana neutrino $\nu_R$, singlet under the SM gauge group. 
The DM particle $\chi$ is taken to be a Majorana right-handed fermion, and can either coincide with $\nu_R$ or represent an independent field.

In this chapter we construct the effective interaction Lagrangian describing the couplings beyond the SM introduced by the new degrees of freedom.
Working in an effective field theory framework, we classify all gauge–invariant operators under the SM gauge group $SU(3)_C\times SU(2)_L\times U(1)_Y$ up to dimension six. 
The resulting operator basis is shown in Tables \ref{tab:oper1+}–\ref{tab:oper2-} and provides the starting point for the diagrammatic analysis carried out in Sec. \ref{sec:DiagEva}.

\section{The Effective Interaction Lagrangian}
 
In this section we aim at building up a minimal model, characterized by two key interactions involving the new mediator $X$.
The first is a coupling of $X$ to electrons, motivated directly by the experimental anomalies observed at PADME and associated with the $X_{17}$ hypothesis; the second is a coupling of $X$ to DM. In this way, the same mediator can interact with SM particles only, with both SM and DM particles or, finally, with only DM (in this last case, $X$ can be thus thought as a portal to the dark sector).
Additional interactions, such as the coupling to neutrinos or to the Higgs-boson, appear as natural byproducts of gauge invariance, and will be treated with care. In particular, we must ensure that they do not introduce additional couplings between DM and other SM fields, which may spoil DM stability.

We focus now on the interaction with the electrons (and positrons).
At low energies, this interaction can be parametrized in a model–independent way as
\begin{equation}
\label{Xeecoupl}
  \mathcal{L}_{X,e}^{\rm SM}
  = g_e\,(\bar e\,\Gamma_X e)\,X + \ldots,
\end{equation}
where, in general, $\Gamma_X = 1,\,\gamma^{5}$ or $\gamma^\mu,\gamma^\mu\gamma^{5}$ depending on whether $X$ is a scalar, pseudoscalar or (axial–)vector state. Pure vector couplings are strongly disfavoured by current data \cite{DiLuzio:2025ojt}.

We now turn to the coupling between $X$ and DM. 
At low energies, this interaction can be parametrized as
\begin{equation}
\label{XDMDMcoupl}
  \mathcal{L}_{X,\chi}^{\rm DM}
  = g_\chi\,(\bar\chi\,\Gamma_\chi \chi)\,X+ \ldots,
\end{equation}
where $\chi$ is a SM–singlet fermion and $\Gamma_\chi$ denotes the Lorentz structure of the coupling.

In addition to the electron and DM couplings, an interaction of $X$ with neutrinos is also phenomenologically interesting and arises naturally in our setup. 
At low energies, a generic coupling to SM neutrinos and DM takes the form:
\begin{equation}
\label{XDMnucoupl}
  \mathcal{L}_{X,\chi,\nu}^{\rm DM}
  = g_{\chi\nu}\,(\bar\chi\,\Gamma_{\chi\nu} \nu)\,X + \ldots\;.
\end{equation}

Our goal for this section is, then, to derive the set of operators that are invariant under the SM gauge group $SU(3)_C\times SU(2)_L\times U(1)_Y$ and that reproduce the low-energy couplings of $X_{17}$ to electrons as in eq.(\ref{Xeecoupl}), as well as its interactions with the rest of the visible and dark sectors, see eqs.\,(\ref{XDMDMcoupl}-\ref{XDMnucoupl}).

\subsection{Gauge-invariant Interactions Classification}
\label{sec:BSMOpe}

We collect all the allowed operators in four tables.
The first two, Tables \ref{tab:oper1+} and \ref{tab:oper2+}, correspond to the case in which $X$ is a scalar ($0^{++}$), while Tables \ref{tab:oper1-} and \ref{tab:oper2-} correspond to the pseudoscalar case ($0^{+-}$).
Tables \ref{tab:oper1+} and \ref{tab:oper1-} describe the scenario in which the right–handed neutrino $\nu_R$ and the dark fermion $\chi$ are independent fields ($\chi \neq \nu_R$), whereas Tables \ref{tab:oper2+} and \ref{tab:oper2-} refer to the case in which they coincide and $\chi$ plays the role of a Majorana right–handed neutrino ($\chi = \nu_R$).

The first two columns of each of the four Tables specify the gauge quantum numbers of $X$ and of the dark fermion $\chi$ under the SM gauge group. 
We adopt the compact notation
\begin{equation*}
(\,d_{SU(3)},\,d_{SU(2)},\,Y\,)
\end{equation*}
to denote the dimensions of the representation under $SU(3)_C$, $SU(2)_L$ and the hypercharge $Y$ under $U(1)_Y$.
\footnote{We always take both $X$ and $\chi$ to be color-singlets, i.e. $d_{SU(3)}=1$. Allowing them to transform non–trivially under $SU(3)_C$ would lead to a much richer (and more constrained) phenomenology, which lies beyond the scope of this thesis.}
Thus, it is clear that each row in the Tables corresponds to a different theoretical model.

All the remaining columns of Tables \ref{tab:oper1+}-\ref{tab:oper2-} show the new interactions beyond the SM that are allowed for each choice of gauge quantum numbers. 
They are grouped according to their field content.

The terms collected under $\mathcal{L}_{X,\phi}^{\rm SM}$, with $\phi \in \{e, H, B_\mu, W^a_\mu\}$, are couplings of $X$ to SM fields.
In particular, they include the interactions with electrons and positrons, as well as couplings to the Higgs doublet and to the electroweak gauge bosons, which in turn may induce interactions with photons after electroweak symmetry breaking. 
All such couplings of $X$ with the SM are allowed by gauge invariance. However, for the purposes of this Thesis, it is sufficient to focus only on $\mathcal{L}_{X,e}^{\rm SM}$ hereafter.

The last two columns, labeled $\mathcal{L}_{X}^{\rm DM}$ and $\mathcal{L}_{X,SM}^{\rm DM}$, contain instead the interactions involving the dark fermion $\chi$. 
The operators in $\mathcal{L}_{X}^{\rm DM}$ describe the desired portal couplings between $\chi$ and $X$, which, as we will see in chapter \ref{cap:DMAbCalcu}, are responsible for production of DM in the early Universe. 
On the contrary, the terms in $\mathcal{L}_{X,SM}^{\rm DM}$ involve both DM and SM fields, and may or may not contain an explicit $X$. 
After electroweak symmetry breaking, these operators might induce effective direct couplings between DM and SM fields that are not purely mediated by the exchange of $X$. In particular, such terms can endanger the DM role of $\chi$ (for instance by inducing fast decays). For this reason, the operators in $\mathcal{L}_{X,SM}^{\rm DM}$ must be treated with care.

\begin{table}[t]
    \centering
    \begin{adjustbox}{max width=\textwidth}
    \begin{tabular}{c|c|c|c|c|c|c|c}
    
        $X(0^{++})$ & $\chi\neq\nu_R$ & $\mathcal{L}_{X,e}^{\rm SM}$ & $\mathcal{L}_{X,H}^{\rm SM}$ & $\mathcal{L}_{X,B}^{\rm SM}$ & $\mathcal{L}_{X,W}^{\rm SM}$ & $\mathcal{L}_{X}^{\rm DM}$ & $\mathcal{L}_{X,SM}^{\rm DM}$\\
        \hline
        
        (1,1,0) & (1,1,0) & 
        \begin{tabular}[t]{@{}c@{}} 
        $\dfrac{1}{\Lambda}X\,\bar{\ell} H e_R$ 
        \end{tabular} &
        \begin{tabular}[t]{@{}c@{}} 
            $XXH^{\dagger}H$ \\ 
            $\lambda \,XH^{\dagger}H$ 
        \end{tabular} &
        $\dfrac{1}{\Lambda} \, X \, B_{\mu\nu} B^{\mu\nu}$ &
        $\dfrac{1}{\Lambda} \, X \, W^a_{\mu\nu} W^{a\,\mu\nu}$ &
        \begin{tabular}[t]{@{}c@{}} 
            $\bar{\chi} \chi X$ \\ 
            $\bar{\chi_L} \nu_R X$ \\
            $\bar{\nu_R^c} \nu_R X$
        \end{tabular} &  
        \begin{tabular}[t]{@{}c@{}} 
            $y \bar{\ell} \tilde{H} \chi_R$ \\ 
            $\lambda \,\bar\chi_L{\nu_R}$\\
            $\dfrac{1}{\Lambda}\,X\,\bar{\ell} \tilde{H} \chi_R$ \\
            $X\,\bar\chi_L{\nu_R}$
        \end{tabular} \\
        \hline
        
        (1,1,0) & (1,2,$\tfrac{1}{2}$) &
        $\dfrac{1}{\Lambda}\,X\,\bar{\ell} H e_R$ &
        \begin{tabular}[t]{@{}c@{}} 
            $XXH^{\dagger}H$ \\ 
            $\lambda \,XH^{\dagger}H$ 
        \end{tabular} &
        $\frac{1}{\Lambda} \, X \, B_{\mu\nu} B^{\mu\nu}$ &
        $\frac{1}{\Lambda} \, X \, W^a_{\mu\nu} W^{a\,\mu\nu}$ &
        $\bar{\chi} \chi X$ & 
        \begin{tabular}[t]{@{}c@{}} 
            $\bar{\chi_L}H \nu_R$\\
            $X\bar{\ell}\chi_L^c $\\
            $\lambda\bar{\ell}\chi_L^c $\\
            $\frac{1}{\Lambda}X\bar{\chi_L}H \nu_R$
        \end{tabular} \\
        \hline 
        
        (1,2,$\tfrac{1}{2}$) & (1,1,0) & 
        \begin{tabular}[t]{@{}c@{}} 
        $\bar{\ell} X e_R$ \\
        $\dfrac{1}{\Lambda} \left( \bar{\ell} X \right)\left( H^\dagger \ell \right)$
        \end{tabular} &  
        \begin{tabular}[t]{@{}c@{}}
        $(H^\dagger H)(X^\dagger X)$  \\ 
        $\lambda^2\,H^\dagger X$
        \end{tabular} &
        $\frac{1}{\Lambda^2} (X^\dagger H) B_{\mu\nu} B^{\mu\nu}$ &
        \begin{tabular}[t]{@{}c@{}}
        $\frac{1}{\Lambda^2} (X^\dagger H) W^a_{\mu\nu} W^{a\,\mu\nu}$\\
         $\frac{1}{\Lambda^2} (X^\dagger\frac{\sigma_a}{2} H) W^a_{\mu\nu}\, B^{\mu\nu}$
        \end{tabular} &
        \begin{tabular}[t]{@{}c@{}} 
            $\dfrac{1}{\Lambda} X^\dagger H \bar{\chi} \chi$ \\ 
            $\dfrac{1}{\Lambda} X^\dagger H \bar{\chi_L} \nu_R$ \\
            $\dfrac{1}{\Lambda} X^\dagger H \bar{\nu_R^c} \nu_R$
        \end{tabular} & 
        \begin{tabular}[t]{@{}c@{}} 
            $ \bar{\ell} \tilde{H} \chi_R$ \\ 
            $\bar{\ell} \tilde{X} \chi_R$ \\
            $\lambda\,\bar\chi_L{\nu_R}$\\
            $\frac{1}{\Lambda}X^\dagger H\,\bar\chi_L{\nu_R}$
        \end{tabular} \\
        \hline
        
        (1,2,$\tfrac{1}{2}$) & (1,2,$\tfrac{1}{2}$) & 
        \begin{tabular}[t]{@{}c@{}} 
        $\bar{\ell} X e_R$ \\
        $\dfrac{1}{\Lambda} \left( \bar{\ell} X \right)\left( H^\dagger \ell \right)$
        \end{tabular} &  
        \begin{tabular}[t]{@{}c@{}}
        $(H^\dagger H)(X^\dagger X)$  \\ 
        $\lambda^2\,H^\dagger X$
        \end{tabular} &
        $\frac{1}{\Lambda^2} (X^\dagger H) B_{\mu\nu} B^{\mu\nu}$ &
        \begin{tabular}[t]{@{}c@{}}
        $\frac{1}{\Lambda^2} (X^\dagger H) W^a_{\mu\nu} W^{a\,\mu\nu}$\\
         $\frac{1}{\Lambda^2} (X^\dagger\frac{\sigma_a}{2} H) W^a_{\mu\nu}\, B^{\mu\nu}$
        \end{tabular} &
        \begin{tabular}[t]{@{}c@{}} 
            $\bar{\chi}_L X\nu_R$ \\ 
            $\dfrac{1}{\Lambda}\bar{\chi} \chi X^\dagger H$ 
        \end{tabular} &
        \begin{tabular}[t]{@{}c@{}} 
        $\frac{1}{\Lambda}\bar{\ell} \chi_R X^\dagger \tilde{H}$ \\
        $\bar{\chi_L}H \nu_R$\\
        $\lambda\,\bar{\ell}\chi_L^c$\\
        $\frac{1}{\Lambda}X^\dagger  H \bar{\ell}\chi_L^c$
        
        \end{tabular} \\
        \hline

    \end{tabular}
    \end{adjustbox}
    \caption{
Gauge quantum numbers and effective operators for a scalar mediator $X(0^{++})$ in the case $\chi \neq \nu_R$. 
The first two columns specify the $(d_{SU(3)},d_{SU(2)},Y)$ assignments of $X$ and of the dark fermion $\chi$. 
The columns $\mathcal{L}^{\rm SM}_{X,\phi}$ collect the gauge–invariant couplings of $X$ to SM fields $\phi$, while $\mathcal{L}^{\rm DM}_{X}$ and $\mathcal{L}^{\rm DM}_{X,SM}$ contain, respectively, portal interactions involving $X$ and $\chi$, and additional operators with both $\chi$ and SM fields.
}
    \label{tab:oper1+}
\end{table}

\begin{table}[t]
    \centering
    \begin{adjustbox}{max width=\textwidth}
    \begin{tabular}{c|c|c|c|c|c|c|c}
    
        $X(0^{++})$ & $\chi=\nu_R$ & $\mathcal{L}_{X,e}^{\rm SM}$ & $\mathcal{L}_{X,H}^{\rm SM}$ & $\mathcal{L}_{X,B}^{\rm SM}$ & $\mathcal{L}_{X,W}^{\rm SM}$ & $\mathcal{L}_{X}^{\rm DM}$ & $\mathcal{L}_{X,SM}^{\rm DM}$\\
        \hline
        
        (1,1,0) & (1,1,0) & 
        $\frac{1}{\Lambda}X\,\bar{\ell} H e_R$ & 
        \begin{tabular}[t]{@{}c@{}} 
            $XXH^{\dagger}H$ \\ 
            $\lambda \,XH^{\dagger}H$ 
        \end{tabular} &
        $\frac{1}{\Lambda} \, X \, B_{\mu\nu} B^{\mu\nu}$ &
        $\frac{1}{\Lambda} \, X \, W^a_{\mu\nu} W^{a\,\mu\nu}$ &
        $\bar{\chi^c} \chi X$ &  
        \begin{tabular}[t]{@{}c@{}} 
            $y \bar{\ell} \tilde{H} \chi$ \\ 
            $  \dfrac{1}{\Lambda}\,X\bar{\ell} \tilde{H} \chi$ 
        \end{tabular} \\
        \hline
        
        
        (1,2,$\tfrac{1}{2}$) & (1,1,0) & 
        \begin{tabular}[t]{@{}c@{}}
        $\bar{\ell} X e_R$ \\ 
        $\dfrac{1}{\Lambda} \left( \bar{\ell} X \right)\left( H^\dagger \ell \right)$
        \end{tabular} & 
        \begin{tabular}[t]{@{}c@{}}
        $(H^\dagger H)(X^\dagger X)$  \\ 
        $\lambda^2\,H^\dagger X$
        \end{tabular} & 
        $\frac{1}{\Lambda^2} (X^\dagger H) B_{\mu\nu} B^{\mu\nu}$ &
        \begin{tabular}[t]{@{}c@{}}
        $\frac{1}{\Lambda^2} (X^\dagger H) W^a_{\mu\nu} W^{a\,\mu\nu}$\\
         $\frac{1}{\Lambda^2} (X^\dagger\frac{\sigma_a}{2} H) W^a_{\mu\nu}\, B^{\mu\nu}$
        \end{tabular} &
        $\dfrac{1}{\Lambda} X^\dagger H \bar{\chi^c} \chi$ & 
        \begin{tabular}[t]{@{}c@{}} 
            $ \bar{\ell} \tilde{H} \chi$ \\ 
            $\bar{\ell} \tilde{X} \chi$ 
        \end{tabular} \\
        \hline
        

    \end{tabular}
    \end{adjustbox}
    \caption{
    Same as tab. \ref{tab:oper1+}, but for the case in which the dark fermion coincides with the right–handed neutrino, $\chi = \nu_R$, so that only one singlet fermion appears in the dark sector.
    }
    \label{tab:oper2+}
\end{table}
It is worth emphasize on some final comments regarding the tables:

\begin{itemize}
    \item Note that an $SU(2)_L$ doublet field contains one electrically neutral component regardless of whether its hypercharge is $+\tfrac{1}{2}$ or $-\tfrac{1}{2}$. This applies both to $X$ and to $\chi$. Since the interactions that appear in the Lagrangian are essentially the same in both cases, up to the use of a conjugated doublet (e.g., $\tilde{X}$ instead of $X$), we chose not to explicitly list the multiplicity of hypercharge assignments in the table for clarity.
    
    It is worth stressing, however, that whenever either $X$ or $\chi$ is embedded in an $SU(2)_L$ doublet, the spectrum necessarily contains additional electrically charged partners.
    As a consequence, such model realizations entail the presence of extra degrees of freedom beyond the neutral state responsible for the observed anomalies or for the DM phenomenology.
    This introduces the additional theoretical and phenomenological requirement of accounting for the effects induced by these charged states, and of ensuring their consistency with existing experimental constraints.

    \item In the cases where $X \sim (1,2,1/2)$, the field shares the same quantum numbers as the SM Higgs doublet. As a result, terms involving combinations or permutations of $X$ and $H$ in operators containing both fields are, in principle, allowed by gauge invariance. However, these additional terms have not been explicitly listed in the table for simplicity.

    \item Finally, we stress that all operators with mass dimension $d>4$ in tables \ref{tab:oper1+}–\ref{tab:oper2-} are written in an Effective Field Theory (EFT) form.  
    A generic operator $\mathcal{O}_d$ of dimension $d$ appears in the Lagrangian as ${c_d}/{\Lambda^{\,d-4}}\,\mathcal{O}_d\,$, so that higher–dimensional interactions are suppressed by inverse powers of the heavy scale $\Lambda$.  
    In other words, the larger the operator dimension, the more strongly its contribution is suppressed at energies well below $\Lambda$. We also note that, differently from the EFT operators discussed above, the parameter $\lambda$ denotes a genuine dimensionful coupling, it appears in operators of mass dimension $d=3$ and $d=2$.

    \item In the tables, $\chi_L$ and $\chi_R$ denote the left-- and right--handed components of the fermion field $\chi$, defined through the chiral projection operators
    \begin{equation}
    \chi_{L,R} \equiv P_{L,R}\,\chi,
    \qquad
    P_{L,R} = \frac{1}{2}\left(1 \mp \gamma^5\right).
    \end{equation}
    And $\sigma_a$ are the pauli matrices.

\end{itemize}

\begin{table}[t]
    \centering
    \begin{adjustbox}{max width=\textwidth}
    \begin{tabular}{c|c|c|c|c|c|c|c}
    
        $X(0^{+-})$ & $\chi\neq\nu_R$ & $\mathcal{L}_{X,e}^{\rm SM}$ & $\mathcal{L}_{X,H}^{\rm SM}$ & $\mathcal{L}_{X,B}^{\rm SM}$ & $\mathcal{L}_{X,W}^{\rm SM}$ & $\mathcal{L}_{X}^{\rm DM}$ & $\mathcal{L}_{X,SM}^{\rm DM}$\\
        \hline
        
        (1,1,0) & (1,1,0) & 
        $\dfrac{1}{\Lambda}\,i\,X\bar{\ell} H  e_R$ & 
        $XXH^{\dagger}H$&
        $\dfrac{1}{\Lambda}X \, B_{\mu\nu} \tilde{B}^{\mu\nu}$ &
        $\dfrac{1}{\Lambda}X \, W^a_{\mu\nu} \tilde{W}^{a\,\mu\nu}$ &
        \begin{tabular}[t]{@{}c@{}} 
            $\,i\,\bar{\chi} \chi X$ \\ 
            $\,i\,\bar{\chi_L} \nu_R X$ \\
            $\,i\,\bar{\nu_R^c}\nu_R X$
        \end{tabular} &  
        \begin{tabular}[t]{@{}c@{}} 
            $y \bar{\ell} \tilde{H} \chi_R$ \\ 
            $\lambda\,\bar\chi_L{\nu_R}$\\
            $  \dfrac{1}{\Lambda}\,i\,X\bar{\ell} \tilde{H}  \chi_R$ \\
            $\,i\,X\,\bar\chi_L {\nu_R}$
        \end{tabular} \\
        \hline
        
        (1,1,0) & (1,2,$\tfrac{1}{2}$) &
        $\dfrac{1}{\Lambda}\,i\,X\bar{\ell} H e_R$ & 
        $XXH^{\dagger}H$&
        $\dfrac{1}{\Lambda}X \, B_{\mu\nu} \tilde{B}^{\mu\nu}$ &
        $\dfrac{1}{\Lambda}X \, W^a_{\mu\nu} \tilde{W}^{a\,\mu\nu}$ &
        $\,i\,\bar{\chi} \chi X$ & 
        \begin{tabular}[t]{@{}c@{}} 
            $\bar{\chi_L}H \nu_R$ \\
            $\lambda\,\bar{\ell}\chi_L^c$\\
            $\,i\,X\,\bar{\ell}\chi_L^c$\\
            $\frac{1}{\Lambda}\,i\,X\bar{\chi_L}H \nu_R$
        \end{tabular} \\
        \hline 
        
        (1,2,$\tfrac{1}{2}$) & (1,1,0) & 
        $\,i\,\bar{\ell} X  e_R$ & 
        \begin{tabular}[t]{@{}c@{}}
        $(H^\dagger H)(X^\dagger X)$  \\ 
        \end{tabular} &
        $\dfrac{1}{\Lambda^2}(X^\dagger H) B_{\mu\nu} \tilde{B}^{\mu\nu}$ &
        \begin{tabular}[t]{@{}c@{}}
        $\frac{1}{\Lambda^2} (X^\dagger H) W^a_{\mu\nu} \tilde{W}^{a\,\mu\nu}$\\
         $\frac{1}{\Lambda^2} (X^\dagger\frac{\sigma_a}{2} H) W^a_{\mu\nu}\, \tilde{B}^{\mu\nu}$
        \end{tabular} &
        \begin{tabular}[t]{@{}c@{}} 
            $\dfrac{1}{\Lambda}\,i\,X^\dagger H \bar{\chi} \chi$ \\ 
            $\dfrac{1}{\Lambda}\,i\,X^\dagger H \bar{\chi_L} \nu_R$ \\
            $\dfrac{1}{\Lambda}\,i\,X^\dagger H \bar{\nu_R^c} \nu_R$
        \end{tabular} & 
        \begin{tabular}[t]{@{}c@{}} 
            $ \bar{\ell} \tilde{H} \chi_R$ \\ 
            $\lambda\,\bar\chi_L{\nu_R}$\\
            $\,i\,\bar{\ell} \tilde{X} \chi_R$ \\
            $\frac{1}{\Lambda}\,i\,X^\dagger H\bar\chi_L{\nu_R}$
        \end{tabular} \\
        \hline
        
        (1,2,$\tfrac{1}{2}$) & (1,2,$\tfrac{1}{2}$) & 
        $\,i\,\bar{\ell} X  e_R$ & 
        \begin{tabular}[t]{@{}c@{}}
        $(H^\dagger H)(X^\dagger X)$  \\ 
        \end{tabular} &
        $\dfrac{1}{\Lambda^2}(X^\dagger H) B_{\mu\nu} \tilde{B}^{\mu\nu} $ &
        \begin{tabular}[t]{@{}c@{}}
        $\frac{1}{\Lambda^2} (X^\dagger H) W^a_{\mu\nu} \tilde{W}^{a\,\mu\nu}$\\
         $\frac{1}{\Lambda^2} (X^\dagger\frac{\sigma_a}{2} H) W^a_{\mu\nu}\, \tilde{B}^{\mu\nu}$
        \end{tabular} &
        \begin{tabular}[t]{@{}c@{}} 
            $\,i\,\bar{\chi}_L X \nu_R$ \\ 
            $\dfrac{1}{\Lambda}\,i\,\bar{\chi}  \chi X^\dagger H$ 
        \end{tabular} & 
        \begin{tabular}[t]{@{}c@{}} 
        $\dfrac{1}{\Lambda}\,i\,\bar{\ell} \chi_R X^\dagger \tilde{H}$ \\
        $\bar{\chi_L}H \nu_R$ \\
        $\lambda\,\bar{\ell}\chi_L^c $\\
        $\frac{1}{\Lambda}X^\dagger  H \bar{\ell}  \chi_L^c$
        
        \end{tabular} \\
        \hline

    \end{tabular}
    \end{adjustbox}
    \caption{
    Gauge quantum numbers and effective operators for a pseudoscalar mediator $X(0^{+-})$ in the case $\chi \neq \nu_R$. 
    The structure of the table is analogous to tab. \ref{tab:oper1+}, with some replacements in structures induced by the pseudoscalar nature of $X$ appropriate to a pseudoscalar coupling (e.g.\ $\,i\,$ terms and dual field strengths $\tilde B_{\mu\nu}$, $\tilde W^a_{\mu\nu}$).
    }
    \label{tab:oper1-}
\end{table}

\begin{table}[t]
    \centering
    \begin{adjustbox}{max width=\textwidth}
    \begin{tabular}{c|c|c|c|c|c|c|c}
    
        $X(0^{+-})$ & $\chi=\nu_R$ & $\mathcal{L}_{X,e}^{\rm SM}$ & $\mathcal{L}_{X,H}^{\rm SM}$ & $\mathcal{L}_{X,B}^{\rm SM}$ & $\mathcal{L}_{X,W}^{\rm SM}$ & $\mathcal{L}_{X}^{\rm DM}$ & $\mathcal{L}_{X,SM}^{\rm DM}$\\
        \hline
        
        (1,1,0) & (1,1,0) & 
        $\frac{1}{\Lambda}\,i\,X\bar{\ell} H  e_R$ &
        \begin{tabular}[t]{@{}c@{}} 
        $XXH^{\dagger}H$
        \end{tabular} &
        $\dfrac{1}{\Lambda}X \, B_{\mu\nu} \tilde{B}^{\mu\nu}$ &
        $\dfrac{1}{\Lambda}X \, W^a_{\mu\nu} \tilde{W}^{a\,\mu\nu}$ &
        $\,i\,\bar{\chi^c} \chi X$ &  
        \begin{tabular}[t]{@{}c@{}} 
            $y \bar{\ell} \tilde{H} \chi$ \\ 
            $ \dfrac{1}{\Lambda}\,i\,X \bar{\ell} \tilde{H}  \chi$ 
        \end{tabular} \\
        \hline
        
        
        (1,2,$\tfrac{1}{2}$) & (1,1,0) & 
        $\,i\,\bar{\ell} X  e_R$ & 
        \begin{tabular}[t]{@{}c@{}} 
            $(H^\dagger H)(X^\dagger X)$ \\
        \end{tabular} & 
        $\frac{1}{\Lambda^2} (X^\dagger H) B_{\mu\nu} \tilde{B}^{\mu\nu}$ &
        \begin{tabular}[t]{@{}c@{}}
        $\frac{1}{\Lambda^2} (X^\dagger H) W^a_{\mu\nu} \tilde{W}^{a\,\mu\nu}$\\
         $\frac{1}{\Lambda^2} (X^\dagger\frac{\sigma_a}{2} H) W^a_{\mu\nu}\, \tilde{B}^{\mu\nu}$
        \end{tabular} &
        $\dfrac{1}{\Lambda}\,i\, X^\dagger H \bar{\chi^c} \chi$ & 
        \begin{tabular}[t]{@{}c@{}} 
            $ \bar{\ell} \tilde{H} \chi$ \\ 
            $\,i\,\bar{\ell} \tilde{X} \chi$ 
        \end{tabular} \\
        \hline
        

    \end{tabular}
    \end{adjustbox}
    \caption{Same as tab. \ref{tab:oper1-}, but for the case $\chi = \nu_R$, where the dark fermion is identified with the right–handed neutrino.}
    \label{tab:oper2-}
\end{table}

Having now listed our operators in full detail, we can directly examine the
terms in the last columns, which are the ones that could potentially
jeopardize the stability of $\chi$ by inducing fast decay channels.

Schematically, the operators in $\mathcal{L}_{X,SM}^{\rm DM}$ fall into two classes: 
mixing terms between $\chi$ and $\nu$, and interaction terms of the type $\chi\nu_R X$ (or the ones involving the SM neutrino, such as $\bar{\ell}\tilde H \chi$). 
These interactions can, indeed, induce DM decays into SM states, potentially ruling out $\chi$ as a viable DM candidate. 
We will return to this issue in the appendix \ref{app:NeuCou}.

In what follows we will adopt as benchmark scenario the minimal configuration corresponding to the first row of tab. \ref{tab:oper2+}, in which the mediator is a real scalar
$X_{17}(0^{++}) \sim (1,1,0)$ and the dark fermion coincides with a single
right–handed Majorana neutrino, $\chi = \nu_R \sim (1,1,0)$. This set-up is, in fact, the one which already captures all the ingredients relevant for our analysis: a coupling of $X_{17}$
to electrons, a portal coupling to DM, and the minimal set of
additional interactions implied by gauge invariance, without introducing extra degrees of freedom in addition to $X$ and $\chi$. 

\subsection{Model Limitations and Future Extensions}
\label{sec:FutDev}

Before proceeding to the detailed calculations, it is useful to highlight three 
non-trivial features of our model. 
First, in the present work, we  omit quark operators in order to keep
 our model (e.g. Tables \ref{tab:oper1+}--\ref{tab:oper2-}) as minimal as possible and to focus exclusively on the electron
excess observed at PADME. Couplings of an $X_{17}$-like mediator to quarks are, however, not forbidden by
any fundamental principle; on the contrary, they are expected to play a role in
a complete description of the ATOMKI anomalies discussed in
Chapter \ref{cap:MotAndResObj}, which originate from nuclear transitions. 
A more comprehensive analysis that incorporates
quark couplings may be developed in future extensions of this framework.

Another relevant aspect of our model concerns its flavour structure. In the
discussion above, we have treated $\nu_R$ (and, when applicable, $\chi$) as a
single right-handed Majorana state. In a more realistic setup, one would
promote this field to a flavour multiplet, for example a triplet of
right-handed neutrinos, in order to account for the observed pattern of light
neutrino masses and mixings via a seesaw-like mechanism. Such an extension
would generically induce additional couplings of $X_{17}$ to muons and taus,
as well as a non-trivial flavour structure within the dark sector. While
technically straightforward, this generalization would introduce an additional
 complexity that is largely orthogonal to the questions
addressed in this thesis. For this reason, we keep the flavour structure
as minimal as possible and restrict ourselves to a single effective right-handed state. 

Finally, the available experimental data do not allow for a definitive
determination of the spin of the $X_{17}$. In the tables, we restricted
our analysis to the simpler case of a (pseudo)scalar mediator. As discussed in
sec.~\ref{sec:X17}, recent analyses in \cite{Barducci:2025hpg} of the
experimental anomalies associated with the ATOMKI measurements indicate that a
scalar, CP-even interpretation of the new particle is currently favored, once
theoretical uncertainties and nuclear-structure effects are properly taken into
account.
In the near future, however, PADME runs will probe the possible coupling of the $X_{17}$
mediator to photons with significantly sensitivity, 
providing then a direct information on its spin. Indeed, according to the Landau--Yang
theorem \cite{Landau1948, Yang1950}, a massive spin-1 particle cannot decay into two photons. Therefore,
the information of the $X_{17}\to \gamma\gamma$ decay would favor or not a scalar interpretation of $X_{17}$.

\section{Relic Abundance Calculation: Diagram Evaluation}
\label{sec:DiagEva}

In the previous section we classified all the effective operators allowed by gauge invariance and identified a minimal benchmark scenario, corresponding to the first row of tab. \ref{tab:oper2+}.

In this section we move from the operator level to the evaluation of the corresponding Feynman diagrams.
First, we extract the Feynman rules starting from the Lagrangian terms discussed in the previous section and, then, we compute the squared
amplitudes for the decay and scattering processes of interest, such as
$X\to\chi\chi$ and $e^+ e^- \to \chi\chi$, which will be relevant for the computations in chapter \ref{cap:DMAbCalcu}.
We recall that the total Lagrangian is given by

\begin{equation}
    \mathcal{L}_{tot}=\mathcal{L}_{SM}+\mathcal{L}_{DS} \,.
\end{equation}
Here, $\mathcal{L}_{DS}$ is the dark sector Lagrangian
\begin{equation}
  \begin{aligned}
    \mathcal{L}_{DS} \equiv\;&
    \frac{1}{2}(\partial_\mu X)(\partial^\mu X)
    - \frac{1}{2}m_X^2 X^2
    - \frac{1}{3}\,\kappa_X X^3
    - \frac{1}{4}\,\lambda_X X^4
    + \frac{1}{2}\bar{\chi}\,i\slashed{\partial}\,\chi
    - \frac{1}{2}\hat m\,\bar{\chi}^c \chi  \\[4pt]
    &- \frac{g_{Xee}}{\Lambda}\,X\,\bar{\ell} H e_R+\, g_{XDM}\,X\bar{\chi}^c \chi 
    - y_{mix}\,\bar{\ell}\tilde{H}\chi
    - \frac{g_{XDM\nu}}{\Lambda}\,X\,\bar{\ell}\tilde{H}\chi \\[4pt]
    &
    - g_H\,X X H^\dagger H
    - \tilde{g}_H\,X H^\dagger H
    - \frac{g_{XBB}}{\Lambda}\,X B_{\mu\nu} B^{\mu\nu}
    - \frac{g_{XWW}}{\Lambda}\,X W^a_{\mu\nu} W^{a\,\mu\nu}
    + \text{h.c.}\;.
  \end{aligned}
  \label{eq:L1}
\end{equation}

In what follows we will retain only the interaction terms that are directly relevant for the phenomenology discussed in this thesis, namely
\begin{equation}
  \mathcal{L}_{int} \equiv\frac{g_{Xee}}{\Lambda}\,X\,\bar{\ell} H e_R
  + g_{XDM}\,X\bar{\chi}^c \chi 
  + y_{mix}\,\bar{\ell}\tilde{H}\chi
  + \frac{\hat{g}_{XDM\nu}}{\Lambda}\,X\,\bar{\ell}\tilde{H}\chi\,+\,\text{h.c.}\;.
\label{eq:LintUV} 
\end{equation}

After electroweak symmetry breaking and working in the unitary gauge, i.e.
\begin{equation}
  H = 
  \begin{pmatrix}
    0 \\
    \dfrac{v_h+h}{\sqrt{2}}
  \end{pmatrix},
  \qquad
  \tilde H =
  \begin{pmatrix}
    \dfrac{v_h+h}{\sqrt{2}} \\
    0
  \end{pmatrix},
  \qquad
  \ell =
  \begin{pmatrix}
    \nu_L \\ e_L
  \end{pmatrix},
\end{equation}
where $v_h$ is the Higgs vacuum expectation value, defined in sec. \ref{sec:SM}, and $h$ is the physical Higgs field, the interaction Lagrangian becomes:
\begin{equation}
  \mathcal{L}_{\rm int}^{\rm EWSB}
  =
  -\frac{v_h}{\sqrt{2}}\frac{g_{Xee}}{\Lambda}\,
    X\,\bar e_L e_R
  - g_{XDM}\,X\bar{\chi}^c \chi
  - y_{mix}\,\frac{v_h}{\sqrt{2}}\,\bar{\nu}_L \chi
  - \frac{v_h}{\sqrt{2}}\frac{\hat{g}_{XDM\nu}}{\Lambda}\,
    X\,\bar{\nu}_L \chi
  + \text{h.c.}\,,
\label{eq:LintBrutta}
\end{equation}
(\ref{eq:LintBrutta}) contains explicit couplings of $X$ to electrons and neutrinos, as well as mass and mixing terms for the dark fermion $\chi$.

It is convenient to parametrize the mixing between the active neutrino and
the dark fermion in terms of a small dimensionless mixing angle $\theta$, defined as 
\begin{equation}
    \theta\equiv \frac{y_{\rm mix}\,v_h}{\sqrt{2}\, m_\chi}\,.
\label{eq:theta}
\end{equation}

For later convenience it is also useful to rewrite the interaction terms in
terms of effective low–energy couplings.  
We therefore define:
\begin{equation}
\begin{aligned}
    g_e &\equiv \frac{v_h}{\sqrt{2}\Lambda}\,g_{Xee}\,,\\
    g_\nu &\equiv \frac{v_h}{\sqrt{2}\Lambda}\,g_{XDM\nu}\,,\\
    g_\chi &\equiv 2\,g_{XDM}\,,\\
\end{aligned}
\end{equation}
so that $g_e$ and $g_\nu$ directly parametrise the strength of the
$X\bar e e$ and $X\bar\chi\nu$ vertices after EWSB, while the extra
factor of $2$ in $g_\chi$ accounts for the Majorana nature of $\chi$ in
the $X\chi\chi$ vertex. These effective parameters $g_e$, $g_\nu$ and $g_\chi$ are precisely the low–energy couplings that appeared earlier in the interactions (\ref{Xeecoupl})–(\ref{XDMnucoupl}).  

The interaction Lagrangian relevant for our analysis can be re-written in terms of such effective couplings $(g_e,\,g_\nu,\,g_\chi)$ as
\begin{equation}
  \mathcal{L}_{\rm int}^{\rm EWSB}
  =
  -g_e\,X\,\bar e_L e_R
  - \frac{g_\chi}{2}\,X\bar{\chi}^c \chi
  - g_\nu\,X\,\bar{\nu}_L \chi
  + \text{h.c.}\;.
\label{eq:Lint}
\end{equation}

As anticipated in the previous section, after electroweak symmetry breaking the  Lagrangian is effectively controlled by the two couplings we are interested in, namely $g_e$ and $g_\chi$.
At the same time, one additional interaction involving neutrinos, proportional to $g_\nu$, is still present. As previously mentioned, this "unwanted" term may destabilize $\chi$, and therefore require a dedicated discussion in appendix \ref{app:NeuCou}.

Starting from the interaction Lagrangian in eq. (\ref{eq:Lint}), the corresponding
Feynman rules read:

\begin{figure}[H]
    \centering
    \includegraphics[width=0.9\linewidth]{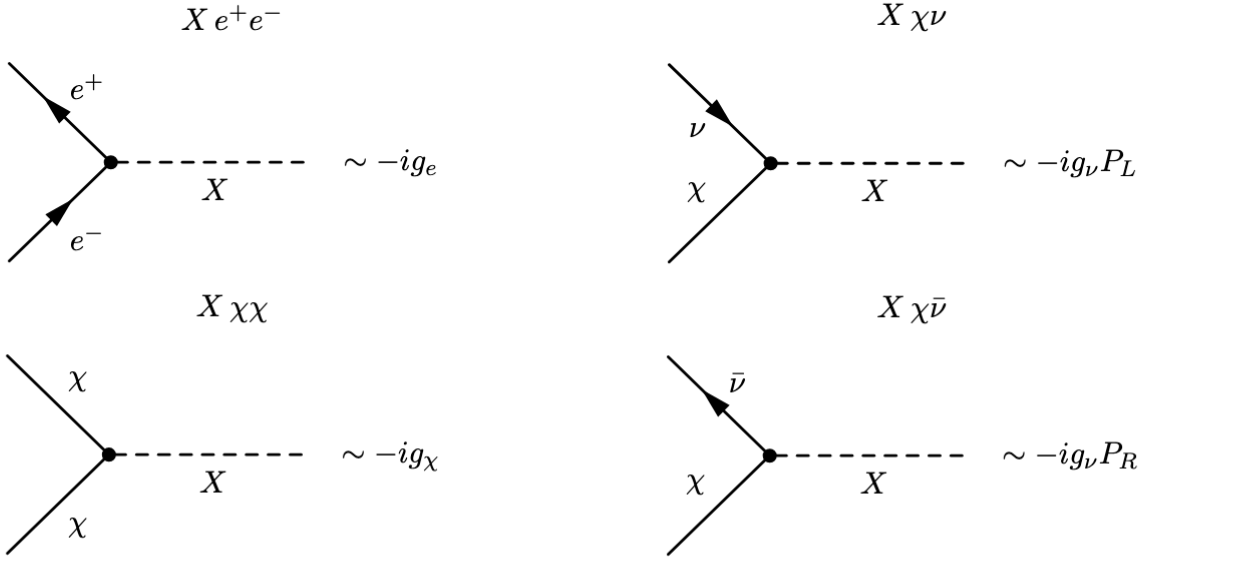}
\end{figure}

We then compute explicitly the squared amplitudes for the processes relevant for our computations of the DM production, as it will be discussed in the next chapter.  
Below, we list the squared moduli of the amplitudes, summed over all
initial and final state polarisations (and \emph{not} averaged over the initial state polarisations):

\begin{align}
  \sum_{\rm pols}\!\left|\mathcal{A}(X \to \chi\chi)\right|^{2}
  &= 2\,g_{\chi}^{2}\,\bigl(m_{X}^{2}-4m_{\chi}^{2}\bigr),
  \label{eq:MXtocc}\\[4pt]
  \sum_{\rm pols}\!\left|\mathcal{A}(X \to \chi\nu)\right|^{2}
  &= g_{\nu}^{2}\,\bigl(m_{X}^{2}-m_{\chi}^{2}-m_{\nu}^{2}\bigr),
  \label{eq:MXtocn}\\[6pt]
  \sum_{\rm pols}\!\left|\mathcal{A}(e^{+}e^{-}\to \chi\nu)\right|^{2}
  &= 
  \frac{2\,g_{\nu}^{2}g_{e}^{2}\,\bigl(s-4m_{e}^{2}\bigr)\,
        \bigl(s-m_{\chi}^{2}-m_{\nu}^{2}\bigr)}
       {\bigl(m_{X}^{2}-s\bigr)^{2}},
  \label{eq:Meetocn}\\[6pt]
  \sum_{\rm pols}\!\left|\mathcal{A}(e^{+}e^{-}\to \chi\chi)\right|^{2}
  &= 
  \frac{4\,g_{\chi}^{2}g_{e}^{2}\,\bigl(s-4m_{e}^{2}\bigr)\,
        \bigl(s-4m_{\chi}^{2}\bigr)}
       {\bigl(m_{X}^{2}-s\bigr)^{2}}.
  \label{eq:Meetocc}
\end{align}

The last two amplitudes arise from the following diagrams:

\begin{figure}[H]
    \centering
    \includegraphics[width=0.7\linewidth]{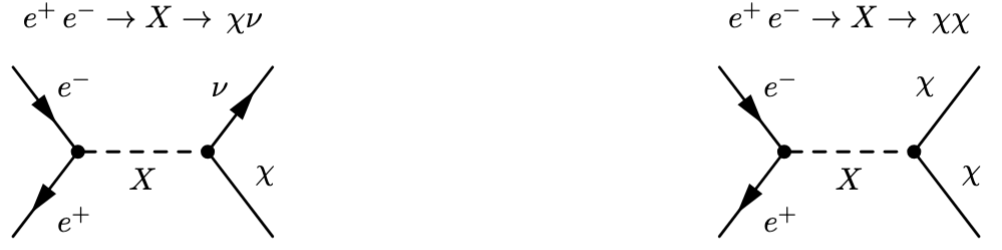}
\end{figure}

The inclusion of the finite width of the mediator in the $s$–channel propagator brings to the modification
\begin{equation*}
\left(m_{X}^{2}-s\right) \to \left(m_{X}^{2}-s+i\,m_{X}\Gamma_{X}\right),
\end{equation*}
so that the last two expressions become
\begin{align}
  \sum_{\rm pols}\!\left|\mathcal{A}(e^{+}e^{-}\to \chi\nu)\right|^{2}
  &= 
  \frac{2\,g_{\nu}^{2}g_{e}^{2}\,\bigl(s-4m_{e}^{2}\bigr)\,
        \bigl(s-m_{\chi}^{2}-m_{\nu}^{2}\bigr)}
       {\bigl(m_{X}^{2}-s\bigr)^{2}+m_{X}^{2}\Gamma_{X}^{2}},
  \label{eq:MeetocnD}\\[6pt]
  \sum_{\rm pols}\!\left|\mathcal{A}(e^{+}e^{-}\to \chi\chi)\right|^{2}
  &= 
  \frac{4\,g_{\chi}^{2}g_{e}^{2}\,\bigl(s-4m_{e}^{2}\bigr)\,
        \bigl(s-4m_{\chi}^{2}\bigr)}
       {\bigl(m_{X}^{2}-s\bigr)^{2}+m_{X}^{2}\Gamma_{X}^{2}}.
  \label{eq:MeetoccD}
\end{align}

Before concluding, we stress two considerations:
\begin{itemize}
    \item We have not listed \emph{all} possible processes involving the dark fermion $\chi$, but only those that are relevant for our purposes. In view of the DM-production related calculation in chapter~\ref{cap:DMAbCalcu}, the amplitudes above are precisely the ones for which the number of DM particles differs between the initial and final states at least by one unit.
    
    \item During the rest of this thesis, we will effectively neglect the $X$-mediated interactions between $\chi$ and active neutrinos, since they are expected to be phenomenologically suppressed for several independent reasons.

\begin{enumerate}[(i)]
    \item Couplings involving neutrinos generically open decay channels of the form $\chi \to \nu + \cdots$.
    Requiring DM to be stable on cosmological time scales therefore strongly suppresses such interactions.

    \item Neutrino couplings typically lead to enhanced annihilation or decay channels producing high--energy neutrinos.
    Existing indirect detection searches place stringent bounds on these processes, making sizable neutrino couplings disfavored.

    \item \textbf Interactions involving neutrinos can generate monochromatic or sharply peaked neutrino signals.
    The absence of observed neutrino lines in current experiments further constrains these couplings.
\end{enumerate}

For similar reasons, we also neglect the effects induced by active--sterile mixing between $\chi$ and the SM neutrinos.
In summary, the active-sterile mixing angle $\theta$ defined in eq. \ref{eq:theta} is forced to be very small by three independent considerations.
\begin{enumerate}[(i)]
  \item The correction to the light neutrino mass induced by the seesaw--like
  structure, $\Delta m_\nu \simeq \theta^2 m_\chi$, must satisfy the cosmological
  bound on $\sum_\nu m_\nu$.

  \item DM stability on cosmological time--scales requires the total
  decay rate induced by the mixing to be much smaller than $T_U^{-1}$.

  \item Radiative decays $\chi \to \nu\gamma$ are additionally constrained by
  X--ray observations, which demand an extremely long lifetime for this specific
  channel.
\end{enumerate}

A more detailed discussion of the suppression mechanisms associated with neutrino couplings is deferred to the app. \ref{app:NeuCou}. 
\end{itemize}

In this chapter we have constructed a theoretical framework linking a light scalar mediator $X$ to a fermionic dark-matter particle $\chi$. We have then derived the squared amplitudes for the key decay and scattering processes. On this basis, we are now ready to move to Chapter \ref{cap:DMAbCalcu}, where these results will be used to reproduce the observed dark–matter relic abundance.

\chapter{Our Model: Dark Matter Relic Abundance}
\label{cap:DMAbCalcu}

In this chapter we will study DM production to test our model, outlined in cap.~\ref{cap:TheFra}. 
After an introductory discussion on the structure of the Boltzmann equation and two DM production mechanisms, we will compute the relic abundance of DM. 
We will exploit controlled analytic limits to expose the parametric behavior, and identify the regions of the parameter space compatible with the observed relic abundance.

\section{Boltzmann Equation}
\label{sec:Bol}
In cap. \ref{cap:TheFra} we built up a theoretically self-consistent DM model, in which the $X$-boson acts as a mediator of the interactions between the DM particle and the SM ones. 
In order to check its viability, we need to describe how to compute in practice the abundance of DM.

For a DM model to be viable, it must reproduce coherently the current DM abundance observed in the universe ($\Omega_{DM}\,h^2$), parameter introduced in sec. \ref{sec:DM}.
Our aim is, thus, to figure out how the dynamics of our specific particle-physics model may explain DM abundance we observe today. 
Boltzmann equation has precisely this aim, since it connects the DM particle number density ($n_{DM}$), a parameter strictly tied to abundance, to the amplitudes of particle-physics processes.

Consider the simplest case, in which a particle species $A$ has no interactions at all. 
The number density of particles $A$ ($n_{A}$) in a comoving volume $V$ will then remain constant
\begin{equation}
    \frac{d}{dt}\bigl(n_{A} V\bigr) =0\,.
\end{equation}

On the contrary, let us assume that a particle species $A$ may interact with a particle species $B$ via the processes:
\begin{equation}
\begin{aligned}
    B\rightarrow A \,A\,,\\
    A\,A\rightarrow B\,.
\end{aligned}
\end{equation}
The number density of $A$ particles in a volume V will vary in time as
\begin{equation}
\begin{aligned}
    \frac{d}{dt}\bigl(n_{A} V\bigr) =& \,V\,S_f\sum_{dofs}\int d\Pi_B\!\int d\Pi_{{A}_1}\!\int d\Pi_{{A}_2}\,(2\pi)^{4}\,\delta^{(4)}(p_{B}-p_{A_1}-p_{{A}_2})\,
    \\&\times\Bigl[ \,2|\mathcal{A}_{B\to {A}{A}}|^{2}\, f_{B}\,(1 \pm f_{{A}})^2-2|\mathcal{A}_{{A}{A}\to B}|^{2}\,(1\pm f_{B})\,f_{{A}}^2\Bigr],
\end{aligned}
\label{eq:BolAAB}
\end{equation}
where $S_f=1/2!$ is the overall symmetry factor, accounting for the presence of identical particles in the initial and/or final state of the reaction. This factor ensures that the phase–space integrals do not overcount physically equivalent configurations.
Here $d\Pi_a \;\equiv\; \frac{d^3 p_a}{(2\pi)^3\,2E_a}$ is the relativistic phase space of particle species $a$ $(a=B,A_1,A_2)$.  
$|\mathcal{A}_{{B}\to {A}{A}}|^2$ and $|\mathcal{A}_{{A}{A}\to B}|^2$ are the squared amplitudes of the processes we are considering. The sum in eq. (\ref{eq:BolAAB}) runs over the initial and final states internal degrees of freedom $(dofs)$. $f_i$ $(i=A,\,B)$ are the phase space densities of the various particle species, namely 
\begin{equation}
    n_i =\hat g_i \int \frac{\mathrm{d}^3 p}{(2\pi)^3} \, f_i\,,
\label{eq:ni}
\end{equation}
$\hat g_a$ counts the internal degrees of freedom such as polarizations and multiplet components.

Each process is weighted by relative phase space density $f_i$ of each of the initial state particles, while terms like $(1 \pm f_i)$ takes into account stimulated emission effects for bosonic final state particles (plus sign) or Pauli blocking for fermionic ones (minus sign). Finally, pre-factors $\pm2$ take into account the fact that the process $B\rightarrow {A}\,{A}$  is adding two $A$ particles to the bath, while $A\,A\rightarrow B$ is removing two. (In what follows, we will use the shorthand notation $B\Leftrightarrow A\, A$ to include both processes) 

Notice that here $A_1$ and $A_2$ denote the two identical particles in the final state: they carry, in general, different momenta in the phase–space integration, but share
the same phase space densities $f_A$ since they belong to the same species.

The same reasoning applies to a generic $2\to2$ scattering or annihilation process, 
where the species $A$ interacts with other particles $B$, $C$ and $D$ through
\begin{equation}
    A\,B \;\Longleftrightarrow\; C\,D\,.
\label{eq:AB<->CD}
\end{equation}
Repeating the steps outlined above, one finds that the Boltzmann equation for 
the number density of $A$ in a comoving volume $V$ takes the form 
\begin{equation}
\begin{aligned}
\frac{d}{dt}\bigl(n_A V\bigr)
=&\; V\,S_i\,S_f\, \sum_{\text{dofs}}
   \int d\Pi_A\,d\Pi_B\,d\Pi_C\,d\Pi_D\,
   (2\pi)^4\,\delta^{(4)}(p_A+p_B-p_C-p_D)
\\[4pt]
&\times\Bigl[
   |\mathcal{A}_{CD\to AB}|^2\, f_C f_D\,(1\pm f_A)(1\pm f_B)
   \;-\;
   |\mathcal{A}_{AB\to CD}|^2\, f_A f_B\,(1\pm f_C)(1\pm f_D)
   \Bigr]\,.
\end{aligned}
\label{eq:BoltzmannABCD}
\end{equation}
All symbols appearing in eq. (\eqref{eq:BoltzmannABCD}) have the same meaning as in the previous case. Notice that in this case, the symmetry factors $S_i$ and $S_f$ are nothing but the same kind of combinatorial factors discussed above, now written in a fully general form for arbitrary initial and final states:
\begin{equation}
S_i =
\begin{cases}
\dfrac{1}{2!} & \text{for } A = B\,,\\[6pt]
1             & \text{for } A \neq B\,,
\end{cases}
\qquad
S_f =
\begin{cases}
\dfrac{1}{2!} & \text{for } C = D\,,\\[6pt]
1             & \text{for } C \neq D\,.
\end{cases}
\end{equation}

This makes clear that the structure of the Boltzmann equation is completely general: for a process with $n$ incoming and $m$ outgoing particles, the Boltzmann equation always contains a product of $n$ distribution functions $f_i$ for the initial states and $m$ statistical factors $(1\pm f_i)$ for the final states.

After these two examples, it is useful to introduce a few general quantities, that will be key ingredients in the following analysis.

From the physical point of view, the expansion of the universe dilutes the concentration of particles. In the Boltzmann equation (\ref{eq:BolAAB}) this can be made evident by expanding the time derivative in the LHS
\begin{equation}
    \frac{1}{V}\frac{d}{dt}\bigl(n_{A}V\bigr)= \dot{n}_{A}+3\,n_{A}\,H=
    \dot{Y}_{A}s\,,
    \label{Yabundance}
\end{equation}
where $H$ is the Hubble expansion rate. 
The term $3\,n_A\,H$ is taking into account the dilution effect. 
The abundance is defined as $Y\equiv n/s$, where $s$ is the entropy density. 
The entropy density and the Hubble parameter vary with the temperature of the Universe as:
\begin{equation}
    \begin{aligned}
        s(T) &= \frac{2\pi^2}{45}\,g^*_S(T)\,T^3\,,\\[4pt]
        H(T) &= 1.66\,\sqrt{g^*_\rho(T)}\,\frac{T^2}{M_{\rm Pl}}\,,
    \end{aligned}
\label{eq:sandH}
\end{equation}
where $M_{Pl}$ is the Planck mass. The functions $g^*_S(T)$ and $g^*_\rho(T)$ quantify the total number of relativistic
degrees of freedom contributing to the entropy and energy density of the Universe.
Their values decrease whenever a particle species becomes non–relativistic.
They are defined as:
\begin{equation}
    g^*_\rho(T) 
    = \sum_{i\,\in\,\text{bosons}} 
        \hat g_i \left(\frac{T_i}{T}\right)^4 
      + \frac{7}{8}\sum_{i\,\in\,\text{fermions}} 
        \hat g_i \left(\frac{T_i}{T}\right)^4 ,
\label{eq:gr}
\end{equation}
\begin{equation}
    g^*_S(T) 
    = \sum_{i\,\in\,\text{bosons}} 
        \hat g_i \left(\frac{T_i}{T}\right)^3 
      + \frac{7}{8}\sum_{i\,\in\,\text{fermions}} 
        \hat g_i \left(\frac{T_i}{T}\right)^3 ,
\label{eq:gs}
\end{equation}
where, again, $\hat g_i$ counts the internal degrees of freedom of species $i$ and $T_i$ is its
temperature.

\begin{figure}[tb]
    \centering
    \includegraphics[width=0.7\linewidth]{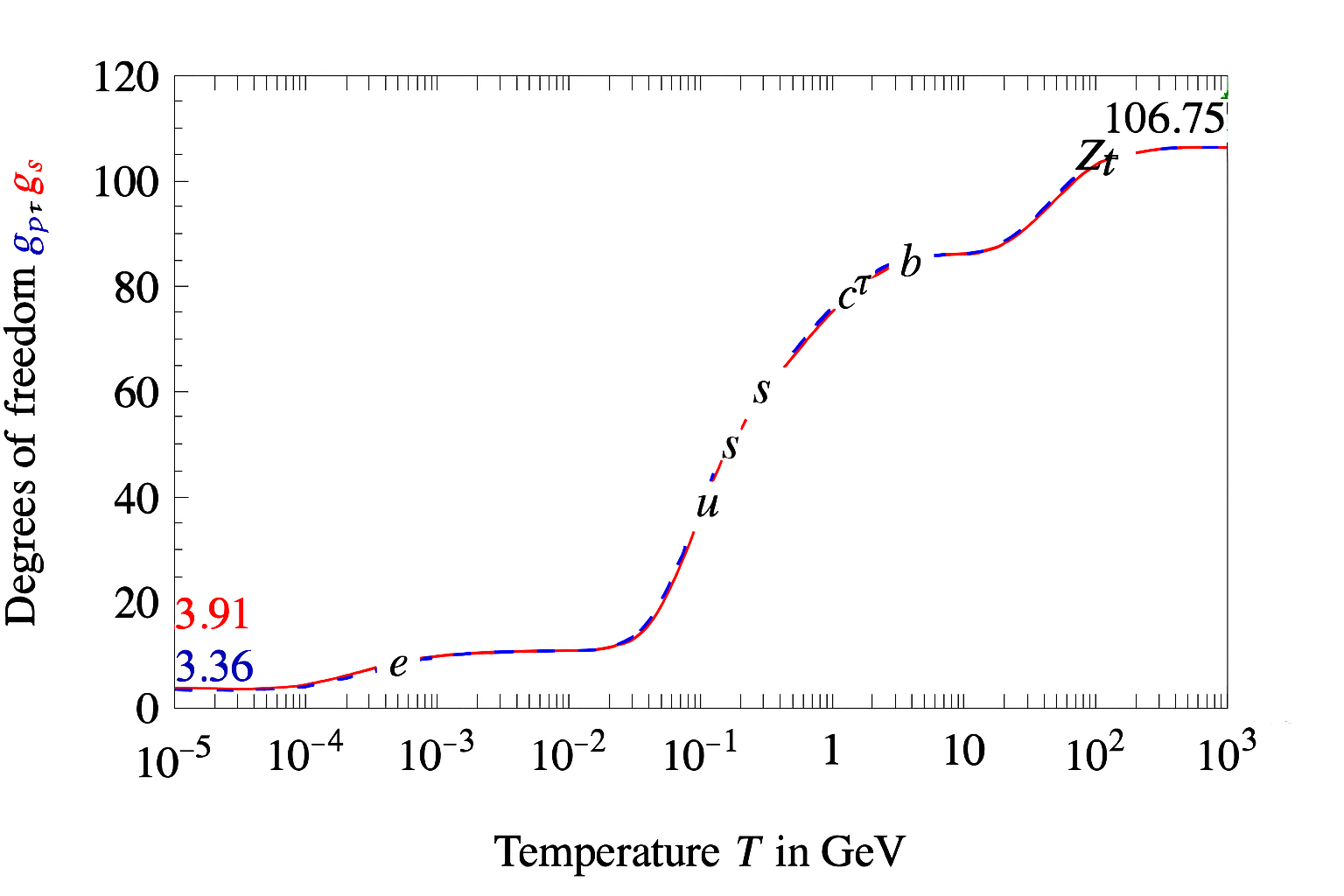}
    \caption{Temperature evolution of the effective numbers of degrees of freedom. Image taken from Ref. \cite{Cirelli:2024ssz}.}
    \label{fig:grhogs}
\end{figure}

The evolution of $g^*_\rho(T)$ and $g^*_S(T)$ across the thermal history of the Universe is shown in fig.\ref{fig:grhogs}.

Before moving to specific models, it is also convenient to introduce the thermally averaged cross section, which will be useful in the following. 
For a generic $2\to2$ process $A\,B \to C\, D$, we define $\langle\sigma v\rangle$ as:

\begin{equation}
\langle \sigma v \rangle \;\equiv\;
\int \frac{d\Pi_A\,d\Pi_B}{n_A^{\text{eq}}\,n_B^{\text{eq}}}\,d\Pi_C\,d\Pi_D\,
f_A^{\rm eq}(p_A)\,f_B^{\rm eq}(p_B)\,
(2\pi)^4\,
\delta^{(4)}(p_A+p_B-p_C-p_D)
|\mathcal{A}_{AB\to C D}|^2 \, ,
\label{eq:TheAveCS}
\end{equation}
where $f_{A,B}^{\rm eq}$ are the equilibrium distribution functions. 
A species is in equilibrium when the processes that create and annihilate it occur at the same rate, so that on average its number density tracks the equilibrium value $n_i^{\rm eq}(T)$ associated with the Maxwell–Boltzmann distribution $f_i^{\rm eq}(p,T)\propto e^{-E(p)/T}$.

A more detailed discussion of equilibrium, and of the conditions under which a species follows or departs from $n^{\rm eq}(T)$ will be given in the next section.

Let us highlight that Boltzmann equation allows us to compute the DM abundance, but being a differential equation, is highly dependent from initial state conditions. From a physical point of view, this means that the same abundance of DM we observe today can be obtained through very different production mechanisms, corresponding to radically different dynamics in the early Universe. In the next section, we will discuss two DM production scenarios which differ from each other in this sense: freeze-in and freeze-out.

\section{Production Mechanisms}
\label{sec:FIFO}

We will now discuss the main features of two of the possible DM production scenarios.

\subsection{Freeze-out}
Historically, freeze-out is the standard framework for DM production. 
Freeze-out has two main assumptions: DM was already present in the early Universe after the Big Bang, and was in equilibrium with the SM particles present in the hot plasma, namely the thermal bath.

\begin{figure}[t]
    \centering
    \includegraphics[width=0.7\linewidth]{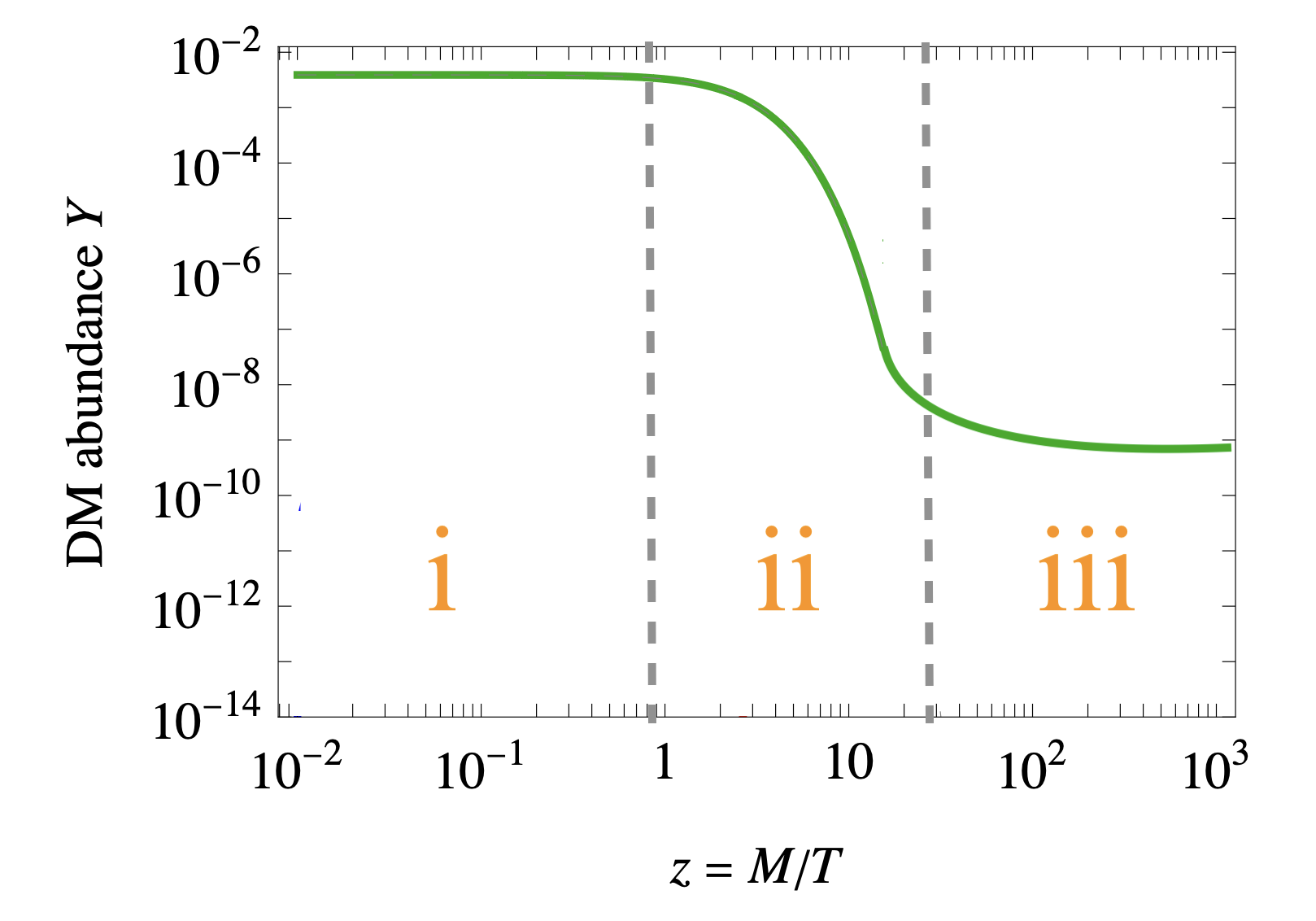}
    \caption{Evolution of the DM abundance $Y$ in a freeze-out scenario.  
(i) For $z \ll 1$ ($T \gg m_\chi$) the abundance tracks its equilibrium value and remains approximately constant.  
(ii) Around $z \sim 1$ the DM becomes non–relativistic, annihilations become inefficient, and the abundance rapidly departs from equilibrium.  
(iii) For $z \gg 1$ annihilations freeze out and the abundance settles to its final asymptotic value, determining the relic density.
Figure adapted from Ref. \cite{Cirelli:2024ssz}.}
    \label{fig:FO}
\end{figure}

At a qualitative level, freeze-out has three phases: as long as equilibrium holds, the DM number density closely tracks its equilibrium value, then, once temperature drops below DM mass, it decreases exponentially. 
As the Universe cools further, however, the annihilation rate can no longer keep up with the Hubble expansion, so the number of dark matter particles stops decreasing and a non-zero relic abundance is left over. 
These three regimes are shown in fig.~\ref{fig:FO}, which displays the evolution of the abundance as a function of the dimensionless parameter
$z \equiv M/T $, where $M$ is the DM mass.

Before proceeding, it is convenient to introduce a compact notation for processes that change the number of DM particles.
We denote by $M \;\leftrightarrow\; N$ a reaction in which $M$ DM particles appear in the initial state and $N$ in the final state. For simplicity, in the standard freeze–out analysis one considers the case of a $2\leftrightarrow 0$ process, namely $\chi \,\chi \leftrightarrow SM SM$ annihilations, where $\chi$ denotes a single dark–matter species and “SM” stands for an arbitrary of Standard Model particle. In this setup, the Boltzmann equation takes precisely the form discussed in sec. \ref{sec:Bol}, in eq. (\ref{eq:BoltzmannABCD}).

We will now discuss more in depth the main ingredients and the features involved in freeze-out computations, describing separately the three regimes from fig. \ref{fig:FO}. 
\begin{enumerate}[(i)]
    \item Thermal equilibrium regime

 As long as the temperature is higher than the DM mass, $T\,\gtrsim \,m_\chi$, the typical kinetic energy of particles in the plasma is large enough to efficiently produce and annihilate DM. Indeed, in this regime, both the annihilation processes $\chi \,\chi \to SM\, SM $ and the inverse $SM \,SM \to \chi \,\chi$ are kinematically allowed and unsuppressed. As a result, for every annihilation there is, on average, a corresponding inverse process. This condition is referred to as (chemical) equilibrium. Consequently, the DM number density closely tracks its equilibrium value $n_\chi^{\rm eq}(T)$, corresponding to the usual Maxwell-Boltzmann distribution $f_\chi^{\rm eq}(p,T) \propto e^{-\frac{E(p)}{T}}$. This behavior is clearly illustrated by region (i) in fig. \ref{fig:FO}, 
where the abundance exhibits an extended plateau at early times (small $z$). 
In the plot, up to $z \simeq 1$ ($T \simeq m_\chi$) the abundance remains constant thanks to equilibrium.  

    \item Freeze-out epoch
    
When the temperature drops below the DM mass, $T\lesssim m_\chi$, the situation changes qualitatively. Inverse processes like $SM \,SM \to \chi\,\chi$ are fed only by the exponentially rare tail of the Boltzmann distribution, and their rate becomes strongly suppressed compared to annihilations. As the universe cools, DM concentration therefore falls exponentially and the actual DM abundance keeps following the Boltzmann–suppressed equilibrium curve down, as shown by the behavior in region (ii) of fig. \ref{fig:FO}.

    \item Frozen relic regime
    
Decoupling is the final key aspect: provided again that DM is coupled to SM, interactions between the two sectors will occur with a annihilation rate $\Gamma$. If the rate is much larger than the Hubble rate, namely
\begin{equation}
    \Gamma \gg H\,,
\label{eq:eqcond}
\end{equation}
the typical timescale between interactions, $\Gamma^{-1}$, is much shorter than the expansion timescale, $H^{-1}$. In this regime, a DM particle undergoes many interactions before the cosmic expansion can significantly dilute the plasma.
As the universe expands and cools further, the interaction rate between DM particles and the thermal bath decreases. When the temperature reaches a decoupling value \(T_{\rm dec}\), the annihilation rate per particle becomes too small to balance the dilution , i.e. when $\Gamma \lesssim H$. From that point on, the Hubble expansion dominates over annihilations: DM particles become effectively decoupled from the plasma, and their number density freezes to an almost constant value. This determines the so-called freeze-out of the DM abundance, after which the relic abundance remains essentially unchanged, as illustrated by the third slice in fig. \ref{fig:FO}, where the abundance settles to its asymptotic freeze-out value.
\end{enumerate}

Now we will derive a rough estimate for the mass and coupling scalings in this scenario: we will treat the observed relic abundance as a constraint in the Boltzmann equation and we will determine which parameters allow to reproduce it.

Consider the standard annihilation process relevant for freeze--out, $2 \leftrightarrow0$, whose rate per particle is
\begin{equation}
   \Gamma_{\rm ann} (T)\;\equiv\; n_\chi(T)\,\langle\sigma v\rangle (T)\,.
\label{eq:GamAnn}
\end{equation}
The expression above (eq. \ref{eq:GamAnn}) can be understood as follows: for a $2\!\leftrightarrow\!0$ process, the probability that a given DM particle annihilates in a unit time is equal to the probability of encountering another DM particle in the plasma, $n_\chi$, multiplied by the interaction strength, encoded in the thermally averaged cross section $\langle\sigma v\rangle$.  

As mentioned before, freeze-out occurs when, at a temperature $T_{\rm fo}$, annihilations become too slow to compete with the Hubble expansion,
\begin{equation}
   \Gamma_{\rm ann}(T) \;\simeq\; H(T) 
\end{equation}
with $H(T)$ defined in eq. (\ref{eq:sandH}). As discussed in any standard reference on thermal relic dark matter (see for instance \cite{ParticleDataGroup:2024cfk}), the requirement of reproducing the observed relic abundance implies a thermally averaged annihilation cross section of the order
\begin{equation}
   \langle \sigma v \rangle_{\rm DM} \;\sim\; 10^{-9}\,{\rm GeV^{-2}}\sim 10^{-26}\ \mathrm{cm^3/s}\,.
\end{equation}
The cross section derived above is naturally obtained for weak-scale couplings and masses. This numerical coincidence led to the so-called Weakly Interacting Massive Particle (WIMP) miracle: the observation that dark matter could simply be made of particles with approximately electroweak strength interactions,
automatically yielding the correct relic abundance. 

However, despite extensive searches no experiment has yet observed a signal compatible with WIMPs. These persistent null results now strongly constrain the simplest WIMP models, and motivate the study of alternative production mechanisms for dark matter.

\subsection{Freeze-in Mechanism}
\begin{figure}[t]
    \centering
    \includegraphics[width=0.7\linewidth]{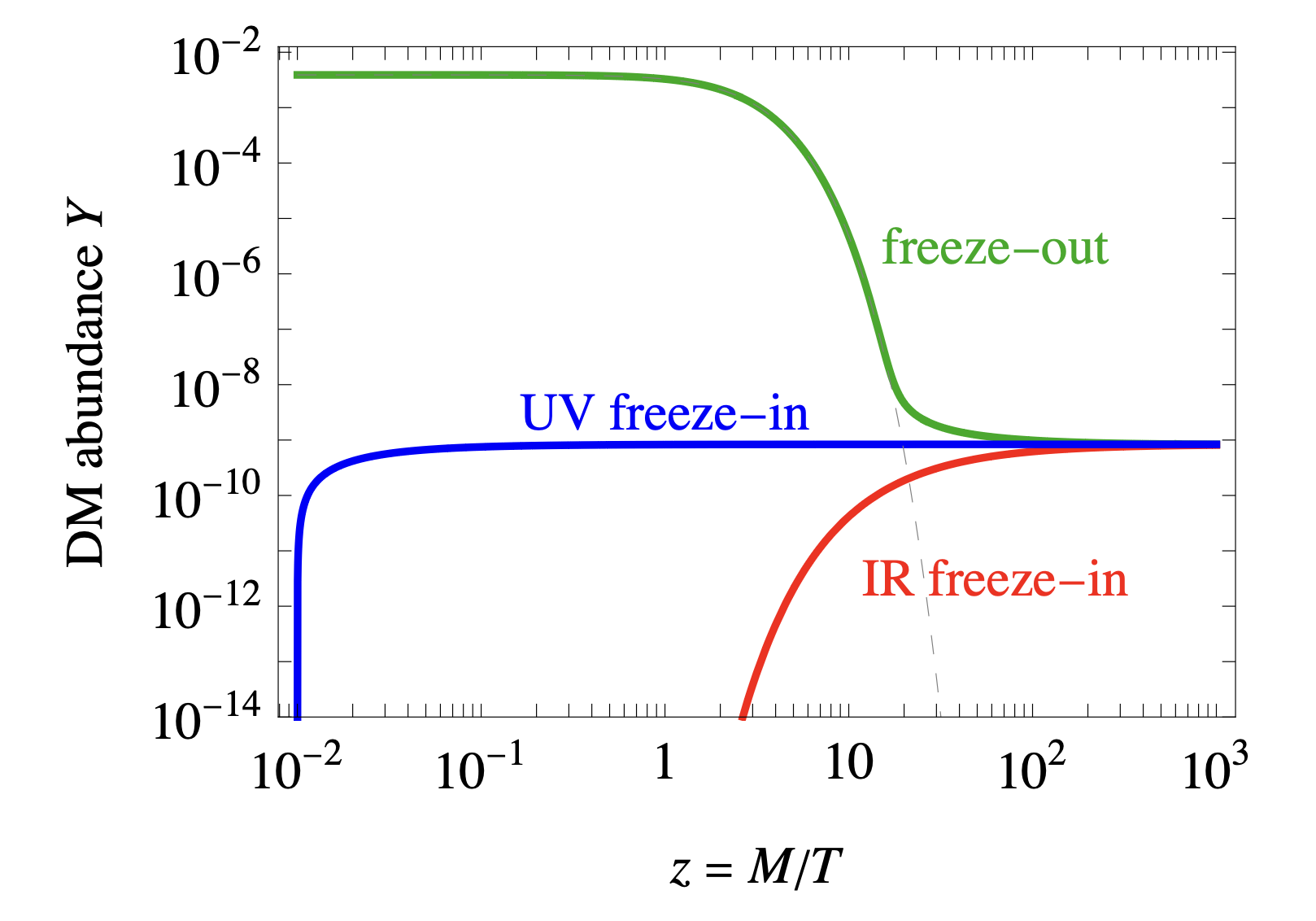}
    \caption{Log-Log plot of the evolution of the DM abundance $Y$ as a function of the dimensionless parameter $z = M/T$, where $M$ denotes the mass scale of the heaviest particle involved in the production process.  
    The green curve shows the standard freeze-out behavior.  
    The blue and red curves illustrate UV and IR freeze-in production, respectively.}
    \label{fig:FOFI}
\end{figure}

During the recent years, another DM production mechanism was brought to light, turning upside down the freeze-out perspective. Firstly introduced in \cite{Hall:2009bx}, freeze-in assumes that after the big bang the DM concentration was negligible, and DM is so feebly coupled with SM particles that it never reached equilibrium with the thermal bath. Although the interactions with SM particles are feeble, they increment the DM abundance gradually over time, with each rare interaction contributing only a small fraction of the final relic density. This behavior is clearly visible in the blue and red curves of fig. \ref{fig:FOFI}. 

Before examining the quantitative evolution of the abundance, it is useful to distinguish the two qualitatively different regimes of freeze-in production: IR and UV freeze-in, both illustrated 
in fig.~\ref{fig:FOFI}.
In the first scenario the production of DM occurs entirely at low energies. As we will show later, the dominant contribution to the abundance comes from temperatures close to the dark matter mass. Most of the abundance is therefore accumulated around $z \sim 1$ (i.e.\ $T \sim m_\chi$), just like the red line in fig. \ref{fig:FOFI}. Notice that in our model, if DM is lighter than $X$, an analogous argument applies to $m_X$: the resonant contribution at temperatures $T \sim m_X$ dominates the production rate, so that this temperature range gives the largest contribution to the final abundance.

UV freeze-in on the other hand is characterized by a high-energy (and high temperature) DM production. This behavior typically arises when the interactions responsible for DM production originate from non-renormalizable operators.
Since the interaction induced by a non-renormalisable operator scales with energy, $\sigma \propto E^{\,2}/\Lambda^{2}$ (or, in general, with an even higher power for operators of  dimension $>5$), the corresponding production rate increases at high temperatures.  
As a result, the dominant contribution to the DM abundance is generated at the earliest, hottest stages of the thermal history, making this channel intrinsically UV–dominated.
This behavior is precisely what is illustrated by the blue curve in fig. \ref{fig:FOFI}, where the DM abundance is built up predominantly at very early times (small $z$), when the temperature is the highest. 

Finally, we comment on the role of other possible production channels.
In the present model, DM production proceeds exclusively through processes of the type $2 \leftrightarrow 0$.
In principle, one could also consider scenarios where freeze-in is driven by $1 \leftrightarrow 0$ processes, corresponding to the decay of a particle in the thermal bath directly into DM.
Such scenarios have been studied in the context of thermal decays of DM.
We do not consider this possibility further here, and refer the reader to refs. \cite{Belfatto:2021ats, Vittorio:2025txp}.

Having outlined the qualitative features of freeze-out and freeze-in production mechanism, we now focus on the scenario relevant for our model. In particular, in our framework the model adopted is the one from the first line of tab.~\ref{tab:oper2+}, and DM is produced through IR freeze-in.  
In what follows we specialize to this case and derive the relic abundance in detail, computing step by step all the quantities entering the Boltzmann equation and identifying the parameter space consistent with observations.

\section{Freeze-in Calculation}
\label{sec:FICal}

In this section we present the core calculation of the DM abundance in our model.  
As mentioned before, we work within a framework where the dark matter particle $\chi$ is a right–handed Majorana fermion, interacting with the visible sector through a light scalar mediator $X$, which acts as the portal between the SM and the dark sector.  
Starting from the general freeze-in formalism discussed above, we now derive explicit analytic expressions for the relic density, and evaluate how it depends on the parameters of the theory.  

Before diving into calculations, we list the main standard approximations for freeze-in, in accordance with those used in classic freeze-in analyses in literature (e.g. \cite{Hall:2009bx}).

First of all, we assume that the $X$ boson is in equilibrium with the SM plasma during the DM production era. This is justified because DM is mainly produced at temperatures $T_{\text{prod}}\sim \mathcal{O}(m_X)$ (as hinted above, and as we will see in sec.~\ref{sec:regimes}), where equilibrium is easily maintained. A rough estimate of the equilibrium condition from eq.~(\ref{eq:eqcond}) gives
\begin{equation}
\frac{\Gamma_X}{H(T = m_X)} \sim \frac{g_e^{2} M_{\text{Pl}}}{m_X} \gg 1.
\label{eq:Xeqcond}
\end{equation}
For the values relevant to our setup, $g_e\sim \mathcal{O}(10^{-4})$ and $m_X\sim \mathcal{O}(10\,\text{MeV})$, this ratio is of the order $10^{13}\gg 1$, so the equilibrium assumption is safely satisfied.

Then we assume the usual Boltzmann phase space distribution for species at equilibrium
\begin{equation*}
    f_X=f_X^{eq}\approx e^{-\frac{E_X}{T}}\,.
\end{equation*}
For detailed definitions and equilibrium criteria, see the standard reference \cite{Kolb:1990vq}.

Finally, in the freeze-in framework, DM concentration remains negligible during the whole production. Hence we set to zero its phase space density: $f_\chi\approx0$.

Our model features several interaction channels that contribute to dark–matter production, all originating from the low‑energy interaction Lagrangian introduced earlier.
In what follows, we analyze each relevant process in turn, insert it into the Boltzmann equation, and compute its individual contribution to the final DM abundance.
The channels of interest are
\begin{enumerate}
\item $X \Longleftrightarrow \chi\chi$
\item $e^+ e^- \Longleftrightarrow \chi\chi$
\item $e^+ e^- \Longleftrightarrow \chi\nu$.
\end{enumerate}
In practice, the third process will be neglected in the subsequent analysis.
As discussed at the end of cap.~\ref{cap:TheFra} and further motivated by the considerations presented in appendix~\ref{app:NeuCou}, interactions involving neutrinos are phenomenologically suppressed and do not play a significant role in determining the DM relic abundance.
For this reason, in the remainder of this chapter  we will focus exclusively on the first two channels, which set the freeze‑in relic abundance.

\subsection[Freeze-in by $X \to \chi \chi$ Decay]{Freeze-in by $X \to \chi\, \chi$ Decay}
\label{sec:FIXcc}

We consider the case where the only possible channel available for DM production  is  $X \leftrightarrow \chi \chi$. 
This case is analogous to the first one discussed in sec. \ref{sec:Bol}, the Boltzmann equation takes the same form as the  expression in eq. (\ref{eq:BolAAB}). 
Once the freeze-in approximations introduced above are applied the formula reduces to
\begin{equation}
    \dot Y_\chi s \approx\frac{1}{2!}\int d\Pi_{X}\,e^{-\frac{E_{X}}{T}}\,\!\int d\Pi_{\chi_1}\!\int d\Pi_{\chi_2}\,(2\pi)^{4}\,\delta^{(4)}(p_{X}-p_{1}-p_{2})\,\times \,2\sum_i|\mathcal{A}_{X\to \chi\chi}|^{2}\,.
\end{equation}
The squared modulus of the amplitude summed over initial and final polarizations ($\sum_i|\mathcal{A}_{X\to \chi\chi}|^{2}$)was derived in sec. \ref{sec:DiagEva}. The factor $\frac{1}{2!}$ is the final state symmetry factor.

Substituting the standard two-body decay rate
\begin{equation}
    \Gamma_{X\to \chi\chi}
=\frac{1}{2m_X}\,
\frac{1}{2!}\!
\int d\Pi_{\chi_1}\,
     d\Pi_{\chi_2}\,
(2\pi)^4\delta^{(4)}(P-p_1-p_2)\;
|\mathcal{A}_{X\to \chi\chi}|^{2}\,,
\label{eq:rateXchichi}
\end{equation}
we obtain:
\begin{equation}
    \dot Y_\chi s\approx2\int d\Pi_{X}\,e^{-\frac{E_X}{T}}\,2\,m_X\Gamma_{X\rightarrow\chi\chi}\,.
\end{equation}

Exploiting spherical coordinates for the 3-momenta and recalling that $|p_X| d|p_X| = E_X dE_X$
\begin{equation}
\begin{aligned}
    \dot Y_\chi s\approx&\dfrac{m_X\,\Gamma_{X\rightarrow\chi\chi}}{\pi^2}\int_{m_X}^\infty dE_X\sqrt{E_X^2-m_X^2}e^{-\frac{E_X}{T}}\\
    =&\dfrac{m_X^2\,\Gamma_{X\rightarrow\chi\chi}}{\pi^2}\,T\,K_1(\frac{m_X}{T})\,,
\end{aligned}
\end{equation}
where $K_1$ is the modified Bessel function of the first type.

As the RHS does not depend explicitly upon $Y_{\chi}$, we simply integrate it in time.
Since $ \dot{T}\approx-H(T)T$, where H(T) is the Hubble parameter discussed in eq. (\ref{eq:sandH}), the integration measure can be rewritten as $dt\approx-{dT}/(H(T)T)$. Indeed, the integration over time is equivalent to an integration over temperature so that
\begin{equation}
\begin{aligned}
    Y_\chi\approx&
    \int_{t_{min}}^{t_{max}}dt\,\frac{m_X^2\,T\,\Gamma_{X\rightarrow\chi\chi} }{\pi^2\,s}K_1\bigl(\frac{m_X}{T}\bigr)
    \\ 
    \approx&\frac{m_X^2\,\Gamma_{X\rightarrow\chi\chi}}{\pi^2}\int_{T_{min}}^{T_{max}}dT\frac{K_1\bigl(\frac{m_X}{T}\bigr)}{H(T)s(T)}\,.
\end{aligned}
\end{equation}
The explicit temperature dependence of $s$ is expressed again in eq. (\ref{eq:sandH}). 

We perform the integral above by using $z\equiv {m_X}/{T}$, which was firstly introduced in sec. \ref{sec:FIFO}. This step allows us to simplify the integral. In fact, since $T_{min}<<m_X$ and $T_{max}>>m_X$, we can take $z_{min}\approx 0$ and $z_{max}\approx \infty$ for the integration range: 
\begin{equation}
\begin{aligned}
    Y_\chi\approx&\,\frac{45}{1.66\cdot2\,\pi^4}\,\frac{M_{Pl}}{g^*_s\,\sqrt{g^*_\rho}} \frac{\Gamma_{X\rightarrow\chi\chi}}{m_X^2}\int_0^\infty dz \,z^3\,K_1(z)\\
    =&\frac{135}{1.66\cdot4\,\pi^3}\,\frac{M_{Pl}}{g^*_s\,\sqrt{g^*_\rho}}\,\frac{\Gamma_{X\rightarrow\chi\chi}}{m_X^2}\,.
\end{aligned}
\label{eq:YXcc0}
\end{equation}
The factor $z^{3}$ in the first line originates from the explicit $T$–dependence of the parameters $H$ and $s$, once rewritten in terms of the dimensionless variable $z=m_{X}/T$.

Regarding the temperature dependence of the effective numbers of relativistic degrees of freedom, $g^{*}_{\rho}(T)$ and $g^{*}_{s}(T)$, can be safely neglected in this context.  
This is due to the general feature of IR freeze–in mentioned in sec. \ref{sec:FIFO}: the production of DM is dominated at most around temperatures of order the mediator mass, $T\sim m_{X}$, which in our setup lies at the level of a few MeV.  
Below $\mathcal{O}(10\,\mathrm{MeV})$ both $g^{*}_{\rho}$ and $g^{*}_{s}$ remain essentially constant throughout the entire production epoch.  
For this reason we evaluate them at $T\simeq m_{X}$ and treat them as fixed parameters in the integral. At these temperatures the only
SM species that are still relativistic are photons, electrons and positrons, and the three active neutrino flavors. As a result, we compute the effective number of relativistic degrees of freedom from eq. (\ref{eq:gr}) and (\ref{eq:gs})
\begin{equation}
  g^* \simeq g^*_\rho \simeq g^*_S
  = \hat g_\gamma + \frac{7}{8}\bigl(2\,\hat g_e + 3\,\hat g_{\nu_i}\bigr)
  = 2 + \frac{7}{8}\bigl(4 + 3\times 2\bigr)
  = 10.75  \,.
\end{equation}

For further clarity, one may look at fig. \ref{fig:grhogs}: below temperatures of order $\mathcal{O}(10\,\mathrm{MeV})$ the effective relativistic degrees of freedom remain essentially constant, confirming that $g^{*}_{\rho}$ and $g^{*}_{s}$ can be treated as temperature independent throughout the IR freeze–in production epoch.

At this point, we can derive the decay rate from eq. (\ref{eq:rateXchichi}) using the amplitude \label{amp:Xcc} derived in sec.~\ref{sec:DiagEva}

\begin{equation}
    \Gamma_{X\rightarrow \chi \chi}=\frac{1}{2!}\frac{\sqrt{m_X^2-4m_\chi^2}}{16\,\pi\,m_X^2}|\mathcal{A}_{X\to \chi\chi}|^{2}=\frac{1}{2!}\frac{g_\chi^2}{8\,\pi\,m_X^2}(\,m_X^2-4\,m_\chi^2)^\frac{3}{2}\,.
\end{equation}

Inserting this in eq. (\ref{eq:YXcc0}) we can finally write the abundance as

\begin{equation}
    Y_\chi=g_\chi^2\frac{135}{1.66\cdot64\,\pi^4}\,\frac{M_{Pl}}{g^*_s\,\sqrt{g^*_\rho}}\,\frac{(m_X^2-4m_\chi^2)^{3/2}}{\,m_X^4}\,.
\label{eq:YXcc}
\end{equation}

The physical parameter constrained by cosmology (see sec.\ref{sec:DMGP}) is the present–day relic density $\Omega_\chi h^2$, which is directly related to $Y_\chi$ through

\begin{equation}
    \Omega_\chi h^2\approx 2.74\times10^5\,m_\chi[MeV]\,Y_\chi\approx\;
\frac{1.09\times 10^{27}}{g^{S}_{*}\,\sqrt{g^{\rho}_{*}}}\;
\frac{m_{\chi}\,\Gamma_{X\rightarrow \chi \chi}}{m_{X}^{2}}\,.
\label{eq:reldenXchichi}
\end{equation}

This concludes the analysis of the freeze-in production via the decay channel $X \rightarrow \chi\chi$.  
We now move to the second class of processes contributing to dark–matter production in our model, the scattering channel $e^+e^-\rightarrow\chi\chi$, which requires a separate treatment within the Boltzmann framework.

\subsection[Freeze-in by $e^+e^-\to \chi \chi$ Scattering]{Freeze-in by $e^+e^-\to \chi \chi$ Scattering}
\label{sec:Scateecc}

In the case of DM interacting with the SM only via $e^+e^-\to \chi \chi$ process, we can start our computation of the frozen-in abundance from eq. (\ref{eq:BoltzmannABCD}).

Working directly with the variable $z$, for the specific case of $e^+e^-\to \chi \chi$, eq. (\ref{eq:BoltzmannABCD}) becomes

\begin{equation}
\begin{aligned}
 s\,H\,z\,\frac{dY_{\chi}}{dz}= \sum_{\text{all}}&
\int d\Pi_{e^{+}}\,d\Pi_{e^{-}} d\Pi_{\chi_1}\,d\Pi_{\chi_2}\, 
(2\pi)^{4}
\delta^{(4)}\!\left(p_{e^{+}}+p_{e^{-}}-p_{\chi_1}-p_{\chi_2}\right)
\,\times\\
&\left(2\,|\mathcal{A}_{e^+e^-\to \chi \chi}|^{2}f_{e^{+}}\,f_{e^{-}}(1-f_\chi)^2-2\,|\mathcal{A}_{\chi \chi\to e^+e^-}|^{2}\,f_\chi^2 (1-f_{e^{-}})(1-f_{e^{+}})\right)\,.
\end{aligned}
\end{equation}

Using the same approximations introduced at the beginning of this section (the very first ones applied in the decay case $X \to \chi\chi$), the Boltzmann equation simplifies to

\begin{equation}
\begin{aligned}
 s\,H\,z\,\frac{dY_{\chi}}{dz}= \sum_{\text{all}}&
\int d\Pi_{e^{+}}\,d\Pi_{e^{-}}e^{-\frac{E_{e^+}}{T}}\,e^{-\frac{E_{e^-}}{T}}\int d\Pi_{\chi_1}\,d\Pi_{\chi_2}\, 
(2\pi)^{4}
\delta^{(4)}\!\left(p_{e^{+}}+p_{e^{-}}-p_{\chi_1}-p_{\chi_2}\right)
\,\times\\
&2\,|\mathcal{A}_{e^+e^-\to \chi \chi}|^{2}\,.
\end{aligned}
\label{eq:Boleeccapp}
\end{equation}
\mg{manca un simmetry factor qua... (da confermare...)}
To proceed, it is convenient to rewrite the phase–space integrals in terms of Lorentz–invariant Mandelstam variables.  
We first introduce:
\begin{equation}
 \hat s = (p_{e^-}+p_{e^+})^2, 
 \qquad 
 \hat t = (p_{e^-}-p_{\chi_1})^2,
 \qquad 
 \hat u = (p_{e^-}-p_{\chi_2})^2 , 
\end{equation}

which encode the full kinematics of the $2\to2$ scattering process.  

In order to change the integration variables from momenta to $(\hat s,\hat t,\hat u)$, we recall that the Mandelstam variables satisfy the constraint:
\begin{equation}
    \hat s + \hat t + \hat u \;=\;2\, m_{e}^2\, +\,2m_{\chi}^2 \,,
\end{equation}
so only two of them are independent variables.  
This allows us to rewrite the phase–space integrals in terms of $(\hat s,\hat t)$ alone, with $\hat u$ fixed by the relation above.

After rewriting the phase–space measure in terms of Mandelstam invariants and using the energy–conserving delta function, the equilibrium factors reorganize into the modified Bessel function $K_{1}(\sqrt{\hat s}/T)$.  
The Boltzmann equation thus reduces to two remaining integrals, over $\hat s$ and $\hat t$.

At this point, the remaining integral over the Mandelstam variable $\hat t$ can be performed explicitly.  
This step isolates the standard cross section $\sigma\,(\hat s)$ defined as follows:
\begin{equation}
    \sigma(\hat s)
=
S_f
\sum_{\rm final}
\int_{t_{min}}^{t_{max}} d\hat t\;
\frac{\langle |\mathcal{A}|^2\rangle_{\rm initial}}
{16\pi\,\hat s^{2}\,\lambda}\, .
\label{eq:sigma2}
\end{equation}

Here $S_f$ is the final–state symmetry factor and $\lambda \equiv \lambda_{\rm Kal}(1,m_e^2/\hat s,m_e^2/\hat s)$ is the normalized usual Källén function.

The integration limits in $\hat t$ are also standard.  
For fixed $\hat s$, the allowed kinematic range of the Mandelstam variable $\hat t$ in the center of mass is obtained from the usual two–body scattering relations (see the PDG kinematics review \cite{ParticleDataGroup:2024cfk}):

\begin{equation}
t_{\min/\max}=-\left(p_{e^-,\mathrm{cm}} \mp p_{\chi_1,\mathrm{cm}}\right)^{2}\,.
\end{equation}

Finally, we highlight that the quantity $\langle|\mathcal{A}|^{2}\rangle_{\rm initial}$ appearing in the integrand (\ref{eq:sigma2}) is the standard spin–averaged squared amplitude from collider physics. In the freeze–in context this choice is merely conventional: the averaging over initial states is exactly compensated later when multiplying by the corresponding initial–state multiplicities $\hat g_{e^{+}}$ and $\hat g_{e^{-}}$(as we will do in eq. (\ref{eq:Boltomand})).  
We keep this convention only to ease the comparison with standard cross–section formulae. 

Putting all these ingredients together, one arrives at the compact expression of eq. (\ref{eq:Boleeccapp})
\begin{equation}
    s\,H\,z\,\frac{dY_{\chi}}{dz}=\,\hat g_{e^{+}}\hat g_{e^{-}}\,
\frac{T}{16\pi^{4}}
\int_{\hat s_{\min}}^{\infty}\! d\hat s\;
\hat s^{3/2}\,\lambda\,\sigma(\hat s)\,
K_{1}\!\left(\frac{\sqrt{\hat s}}{T}\right).
\label{eq:Boltomand}
\end{equation}
The integration boundaries in $\hat s$ follow directly from kinematics.  
For the process we are considering, energy conservation implies that the Mandelstam variable satisfies $\hat s \ge \hat s_{\min} = \max(4m_e^{2},\,4m_\chi^{2})$, while no upper bound exists in general, so that $\hat s_{\max}=\infty$.

The result in \ref{eq:Boltomand} matches with the literature formula (see e.g. Appendix C of \cite{Cirelli:2024ssz}) once specialized to the $e^{+}e^{-}\rightarrow\chi\chi$ channel.

As done for the case of the $X \to \chi\chi$ decay, we solve the differential equation (\ref{eq:Boltomand}) integrating over $z$
\begin{equation}
\begin{aligned}
    Y_\chi&=  \frac{\hat g_{e^{+}}\hat g_{e^{-}}}{16\pi^{4}}\,
\int_{\hat s_{\min}}^{\infty}\! d\hat s\;
\hat s^{3/2}\,\bar\lambda\,
\sigma_{e^{+}e^{-}\to\chi\chi}(\hat s)\,
\int_{z_{min}}^{z_{max}}\! dz\;
\frac{T(z)}{s(z)\,H(z)\,z}\,K_{1}\!\left(\frac{\sqrt{\hat s}}{T(z)}\right)\\
&=\frac{135}{1.66\cdot64\,\pi^5}\frac{M_{Pl}}{g_s^*\sqrt{g_\rho^*}}\,\hat g_{e^{+}}\hat g_{e^{-}}\int_{\hat s_{\min}}^{\infty}\! d\hat s \;
\frac{\bar\lambda\,}{\sqrt{\hat s}}
\sigma_{e^{+}e^{-}\to\chi\chi}(\hat s)\,\,.
\end{aligned}
\end{equation}

Using the squared amplitude derived in sec. \ref{sec:DiagEva} (see eq. (\ref{eq:MeetoccD})) and plugging it into eq. (\ref{eq:sigma2}), we have that

\begin{equation}
    \sigma(\hat s)= \frac{g_\chi^2\, g_e^2\,(\hat s - 4 m_\chi^2)^{3/2}\,\sqrt{\hat s - 4 m_e^2}}
{8\,\pi \,\hat g_{e^{+}}\hat g_{e^{-}}\, \left( \Gamma_X^2 m_X^2 + (m_X^2 - \hat s)^2 \right) \hat s},
\end{equation}
so that
\begin{equation}
Y_\chi=\frac{135}{1.66\cdot64\,\pi^5}\,
\frac{M_{\rm Pl}}{g_{s}^{*}\sqrt{g_{\rho}^{*}}}\,
\frac{g_\chi^{2}g_e^{2}}{8\pi}
\int_{\hat s_{\min}}^{\infty}\! d\hat s\;
\frac{\big(\hat s-4m_\chi^{2}\big)^{3/2}\,\big(\hat s-4m_e^{2}\big)^{3/2}}
{\hat s^{5/2}\,\Big[\Gamma_X^{2} m_X^{2}+\big(m_X^{2}-\hat s\big)^{2}\Big]}\,    \,.
\label{eq:Yeecc}
\end{equation}

The integral in eq. \eqref{eq:Yeecc} has no analytical solution, thus we keep the abundance in its integral representation. The results of a specific numerical study of this integral will be presented in a more detailed way in sec. \ref{sec:regimes} when discussing analytic limits.

\section{Freeze-in Abundance: Mass Regimes and Analytic Limits}
\label{sec:regimes}

In this section we will analyze the freeze-in abundance generated by the scattering process discussed in sec. \ref{sec:Scateecc}.  
Starting from the integral expression in eq. (\ref{eq:Yeecc}), we will derive simple analytic approximations in two complementary DM mass regimes.  
Specifically, we will study the case of DM lighter than the mediator $X$ and the opposite, showing how the relic abundance scales with the DM mass and the couplings in each limit.

\subsection{Resonant Regime}
The resonant regime is the one in which the decay $X \to \chi \, \chi$ is kinematically allowed, namely $m_\chi<m_X/2$, and in the following we will show how the scattering $e^+\,e^-\to \chi \,\chi$  gives the same contribution to the DM abundance. 

To evaluate \ref{eq:Yeecc} in this regime, we will adopt the narrow-width approximation (NWA), which is valid when the decay width of the mediator $X$ is much smaller than its mass, namely
\begin{equation}
    \frac{\Gamma_X}{m_X}\;\ll\;1\,.
\label{eq:NWAjust}
\end{equation}
tiny values of the coupling $g_e$ imply 
\begin{equation}
    \frac{\Gamma_X}{m_X}\;\approx\mathcal{O}\!\left(\frac{g_e^2}{8\pi}\right)\;\ll\;1 \,.
\end{equation}

Notice that among the decay channels of $X$ allowed by the interaction Lagrangian (eq. (\ref{eq:Lint})), we will focus only on $X \to e^+\,e^-$; since the dark-sector couplings governing $X \to \chi\,\chi$ are expected to be tiny (as we will see in what follows), and the ones entering in $X \to \chi\,\nu$ decay are phenomenologically ultra-suppressed, as discussed in app. \ref{app:NeuCou}. Indeed the total decay width can be approximated to:

\begin{equation}
    \Gamma_X\approx\Gamma_{X\rightarrow e^+e^-}=\frac{g_e^2}{8\,\pi\,m_X^2}(m_X^2-4m_e^2)^\frac{3}{2}\,.
\label{eq:GammaX}
\end{equation}

Using the NWA

\begin{equation}
\frac{1}{(\hat s-m^2)^2 + m^2\Gamma^2}
\;\xrightarrow{\Gamma \ll m}\;
\frac{\pi}{m\,\Gamma}\,\delta(\hat s-m^2)\,,
\label{eq:BW}
\end{equation}

so that the abundance can be expressed as 

\begin{equation}
         Y_\chi=g_\chi^2\,\frac{135}{1.66 \cdot 64 \pi^4} 
\frac{M_{Pl}}{g_s^* \sqrt{g_\rho^*}} 
\,\frac{g_e^2(m_X^2 - 4 m_\chi^2)^{3/2}\,(m_X^2 - 4 m_e^2)^{3/2}}
{8\,\pi\,\Gamma_X\, m_X^6}\,.
\label{eq:YscattpreGam}
\end{equation}

We stress that, in the presence of a thermal bath, the validity of the NWA is more subtle than the one proposed in equation~\ref{eq:NWAjust}.
In appendix~\ref{app:ConNum} we report both a comparison with the full numerical integration of the abundance integral in eq.~\eqref{eq:Yeecc} and the discussion of the validity domain of the NWA developed in ref.~\cite{Heeck:2014zfa}.

Substituting the decay rate for the $X\to e^+e^-$ from equation \ref{eq:GammaX} in equation \ref{eq:YscattpreGam} we obtain
\begin{equation}
         Y_\chi=g_\chi^2\,\frac{135}{1.66 \cdot 64 \pi^4} 
\frac{M_{Pl}}{g_s^* \sqrt{g_\rho^*}} 
\,\frac{(m_X^2 - 4 m_\chi^2)^{3/2}}
{m_X^4}\,.
\end{equation}

Comparing the resonant result above with eq.~(\ref{eq:YXcc}), one immediately sees that the two expressions coincide. In the following, we will clarify the physical reason for this agreement and how it is linked to the treatment of the on–shell $X$ in thermal equilibrium.
The NWA, eq.~(\ref{eq:BW}), makes explicit that the resonant contribution to the scattering is dominated by the production of an on–shell $X$ from the $e^+e^-$ pair, followed by its decay into DM. In practice, the main contribution to the abundance comes from scatterings where the center–of–mass energy satisfies $s \simeq m_X^2$, where the available energy is just enough to produce the mediator on–shell. For this reason, the process $e^+ e^- \to \chi\chi$ is effectively described as a two–step sequence: the pure annihilation $e^+ e^- \to X$ and the decay $X \to \chi\chi$. This factorization of the resonant contribution is the usual optical–theorem picture: the imaginary part of the propagator picks up the on–shell intermediate state and relates the scattering amplitude to the decay width of $X$.

The key point is that, in our setup, the $X$ boson is assumed to remain in equilibrium with the SM bath. This means that its number density is fixed by equilibrium and does not depend on the details of how efficiently $e^+e^-$ can resonantly produce $X$. As a consequence, the only relevant ingredient for the dark matter abundance is the decay $X \to \chi\chi$ itself, evaluated for an equilibrium population of $X$ bosons. This is precisely the contribution we computed in the decay picture in sec. \ref{sec:DiagEva}, and it explains why the resonant scattering result in eq. (\ref{eq:YXcc}) reproduces the same expression for $Y_\chi$.
Including both the decay channel and the resonant part of the scattering channel would therefore amount to double counting the same physical process.  
For this reason, in what follows we retain only the decay contribution.

We stress that the resonant expression above applies to both mass hierarchies $m_\chi > m_e$ and $m_\chi < m_e$. Indeed, as long as
$m_\chi < m_X/2$, the lower limit of the integral,
$s_{\min} = \max\big(4m_e^2,4m_\chi^2\big)$, always lies below $m_X^2$.
Therefore, the integration domain includes the on-shell pole at
$s \simeq m_X^2$ picked out by the narrow-width $\delta$-function.

\subsection{Non-Resonant Regime}

In this case only $e^+\,e^-\to \chi \,\chi$ is kinematically open, and the $X$ never goes on-shell. As a consequence, there is no resonant peak in the propagator, and the scattering contribution is much smaller than in the previous case.

In this regime $m_\chi > m_X/2$, thus the lower limit of the integral in eq. (\ref{eq:Yeecc}) is $\hat s_{\min} = 4m_\chi^2$.
Since then $m_\chi \gg m_e$, throughout the integration domain one has $\hat s > 4m_\chi^2 \gg 4m_e^2$, so the electron mass can be safely neglected, which simplifies the integral in eq. (\ref{eq:Yeecc}) to

\begin{equation}
Y_\chi\approx\frac{135}{1.66\cdot64\,\pi^5}\,
\frac{M_{\rm Pl}}{g_{s}^{*}\sqrt{g_{\rho}^{*}}}\,
\frac{g_\chi^{2}g_e^{2}}{8\pi}
\int_{\hat s_{\min}}^{\infty}\! d\hat s\;
\frac{\big(\hat s-4m_\chi^{2}\big)^{3/2}}
{\hat s\,\Big[\Gamma_X^{2} m_X^{2}+\big(m_X^{2}-\hat s\big)^{2}\Big]}\, \,.
\label{eq:Ymeto02}
\end{equation}

eq.\ref{eq:Ymeto02} can be re-written by computing the integral analitically, obtaining

\begin{equation}
\begin{aligned}
   & Y_\chi\approx\dfrac{135}{1.66\cdot64\,\pi^5}\,
\frac{M_{\rm Pl}}{g_{s}^{*}\sqrt{g_{\rho}^{*}}}\,
\frac{g_\chi^{2}g_e^{2}}{8\,\Gamma_X\, m_X^{2}\, \big(\Gamma_X^{2} + m_X^{2}\big)}\times\big[8\,\Gamma_X\,m_\chi^{3}
  - \!\left(\Gamma_X^{2} m_X^{2} + \big(-4 m_\chi^{2} + m_X^{2}\big)^{2}\right)^{3/4}
\\
  & \qquad \quad\!\left[
    \Gamma_X \cos\!\left(\frac{3}{2}\,\arctan\!\big(\frac{4 m_\chi^{2}-m_X^{2}}{\,\Gamma_X m_X}\big)\right)
    + m_X \sin\!\left(\frac{3}{2}\,\arctan\!\big(\frac{4 m_\chi^{2}-m_X^{2}}{\,\Gamma_X m_X}\big)\right)
  \right]\big]\,.
\end{aligned}
\label{eq:Ymeto0}
\end{equation}
As in the resonant case, we explicitly compare the analytic result with the full numerical integration in app. \ref{app:ConNum}, thereby confirming the validity of the approximation in the non-resonant regime.

It is useful to look at the asymptotic behavior of eq. (\ref{eq:Ymeto0}) in the limit of very heavy dark matter. 
Expanding for $m_\chi \gg m_X$ one finds, at leading order in $1/m_\chi$,
\begin{equation}
Y_\chi \;\simeq\; \frac{135}{1.66 \cdot 64 \,\pi^{5}} \,\frac{3}{16 \cdot 8}\;
\frac{g_\chi^{2}\,g_e^{2}\,M_{\rm Pl}}
     {g_{s}^{*}\,\sqrt{g_{\rho}^{*}}\,m_\chi}
\;+\;\mathcal{O}\!\left(\frac{1}{m_\chi^{3}}\right).
\label{eq:Ychi_asympt}
\end{equation}

\subsection{Relic-Abundance Condition and Freeze-In Parameter Space}

After deriving the analytic expressions for the freeze-in abundance in the different kinematic regimes, we can now combine the results and determine the region of parameter space that reproduces the observed DM relic density, using the standard formula
\begin{equation}
    \Omega_\chi h^2
= \frac{m_\chi\,s_0\,Y_\chi}{\rho_c/h^2}\,,
\label{eq:Ytochi}
\end{equation}
where $s_0 \simeq 2.89\times 10^{3}\,\text{cm}^{-3}$ and 
$\rho_c/h^2 \simeq 1.05\times 10^{-5}\,\text{GeV}\,\text{cm}^{-3}$ 
are standard cosmological parameters, for which we adopt the values reported in Ref.\cite{ParticleDataGroup:2024cfk}.

\begin{figure}[t]
    \centering
    \includegraphics[width=0.7\linewidth]{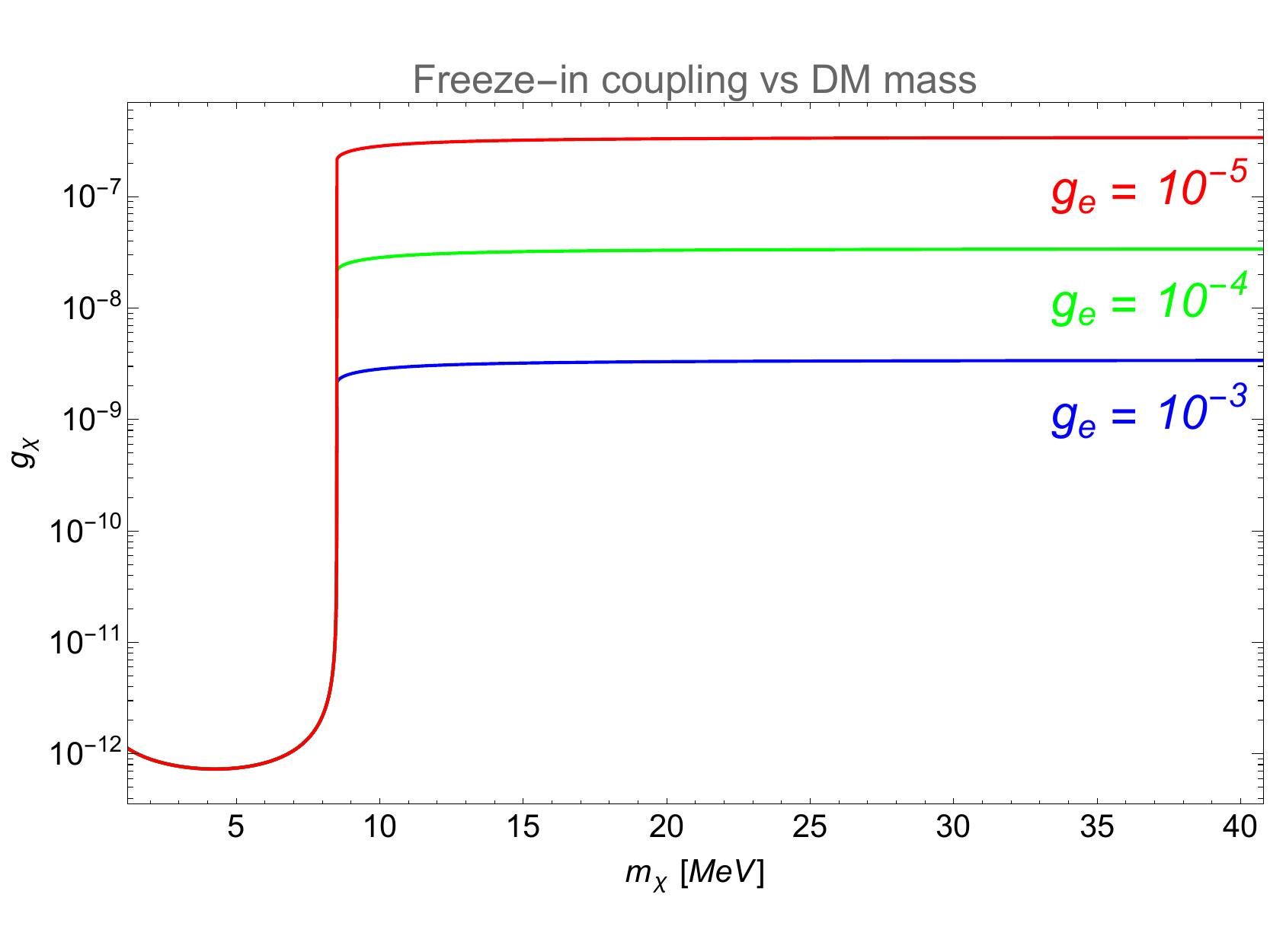}
    \caption{Freeze--in parameter space in the plane $(m_\chi,\,g_\chi)$ reproducing the observed DM relic abundance within the model discussed in cap.~\ref{cap:TheFra}.
    The curves are obtained by numerically evaluating the freeze--in abundance from decays and scattering processes and imposing $\Omega_{\rm DM} h^{2}=0.12$.
    Different regions correspond to the resonant and non-resonant regimes discussed in sec.~\ref{sec:regimes}. In this plot we fix the mediator mass to $m_X = 17~\mathrm{MeV}$ and consider three representative benchmark values of the electron coupling, $g_e = 10^{-3},\,10^{-4},\,10^{-5}$.
    }
    \label{fig:FIpar}
\end{figure}

In particular, within the framework outlined in cap. \ref{cap:TheFra}, the freeze-in mechanism fixes a relation between the dark-sector coupling $g_\chi$ and the DM mass $m_\chi$.
Imposing the relic-density condition discussed in sec. \ref{sec:DM} 
\begin{equation}
\Omega_{\rm DM}\,h^2\big(g_e, m_X,g_\chi , m_\chi\big)=0.12\,,
\end{equation}
we can then identify the subset of parameters for which our scenario accounts for the present dark–matter abundance. In the present analysis we fix the $X$ mass and its coupling to electrons ($g_e$) according to phenomenologically motivated considerations. In particular, we take $m_X = 17\,\mathrm{MeV}$ , consistently with the $X_{17}$ interpretation, and we consider three representative benchmark values for the electron coupling, namely $g_e = 10^{-3},\,10^{-4},\,10^{-5}$. With these choices, the relic-density condition above determines three curves in the $(g_\chi,m_\chi)$ plane along which the observed value $\Omega_{\rm DM}h^2\simeq 0.12$ is reproduced, and plotted in fig.~\ref{fig:FIpar}.

The behavior shown in fig. \ref{fig:FIpar} is a direct consequence of the freeze-in production mechanism. As discussed in sec. \ref{sec:FIFO}, in this regime the dark sector never thermalizes with the SM bath, and reproducing the observed relic abundance requires extremely suppressed interaction strengths. In particular, the effective coupling ($g_{eff}$) controlling DM production must typically lie around $g_{\rm eff}\sim 10^{-13}$ \footnote{in agreement with the general freeze-in estimates obtained in ref.~\cite{Hall:2009bx}}, which explains why the values of $g_\chi$ obtained in fig. \ref{fig:FIpar} are very small.

The sharp change in behaviour visible around $m_\chi = m_X/2$ can be understood as the transition between the resonant and non-resonant regimes. For $m_\chi < m_X/2$ the process proceeds through an on-shell mediator and we are in the resonant regime, in which DM production is dominated by decays and the effective coupling is simply $g_{\rm eff}= g_\chi \sim 10^{-13}$. For $m_\chi > m_X/2$ the resonance is no longer accessible and the dominant production mechanism becomes scattering. In this case the relevant effective coupling is $g_{\rm eff} = g_\chi\,g_e \sim 10^{-13}$, so that a larger value of $g_\chi$ is required to reproduce the observed relic abundance. This explains the upward shift of the curves in the non--resonant region.

Finally, it is worth stressing what is original about the result shown in fig. \ref{fig:FIpar}. The plot combines three specific ingredients: a light \(\sim 17~\mathrm{MeV}\) mediator motivated by the PADME anomalies and DM production via the freeze-in mechanism. Works that do consider freeze-in typically treat the mediator in a model–independent way and do not match the electron coupling to the PADME–preferred scale. By concentrating on a light scalar mediator and showing that the observed relic abundance can be consistently reproduced in a genuine freeze-in regime, our analysis provides a novel and complementary perspective on DM directly linked to the PADME anomalies.

Having identified the region of parameter space that yields the correct freeze‑in relic abundance, we now turn to its phenomenology. In the next chapter, we confront this region with current astrophysical and experimental bounds to assess the model’s viability and its prospects for future tests.

\chapter{Our Results: Astrophysical and Low‑Energy Probes}
\label{cap:DMExpSig}                

The goal of this chapter is to understand whether our DM model can be further constrained through astrophysical and low-energy observations. In the first case, we will focus on indirect detection (ID), which may be a further important probe on the NP couplings of our model, complementarily to the relic abundance calculation discussed in the previous chapters. 
Then, we will compare our results with additional constraints coming from low–energy particle–physics observables, which are sensitive to light mediators coupled to electrons.

\section{Indirect Detection Searches} 
\label{sec:ID}

ID is a strategy to search for DM by looking at SM particles produced by DM annihilations or decays in astrophysical environments, such as the Milky Way halo.  
These SM particles, if detected, may carry distinct information on their DM origin.

There is, however, a well--known sensitivity gap in indirect--detection searches in the
$\mathcal{O}(0.1-100)\,\mathrm{MeV}$ energy range, where current experiments are
significantly less sensitive than at both higher (GeV--TeV) and lower (keV) energies:
\begin{itemize}
\item charged SM particles are not valid probes below the GeV scale, since solar modulation prevents a reliable detection of low–energy cosmic rays;
\item neutrinos do not provide competitive bounds in this window, due to limited experimental sensitivity (see \cite{Bell:2020rkw}).
\item Thus, photons represent the only and most promising ID channel in this energy range, even though current experiments are still only marginally sensitive to energies $\mathcal{O}(\rm {MeV})$–$\mathcal{O}(\rm {GeV})$.
\end{itemize}
An alternative strategy for ID consists in looking at photons with energies well below the MeV scale.  
Such low-energy photons, specifically X-rays, can be produced as secondary radiation from DM annihilations.
In particular, in the context of our model the dominant annihilation channel will be $\chi\chi \to e^+e^-$ and will lead to the injection of energetic electrons and positrons in the astrophysical environment.
These charged particles may subsequently radiate photons through well-known electromagnetic processes.

This kind of signal originates from two main channels: inverse Compton scattering (ICS), where the produced $e^\pm$ up-scatter ambient low-energy photons in the astrophysical environment, and final state radiation (FSR), associated with photon emission directly from the final state charged particles in the DM annihilation process. 

We will now briefly discuss the formalism related to the production of X-rays from DM annihilations, as discussed in ref.~\cite{Cirelli:2020bpc}. In particular, we will derive the differential flux of such photons.
We will then make use of the results of ref. \cite{Cirelli:2020bpc}, obtained by integrating the
differential flux over the spatial region probed by the \textsc{Integral} telescope \cite{Bouchet_2011}.

A revised and updated version of the analysis of \cite{Cirelli:2020bpc} has subsequently appeared, making use of the same ID strategy but incorporating a broader set of X-ray observations \cite{Cirelli_2023}. While the methodology is essentially unchanged, the updated study combines data from several instruments, namely \textsc{Xmm-Newton} \cite{Foster:2021ngm}, \textsc{Integral}, \textsc{NuStar} \cite{Krivonos:2020qvl,Hong:2016qjq} and \textsc{Suzaku}\cite{Yoshino:2009kv}.
By exploiting this enlarged dataset, the authors significantly strengthen the constraints on sub-GeV dark matter.
These results will place upper bounds on the thermally averaged annihilation cross section, which we will thus compare with the one predicted within our model.

\subsection{X-Ray from Dark Matter Annihilations}

We fix a generic direction of observation from Earth, specified either by the angular separation $\theta$ from the Galactic Center or, equivalently, by the Galactic coordinates $(b,\ell)$. The corresponding line of sight (LOS) is the straight line connecting the observer to points in the Galactic halo along this direction, parametrized by the distance from Earth.
Along this LOS, DM particles annihilate at each point in the halo, producing photons that contribute to the observed signal.
The observed signal is therefore described in terms of a differential photon flux ${\mathrm{d}\Phi_\gamma}/{(\mathrm{d}E_\gamma\,\mathrm{d}\Omega)}$, which is computed by integrating the local flux contributions over the LOS.
 
We start with the contribution to the flux coming from FSR, where a photon is emitted by one of the charged particles in the final state of the annihilation process $\chi\chi \to e^+e^-$.

To compute the local flux contributions, we first need to derive the annihilation rate per unit volume
\begin{equation}
  \Gamma_{\rm ann}(r)
  = \frac{1}{2}\,\langle\sigma v\rangle\,n_\chi^2( r)
  = \frac{1}{2}\,\langle\sigma v\rangle
    \left(\frac{\rho(r)}{m_{\chi}}\right)^2 ,
\end{equation}
where $\langle\sigma v\rangle$ is the thermal-averaged annihilation cross section and
the factor $1/2$ avoids double counting of identical pairs. The radial coordinate $r$ measures the distance from the Galactic Center and can be expressed in terms of the LOS variable $s$ (with $s=0$ corresponding to the position of the Earth) according to $r^2 = s^2 + r_\odot^2 - 2 s r_\odot \cos\theta$. Here $r_\odot$ represents the solar distance to the Galactic Center.

Secondly, let $dN_{FSR\,\gamma}/dE_\gamma$ denote the number of photons produced via FSR per annihilation and per unit energy in the channel of interest. Then
\begin{equation}
  \Gamma_{\rm ann}(r)\,\frac{dN_{FSR\,\gamma}}{dE_\gamma}
  = \frac{1}{2}\,\langle\sigma v\rangle
    \left(\frac{\rho(r)}{m_\chi}\right)^2
    \frac{dN_{FSR\,\gamma}}{dE_\gamma} \, .
\end{equation}

The flux observed from a direction $\theta$ on the sky is obtained by integrating the
quantity just derived along the LOS in that direction and dividing by $4\pi$ (assuming isotropic emission):
\begin{equation}
  \frac{d\Phi_{FSR\,\gamma}}{dE_\gamma\,d\Omega}
  = \frac{1}{4\pi}
    \int_{\text{l.o.s.}} ds\;
    \frac{1}{2}\,\langle\sigma v\rangle
    \left(\frac{\rho\big(r(s,\theta)\big)}{m_\chi}\right)^2
    \frac{dN_{FSR\,\gamma}}{dE_\gamma}\, .
\end{equation}

It is convenient to factor out the local DM density $\rho_\odot$ at the Sun's
position and to define the
dimensionless $J$-factor
\begin{equation}
  J(\theta) \equiv
  \int_{\text{l.o.s.}} \frac{ds}{r_\odot}
  \left(\frac{\rho(\mathbf r(s,\theta))}{\rho_\odot}\right)^2 .
\end{equation}
The flux can then be written as
\begin{equation}
  \frac{d\Phi_{FSR\,\gamma}}{dE_\gamma\,d\Omega}(\theta)
  = \frac{1}{2}\,\frac{r_\odot}{4\pi}
    \left(\frac{\rho_\odot}{m_\chi}\right)^2
    J(\theta)\,
    \langle\sigma v\rangle\,
    \frac{dN_{FSR\,\gamma}}{dE_\gamma} \, .
\label{eq:IDFSR}
\end{equation}
This result is consistent with the standard expression reported in ref. \cite{Cirelli:2020bpc}. For completeness, the Galactic DM distribution is usually modelled with a Navarro-Frenk-White profile \cite{Navarro:1995iw}
\begin{equation}
  \rho_{\rm NFW}(r)
  = \rho_s\,\frac{r_s}{r}\left(1+\frac{r}{r_s}\right)^{-2},
\end{equation}
with $\rho_s = 0.184\,\mathrm{GeV/cm^3}$ and $r_s = 24.42\,\mathrm{kpc}$.

The quantity $\langle\sigma v\rangle$ denotes the thermally averaged annihilation
cross section of the process $\chi\,\chi\to e^+\,e^-$, defined in
Eq. \ref{eq:TheAveCS}. In our setup this is the only model--dependent ingredient
entering Eq. \ref{eq:IDFSR}, while the remaining factors follow the standard treatment
of ref. \cite{Cirelli:2020bpc}.

The last missing ingredient is the spectrum of photons from FSR, for which we adopt the expressions of ref. \cite{PhysRevD.72.114019}:
\begin{equation}
  \frac{dN^{e^+e^-}_{\rm FSR\,\gamma}}{dE_\gamma}
  =
  \frac{\alpha}{\pi\,\beta\,(3-\beta^2)\,m_\chi}
  \left[
    \mathcal{A}\,
    \ln\!\left(\frac{1+R(\nu)}{1-R(\nu)}\right)
    - 2\,\mathcal{B}\,R(\nu)
  \right],
\end{equation}
where
\begin{equation}
  \mathcal{A}
  =
  \left[
    \frac{(1+\beta^2)(3-\beta^2)}{\nu}
    - 2(3-\beta^2)
    + 2\nu
  \right],
\end{equation}
\begin{equation}
  \mathcal{B}
  =
  \left[
    \frac{3-\beta^2}{\nu}(1-\nu) + \nu
  \right].
\end{equation}
Here we have defined
\begin{equation*}
  \nu \equiv \frac{E_\gamma}{m_\chi},
  \qquad
  \beta^2 \equiv 1 - 4\mu^2,
  \qquad
  \mu \equiv \frac{m_\ell}{2m_\chi},
\end{equation*}
together with
\begin{equation*}
  R(\nu) \equiv \sqrt{1-\frac{4\mu^2}{1-\nu}}\,.
\end{equation*}

We now move on from the discussion of final state radiation and turn to ICS, which provides an additional contribution to the photon flux relevant for indirect detection.
As anticipated above, ICS probes photons that have been upscattered by electrons and positrons produced in DM annihilations within the Galaxy.
This contribution is obtained by integrating the emissivity produced in each infinitesimal volume element along the LOS. Unlike the FSR case, the resulting photon flux depends on the local distribution of photons, and therefore on the detailed astrophysical environment at each point in the Galaxy.

The differential photon flux from ICS along a given direction of observation can thus be written as
\begin{equation}
\frac{d\Phi_{\mathrm{IC}\,\gamma}}{dE_\gamma\,d\Omega}
=
\frac{1}{E_\gamma}
\int_{\mathrm{l.o.s.}} ds\;
\frac{j\!\left(E_\gamma,\vec{x}(s,b,\ell)\right)}{4\pi}\,,
\label{eq:IDICS}
\end{equation}
where $j(E_\gamma,\vec{x})$ denotes the photon emissivity at a position $\vec{x}$.

The emissivity depends on the spectrum of injected $e^\pm$, their energy losses in the Galactic medium, and the properties of the interstellar radiation field.
Its explicit computation requires a detailed modeling of the Galactic environment and typically involves a numerical treatment.
All the details have been already developed in the literature and can be found in ref. \cite{Cirelli:2020bpc}.

Finally, combining eq. (\ref{eq:IDFSR}) and (\ref{eq:IDICS}) and integrating over the selected region of observation, the total flux of X-rays coming from DM annihilations is obtained:

\begin{equation}
    \frac{\mathrm{d}\Phi_{\mathrm{DM}\,\gamma}}{\mathrm{d}E_\gamma}
=\int_{b_{\min}}^{b_{\max}}\!\int_{\ell_{\min}}^{\ell_{\max}}
\mathrm{d}b\,\mathrm{d}\ell\,\cos b\,
\left(
\frac{\mathrm{d}\Phi_{\mathrm{FSR}\,\gamma}}{\mathrm{d}E_\gamma\,\mathrm{d}\Omega}
+
\frac{\mathrm{d}\Phi_{\mathrm{IC}\,\gamma}}{\mathrm{d}E_\gamma\,\mathrm{d}\Omega}
\right)
\label{eq:IDflux}
\end{equation}

The integration boundaries $b_{\min,\max}$ and $\ell_{\min,\max}$ specify the region of the sky over which the signal is collected and depend on the particular observational strategy and field of view of the experiment under consideration.

In the next section we will compute our theoretical predictions for  $\langle\sigma v\rangle$ within our model and compare them with the most recent ID constraints.

\subsection{Limits on the Thermally-Averaged Annihilation Cross Section}
\label{sec:LimTerCroSec}

In this section we derive constraints on the thermally averaged annihilation cross section $\langle\sigma v\rangle$ by comparing our prediction with the ID limits.
In order to do so, we first need to specify how the thermally averaged cross section
is evaluated.

In the Galactic environment, DM particles are highly non–relativistic, with typical
velocities of order $v \sim \mathcal{O}(100)\,\rm {km/s}$.
It is therefore well justified to work in the non–relativistic limit when computing
the thermal average of the annihilation cross section.

The thermally averaged cross section (defined in chapter \ref{cap:DMAbCalcu}) can be expressed in terms of the {M{\o}l}ler velocity.
The {M{\o}l}ler velocity is defined as 
\begin{equation}
v_{\text{M{\o}l}}
=
\sqrt{\left|v_1-v_2\right|^2-\left|v_1\times v_2\right|^2}\,.
\end{equation}
Here $v_1$ and $v_2$ denote the velocities of the two incoming particles, and are given by $v_1 = p_1/E_1$ and $v_2 = p_2/E_2$.
Eq. (\ref{eq:TheAveCS}) can thus be expressed in terms of the {M{\o}l}ler velocity as:

\begin{equation}
\langle \sigma v \rangle
=
\frac{\displaystyle \int \sigma\, v_{\text{M{\o}l}}\,
e^{-E_1/T}\,e^{-E_2/T}\, d^3 p_1\, d^3 p_2}
{\displaystyle \int
e^{-E_1/T}\,e^{-E_2/T}\, d^3 p_1\, d^3 p_2}\,.
\end{equation}

In the numerator, $\sigma$ denotes the annihilation cross section for the process under consideration, as computed in the previous Chapter.
This expression for the thermally averaged cross section coincides with the standard equation originally derived in ref. \cite{Gondolo:1990dk}. Moreover, following ref. \cite{Gondolo:1990dk}, in the non-relativistic limit the thermal average can be rewritten in terms of the relative velocity $v_r=|v_1-v_2|$.
One finds that
\begin{equation}
\begin{aligned}
\langle \sigma v \rangle_{n.r.}
=&
\int_{0}^{\infty} dv_r\;
\sqrt{\frac{2}{\pi}}\left(\frac{m_\chi}{2\,T}\right)^\frac{3}{2}v_r^{2}\,
e^{\!\left(-\,\frac{m_\chi}{2T}\,\frac{v_r^{2}}{2 }\right)}\,
\sigma v_r \,  \\
=&\int_{0}^{\infty} dv_r\, F_M(v_r)\,\sigma v_r
\end{aligned}
\label{eq:VelAveSig}
\end{equation}

The physical interpretation of this formula is straight-forward: in this non-relativistic regime, the thermal average reduces to an average over the
relative velocity $v_r$ of the annihilating particles.
The corresponding distribution $F_M(v_r)$ has the same functional form as the
Maxwell-Boltzmann distribution for the absolute velocity, with the particle mass
replaced by the reduced mass $\mu = m_\chi/2$ and $v_r$ replacing the
single-particle velocity.

The computation of the annihilation cross section proceeds in close analogy with the calculation of  $ e^+ e^-\to \chi\,\chi$  performed in sec. \ref{sec:Scateecc}.
The squared amplitude for the process $\chi\,\chi \to e^+ e^-$ is the one derived in eq. \ref{eq:MeetoccD}. Inserting it into the standard $2\to2$ cross–section
formula we obtain, in terms of the Mandelstam variable
$\hat s = (p_{\chi_1}+p_{\chi_2})^2$
\begin{equation}
\sigma(\hat s)
=
\frac{g_\chi^{2}\,g_e^{2}}
{4\pi \hat{g_\chi}^2}\,
\frac{\sqrt{\hat s - 4 m_\chi^{2}}\,
\left(\hat s - 4 m_e^{2}\right)^{3/2}}
{\hat s\,
\Bigl(\Gamma_X^{2} m_X^{2} + (m_X^{2}-\hat s)^{2}\Bigr)} \, .
\label{eq:sigccee}
\end{equation}

The Mandelstam variable $\hat s$ can be rewritten in terms of the relative velocity $v_r$ in the non-relativistic limit.
A common way of doing this (see e.g. \cite{Gondolo:1990dk})is to write $\hat s\simeq 4\,m_\chi^2+m_\chi^2\,v_r^2$. Substituting this relation in eq. (\ref{eq:sigccee}), and expanding around $v_r\approx0$ we obtain

\begin{equation}
\sigma\,v_r
=
\sigma_0+\sigma_1v_r^2 +\mathcal{O}(v_r^4)  \,,
\qquad
\qquad
\left\{
\begin{aligned}
\sigma_0 &\equiv0 \,,\\[4pt]
\sigma_1 &\equiv
\frac{g_\chi^{\,2}\, g_e^{\,2}\,\bigl(m^2_{\chi}-m^2_{e}\bigr)^{3/2}}
{8\,\pi\, m_{\chi}\,\Bigl(\Gamma_{X}^{2}\, m_{X}^{2} + \bigl(m_{X}^{2}-4 m_{\chi}^{2}\bigr)^{2}\Bigr)}.
\end{aligned}
\right.
\end{equation}

We stress that, as expected, the leading $s$-wave contribution vanishes, $\sigma_0 = 0$.

Being $\sigma_1$ velocity-independent, we can derive the non relativistic limit of the thermally averaged cross section from eq. \ref{eq:VelAveSig} as
\begin{equation}
    \langle \sigma v \rangle_{\mathrm{n.r.}}=\sigma_0\,+\,\sigma_1\langle v_r^2 \rangle
\end{equation}

Here $\langle v_r^2 \rangle$ denotes the average of the squared relative velocity,
computed with the Maxwell-Boltzmann distribution for the relative velocity $F_M(v_r)$ introduced above.
\begin{equation}
    \langle v_r ^2\rangle
    =
    \int_{0}^{\infty}
    dv\;
    F_M(v_r)\,v_r^2\,, 
    \qquad
    F_M(v_r)=\sqrt{\frac{2}{\pi}}
\left(\frac{\mu}{T}\right)^{3/2}
v_r^{2}\,
\exp\!\left(-\,\frac{\mu}{T}\,\frac{v_r^{2}}{2}\right) \, .
\end{equation}

The integral can be evaluated analytically, yielding
\begin{equation}
    \langle v_r ^2\rangle
    =
    \int_{0}^{\infty}
    dv\;
    \sqrt{\frac{2}{\pi}}
\left(\frac{\mu}{T}\right)^{3/2}
v_r^{4}\,
e^{-\,\frac{\mu\,v_r^{2}}{2T}}\,=\frac{3\,T}{\mu}\,.
\end{equation}

In a halo the DM particles can be treated, to a good approximation, as a collisionless gas with Maxwell–Boltzmann velocity distribution.  
Associating the mean kinetic energy per particle, we can derive the halo temperature
\begin{equation}
\frac{1}{2}m_\chi v_0^2 \;=\; \frac{3}{2}k_B T_{\rm halo}\,,  
\qquad \Longrightarrow  \qquad
T_{\rm halo} \;=\; \frac{m_\chi v_0^2}{3k_B}\,,
\end{equation}
where $v_0$ is the characteristic velocity scale of the Standard Halo Model (typically $v_0 \simeq 220~\mathrm{km/s}$ for the Milky Way).  
Identifying $T$ with $T_{\rm halo}$ in the expression above the average squared relative velocity becomes $\langle v_r^2 \rangle  = \,2\, v_0^2$.

A final remark concerns the resonant regime. Close to resonance, namely for 
$m_\chi \simeq m_X/2$, the non–relativistic expansion performed above is no longer valid and
the full velocity dependence of the cross section must be retained. 
A detailed treatment of this case has been presented in ref. \cite{Murayama:2025ihg}, 
where the annihilation cross section is written in terms of a narrow Breit--Wigner resonance as
\begin{equation}
    \sigma \;\simeq\; \sigma_0 \;+\; \sigma_R\,\frac{\pi\Gamma}{2}\,\delta(E-E_R)\,,
    \label{eq:MurEq4}
\end{equation}
where $E$ is the center–of–mass energy, $E_R$ the resonant energy, and $\Gamma$ the total width of the mediator.
Performing the thermal average over a Maxwell–Boltzmann velocity distribution, the authors obtain the following analytic expression for the resonant contribution to the annihilation cross section for $m_\chi = m_X/2$ 
\begin{equation}
    \langle \sigma v \rangle_{R}
    \;=\;
    \frac{2\pi^{1/2} v_R^{2}}{v_0^{3}}\,
    e^{-\,v_R^{2}/v_0^{2}}\,
    \sigma_R\,,
\label{eq:MurEq11}
\end{equation}
here $v_R$ is the resonant velocity, defined as in~\cite{Murayama:2025ihg}. For the numerical evaluation of the resonant cross section we adopt the same astrophysical velocity parameters discussed in ref.~\cite{Murayama:2025ihg}.
In particular, for the Milky Way halo we take again the characteristic velocity dispersion to be \(v_0 = 220~\mathrm{km/s}\).
\begin{figure}[tb]
    \centering
    \includegraphics[width=0.8\linewidth]{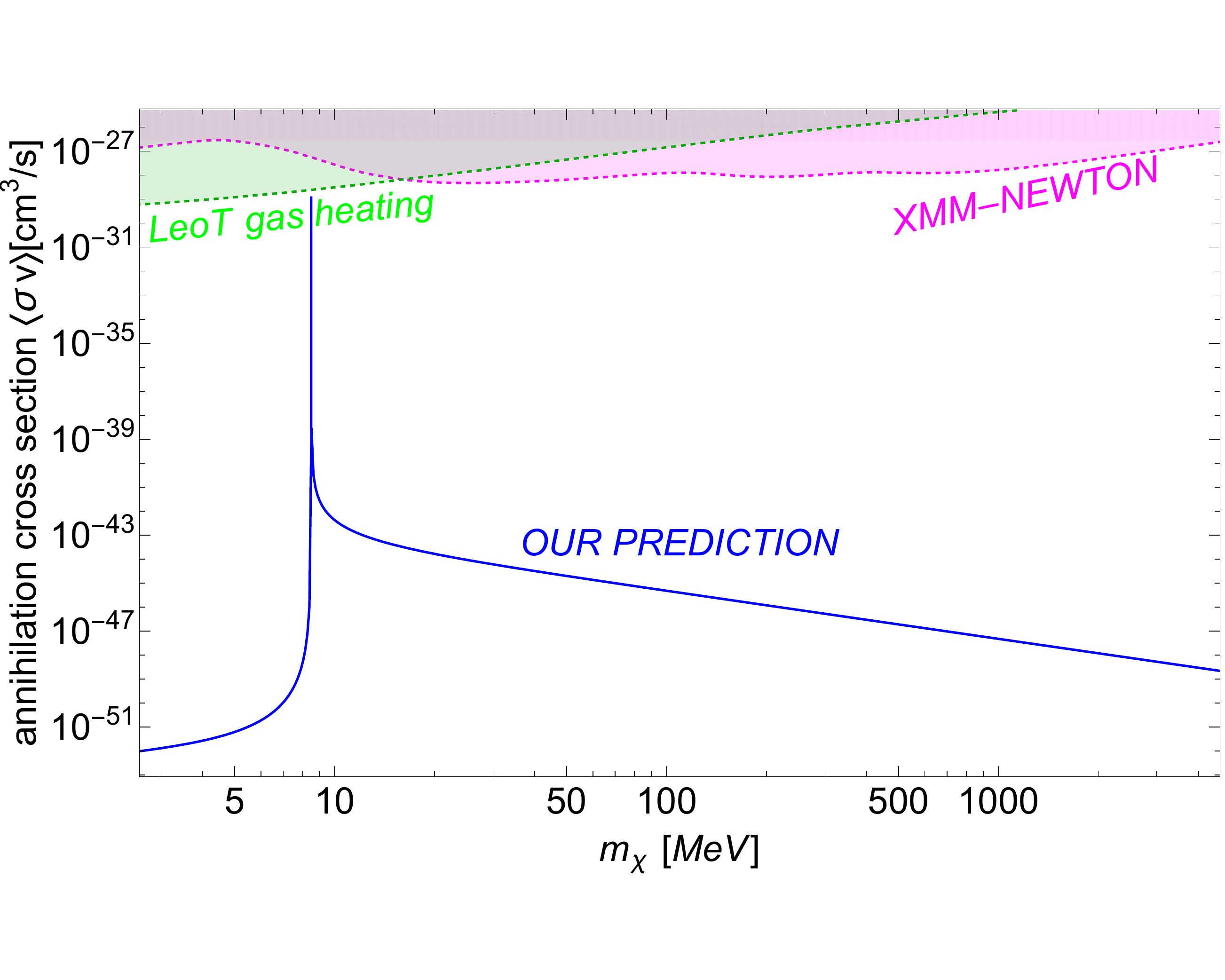}
    \caption{The experimental bound from indirect detection derived in the previous section (pink line) and the annihilation cross section predicted by our model.The green line is the bound derived from gas heating in the Leo-T dwarf galaxy.}
    \label{fig:ID_main}
\end{figure}

We can now evaluate the non–relativistic thermally averaged annihilation cross section and study its dependence on the DM mass.
We plot the resulting cross section and compare it with the experimental
bound derived from indirect detection in the previous section, as shown in
fig. \ref{fig:ID_main}. These limits are taken from the recent analysis of ref. \cite{Cirelli_2023}, where the DM–induced X-ray signal was computed following the same
phenomenological framework outlined in the previous section. The predicted flux is
obtained by integrating the differential photon flux over the region of the sky observed by the instrument; in our notation, this corresponds to the expression in
eq. \eqref{eq:IDflux}, evaluated over the \textit{XMM--Newton} field of view.
In the framework considered here, where the relevant signal originates from DM annihilations into $e^{+}e^{-}$ pairs, not all instruments contribute equally to the final constraint. Among the X–ray telescopes included in the analysis of ref. \cite{Cirelli_2023}, the most constraining limit in the mass range of interest for our model is provided by \textsc{Xmm-Newton}, and therefore sets the dominant bound shown in fig. \ref{fig:ID_main}.

The pink exclusion curve in fig. \ref{fig:ID_main} is precisely the bound derived in
ref. \cite{Cirelli_2023} from \textit{XMM--Newton} data. It is obtained by requiring that the DM–induced X-ray flux predicted in this way does not exceed the observed emission in the corresponding region of the sky. These limits currently represent the most stringent constraints on annihilating DM for $m_\chi\gtrsim 11\,MeV$.

In addition to X-ray constraints, we also show in fig. \ref{fig:ID_main} the limits derived from gas heating in the Leo-T dwarf galaxy. These bounds, indicated by the green curve, are obtained following the analysis of \cite{Wadekar:2021qae}, where the heating of the interstellar medium induced by DM annihilations is required not to exceed the observed gas cooling rate in Leo-T.

For what concerns the theoretical (blue) prediction, it is worth highlighting that it has been obtained by fixing the values of the NP couplings $g_e$ and $g_{\chi}$ in accordance with the final results shown in fig. 3.4. The mass of the mediator has been also fixed as $m_X = 17\,\rm MeV$.

Figure \ref{fig:ID_main} shows that the couplings determined through the freeze-in computations carried out in the previous Chapter are too tiny to be detected by current experiments. We expect that next-generation experiments will not able to close the gap of more than 10 orders of magnitude between the blue and the green/pink curves. For this reason, in the next Section we will investigate whether our results can be, instead, compared with some low-energy data independent of astrophysics and cosmology.

\section{Final Results: Bridging Cosmological with Low‑Energy Data}
\label{sec:CombinedConstraints}

In this section we will compare the results of the relic abundance computation within our theoretical framework with some complementary constraints coming from low-energy data.

\begin{figure}[tb]
    \centering
    \includegraphics[width=0.7
    \linewidth]{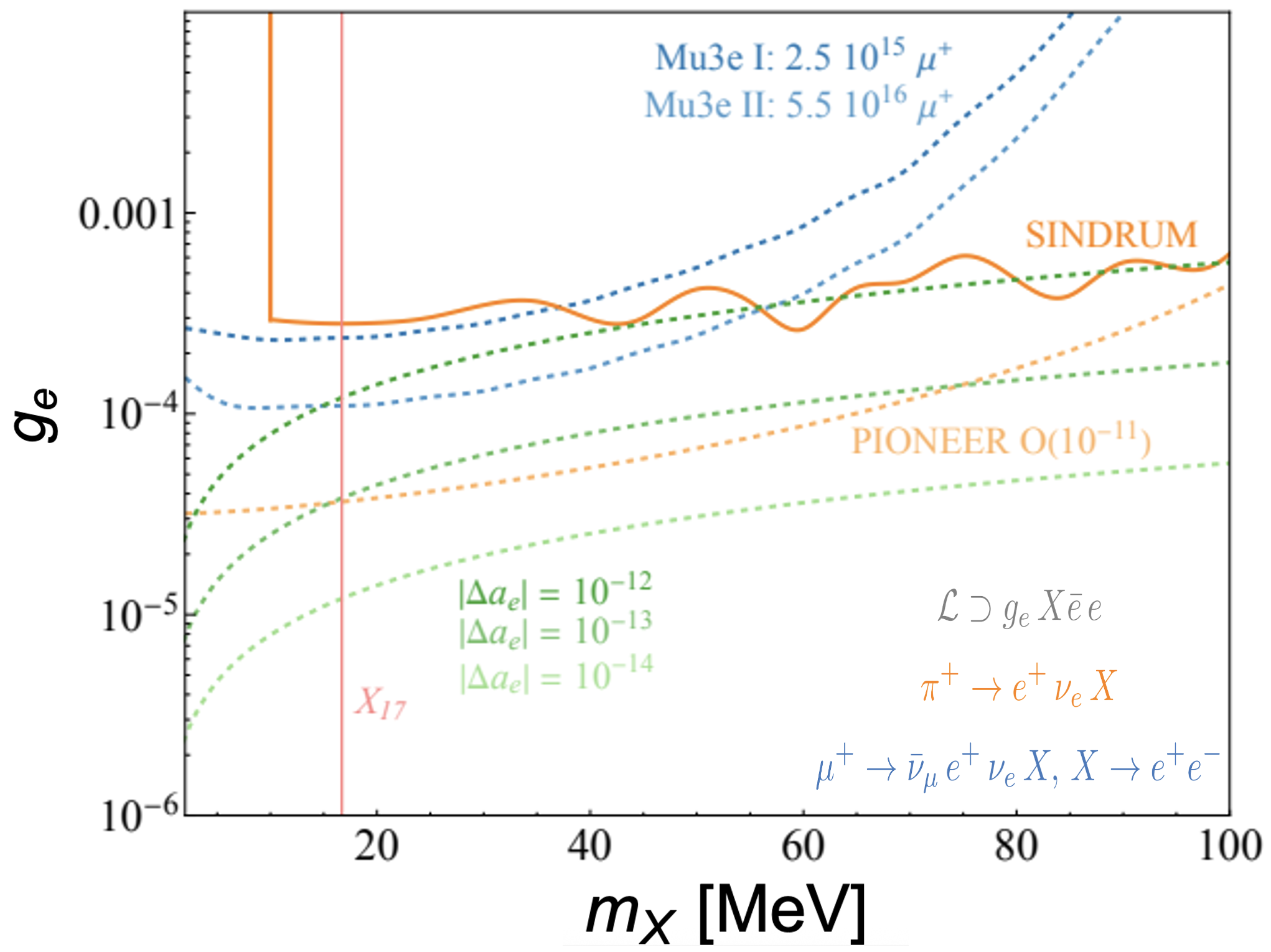}
    \caption{Exclusion plot for a scalar $X$-mediator, taken from
    ref. \cite{DiLuzio:2025ojt}.The solid curve corresponds to the current experimental limit, while the dashed curves indicate projected sensitivities of future experiments.}
    \label{fig:PlotLuztot}
\end{figure}

Ref. \cite{DiLuzio:2025ojt} has recently analyzed several low-energy observables which can, indeed, constrain the coupling of the $X$-mediator to electrons ($g_e$) and its mass ($m_X$) in a largely
model-independent way. Such quantities are related, for instance, to measurements of the anomalous magnetic moment of the electron \cite{Fan:2022eto}, as well as to rare pion and muon decays \cite{SINDRUM:1986klz, SINDRUM:1989qan, PIONEER:2022yag, Mu3e:2020gyw}. Such a study has allowed the authors of ref. \cite{DiLuzio:2025ojt} to obtain several bounds in the plane $(m_X,\,g_e)$, which are reported in fig. \ref{fig:PlotLuztot} for reference and which constrain the parameter space of the hypothetical novel $X_{17}$-boson in a way which is complementary to the direct search carried out at PADME.

The main original point of this thesis is, as already stated in the previous chapters, to relate such $X$-boson particle to DM. As a direct consequence of this, at variance with ref. \cite{DiLuzio:2025ojt}, we have to deal with more parameters, namely $(g_e,\, m_X,\, g_\chi,\, m_\chi)$. However, once some specific values are fixed for the DM mass $m_\chi$ and its coupling to the $X$-mediator $g_\chi$, we will be able to directly compare the predictions of our freeze-in computations with the low-energy constraints computed in ref. \cite{DiLuzio:2025ojt} and shown in fig. \ref{fig:PlotLuztot}.

It is clear that the chosen values of $m_\chi$ and $g_\chi$ have to be consistent with the condition
\begin{equation}
    \Omega_{\rm DM}\, h^2\bigl(g_e, m_X, g_\chi, m_\chi \bigr) = 0.12\,,
\label{eq:ImpAb}
\end{equation}
namely with the requirement that the abundance of DM produced within our framework matches the observations by Planck \cite{Planck:2018vyg}.

As discussed in sec. \ref{sec:regimes}, the freeze-in production of DM in our framework naturally separates into two qualitatively different regimes, depending on the hierarchy between the mediator mass and twice the DM mass. These two regimes lead to distinct phenomenological behaviors and therefore to different implications when the relic-abundance requirement (eq. \eqref{eq:ImpAb}) is combined with the low-energy constraints in fig. \ref{fig:PlotLuztot}.

\subsection*{Non-Resonant Phenomenology}
We begin by considering the non-resonant regime, in which $m_\chi > m_X/2$ and DM production is dominated by the off-shell process $e^{+}e^{-}\rightarrow \chi\chi$ rather than by on-shell mediator decays. This scattering is particularly interesting, since it is controlled not only by the DM coupling $g_\chi$, but also by the electron coupling $g_e$, which is precisely the parameter constrained by the low-energy observables studied in ref. \cite{DiLuzio:2025ojt}.

\begin{figure}[t]
    \centering
    \includegraphics[width=0.7\linewidth]{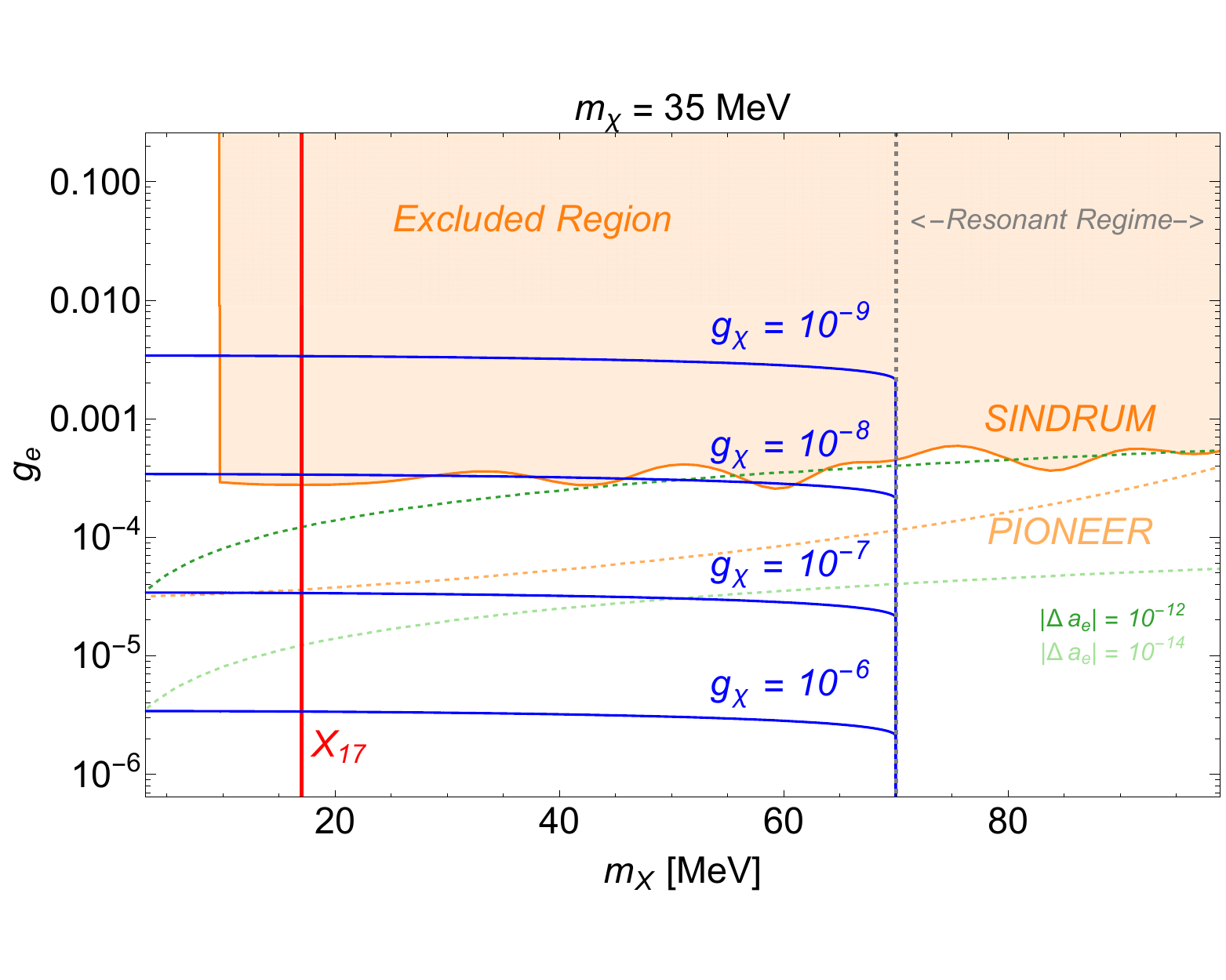}
    \caption{The orange solid band is the exclusion region on the $g_e-m_X$ plane coming from SINDRUM on $\pi^+\to e^+\nu_e X$ experiment, as computed in ref \cite{DiLuzio:2025ojt}. The orange and green dashed lines correspond to exclusion limits expected from next-generation experiments, as discussed and obtained in ref. \cite{DiLuzio:2025ojt}.
     The solid blue lines are the final results of our computations and show the values of
    $g_e$ that reproduce the observed DM relic abundance via freeze-in, by fixing 
    $m_\chi = 35~\text{MeV}$ and by selecting some benchmark values of the coupling $g_\chi$. The most important message of this plot is that, quite interestingly, the combination of the relic abundance and SINDRUM constraint yields a highly non-trivial information: if $m_\chi = 35$ MeV, then a lower bound on the NP coupling $g_\chi$ exists, namely $g_\chi \gtrsim 10^{-8}$. Such a lower bound may be even higher, as the projections by next-generation experiments suggest. The vertical gray dashed line at $m_X = 2\,m_\chi$ marks the transition between the non--resonant
    and resonant production regimes.
    }
    \label{fig:RelicOverlay}
\end{figure}

We proceed as follows: for fixed values of the dark sector coupling, $(m_\chi, g_\chi)$, we impose the relic-abundance condition 
\begin{equation}
    \Omega_{\rm DM}\,h^{2}(m_\chi, g_\chi,m_X,g_e)=0.12\,,
\end{equation}
which identifies a curve in the $(m_X,g_e)$ plane. These curves can then be directly overlaid on the exclusion plot of ref. \cite{DiLuzio:2025ojt}. In this way, the regions already excluded by low-energy measurements immediately translate into excluded portions of the parameter space of our DM model.
In this sense, the low-energy constraints on $g_e$ are effectively projected onto the dark sector, resulting in quantitative bounds on the strength of the $\chi$–$X$ interaction.
The outcome of this procedure is illustrated in fig. \ref{fig:RelicOverlay}, where the curves obtained from the condition above are superimposed onto the low-energy bounds of ref. \cite{DiLuzio:2025ojt}. 

In this plot we fix the DM mass to  $m_\chi = 35\,\mathrm{MeV}$ and consider four representative benchmark values of the DM coupling, namely $g_\chi = 10^{-9},\,10^{-8},\,10^{-7}$ and $10^{-6}$. Each of these values defines a distinct freeze--in contour in the $(m_X,g_e)$ plane, which can then be directly compared with the experimental exclusion regions.

We have also verified that varying the DM mass does not qualitatively modify the structure of the analysis. Changing $m_\chi$ mainly shifts the boundary between the resonant and non-resonant regimes, increasing or reducing the range of  allowed values of the mediator mass $m_X$, but it does not alter the qualitative behavior of the freeze-in curves. As a consequence, the dominant phenomenological quantity in this regime is the DM coupling $g_\chi$, which effectively controls the overall phenomenology of the relic-abundance curves, while $m_\chi$ only plays a secondary role by selecting the available kinematic domain.

The negligible dependence on the DM mass implies that the relevant phenomenological information in the $(m_X,g_e)$ plane can be directly translated into constraints on the coupling $g_\chi$. From the overlaid blue curves in fig. \ref{fig:RelicOverlay} one immediately observes that, in the non-resonant regime, the most stringent constraint currently arises from the SINDRUM bound on $\pi^{+}\!\rightarrow e^{+}\nu_e X$ \cite{SINDRUM:1986klz, SINDRUM:1989qan}. When projected onto our freeze-in curves, this exclusion region translates into a lower limit on the dark-sector coupling
\begin{equation}
    g_\chi \,\gtrsim \,10^{-8}\,.
\end{equation}

Up to now we have implicitly assumed that freeze-in production mediated by the $X$ boson is the sole mechanism responsible for the present DM abundance. This leads to the equality condition $\Omega_{\rm DM}h^{2}=0.12$. However, this hypothesis can be relaxed. If additional DM production mechanisms are present, the only robust requirement is that $X$–mediated freeze–in does not \emph{overproduce} dark matter. In this case the relic-density condition becomes an inequality,
\begin{equation}
    \Omega_{\rm DM}^{(X)}h^{2}(m_\chi, g_\chi,m_X,g_e)\;\leq\;0.12\,,
\label{eq:RelAbIne}
\end{equation}
where $\Omega_{\rm DM}^{(X)}$ denotes the contribution generated via $X$-mediated processes only.
The boundary of this region is still given by the usual freeze–in curve satisfying $\Omega_{\rm DM}h^{2}=0.12$, while the parameter space above it is excluded because it would lead to an excess of dark matter. 

\begin{figure}[t]
    \centering
    \includegraphics[width=0.7\linewidth]{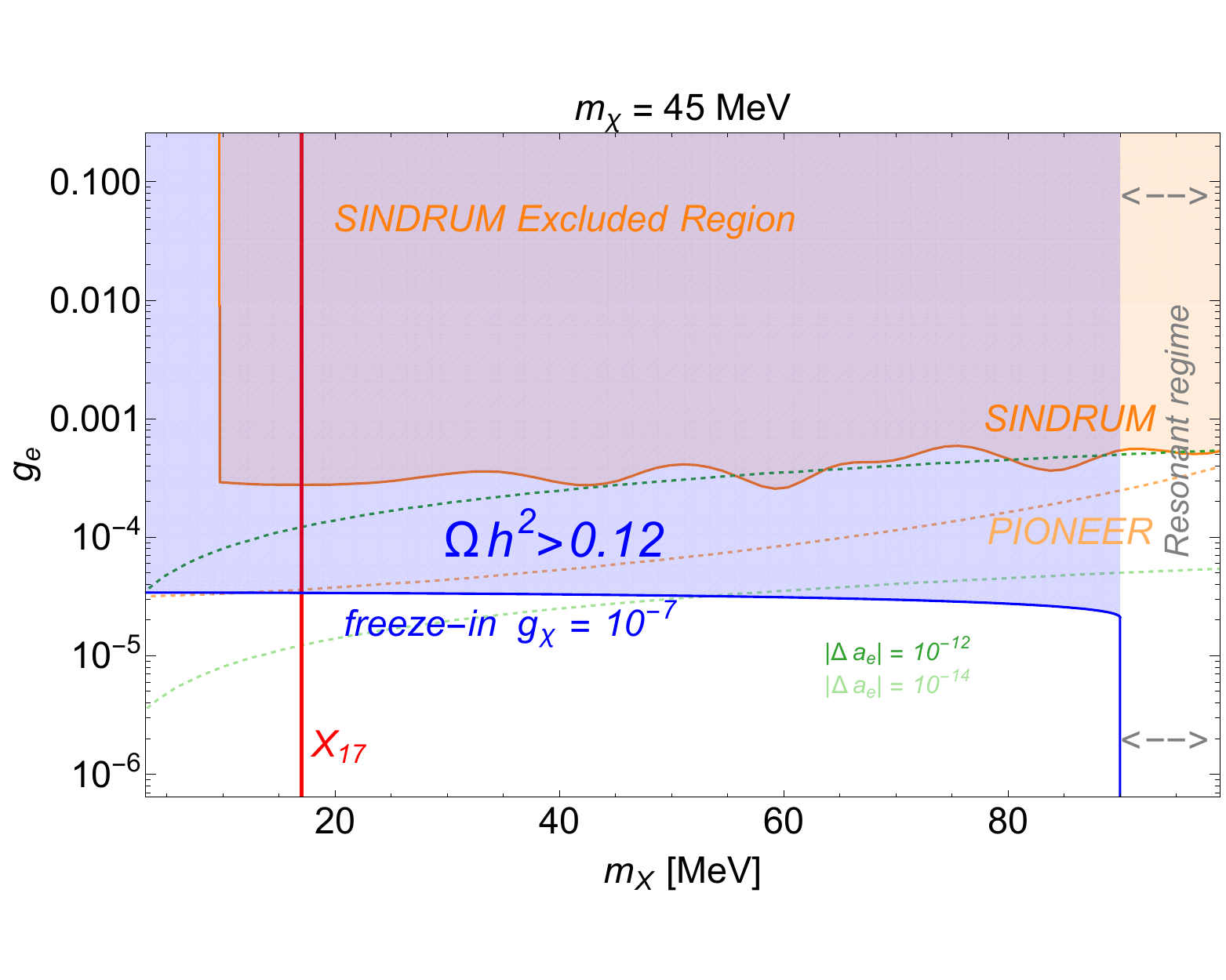}
    \caption{Freeze–in curve reproducing the observed relic abundance for $m_\chi = 45~\mathrm{MeV}$ and $g_\chi = 10^{-7}$ (solid line). The shaded region above the curve is excluded because it would lead to DM overproduction, corresponding to the condition $\Omega_{\rm DM}h^{2}>0.12$.The vertical gray dashed line at $m_X = 2\,m_\chi$ marks the transition between the non--resonant
    and resonant production regimes.}
    \label{fig:OverProd}
\end{figure}

This situation is illustrated in fig. \ref{fig:OverProd}, where, for definiteness, we fix $m_\chi = 45\,\text{MeV}$ and $g_\chi = 10^{-7}$. The solid line corresponds to the parameter combinations reproducing the observed relic abundance, whereas the shaded area indicates the region excluded by DM overproduction. 

It is worth emphasizing that this represents a genuinely new step with respect to ref. \cite{DiLuzio:2025ojt}. While that work constrained the parameter space of a light boson coupled to electrons in a largely model-independent way, here we embed the same mediator into a concrete DM scenario and show how their low-energy observables constrains the DM coupling itself. In other words, we are comparing a DM model not previously explored in the literature, with the set of precision low-energy probes developed in ref. \cite{DiLuzio:2025ojt}.

\subsection*{Resonant Phenomenology}

We now turn to the resonant regime, defined by $m_\chi < m_X/2$, in which DM is predominantly produced via on-shell decays of the mediator, $X \rightarrow \chi\chi$. 
As discussed in sec.~\ref{sec:regimes}, in this regime the scattering contribution is equivalent to production through on-shell decays of the mediator and is therefore not included explicitly in order to avoid double counting. As a consequence, the relic abundance is entirely controlled by the decay width $\Gamma_{X\to\chi \chi}$. An important consequence is that the freeze-in abundance no longer depends on the electron coupling $g_e$, but only on the dark-sector parameters $(m_\chi,\,g_\chi)$ and on the mediator mass $m_X$.

Therefore, imposing the relic-abundance condition (eq. \eqref{eq:ImpAb}) does not select a curve in the $(m_X,g_e)$ plane as in the non-resonant case. Instead, for fixed $(m_\chi,g_\chi)$ we expect a unique value of the mediator mass $m_X$. In the $(m_X,g_e)$ plane this corresponds to a vertical line, reflecting the fact that the relic density is insensitive to $g_e$ in this production regime.

In order to recover phenomenological sensitivity to the low-energy plane $(m_X,g_e)$, we relax again the assumption that freeze-in through the $X_{17}$ mediator is the \emph{only} mechanism responsible for DM production. The relic-density condition therefore becomes the inequality from eq. \eqref{eq:RelAbIne}.

In order to make the phenomenology of this regime fully explicit, it is necessary to examine the functional dependence of the relic abundance on the $X$ mass, in the resonant case, where DM production is dominated by on-shell decays $X\rightarrow \chi\chi$.
Imposing the relic-abundance condition
\begin{equation}
    \Omega_{\rm DM}\,h^{2}(m_X)=0.12
\end{equation}
one finds that, in the resonant regime, this equation typically admits \emph{two} distinct solutions for the mediator mass $m_X$. The origin of this behaviour is not entirely trivial and is worth commenting on it.

Since from eq. \eqref{eq:YXcc} $Y_\chi\propto \Gamma_{X\to \chi \chi}=(m_X^2-4 m_{\chi}^2)^{3/2}/m_X^2$, one may naively expect the relic abundance to increase monotonically with $m_X$ , enhancing the efficiency of DM production. However, besides the growth of the decay rate, one must also take into account the cosmological dilution of the produced particles. To properly assess this effect, it is useful to recall a key feature of the freeze-in mechanism: DM is mainly produced at temperatures of the order of the mass of the heaviest particle involved in the process. Therefore, increasing $m_X$ also shifts the production epoch to earlier times, corresponding to higher temperatures.
The Hubble rate scales as $H(T)\propto T^{2} \sim m_X^2$ in correspondence of the DM production epoch, so production occurring at higher temperatures experiences a stronger Hubble expansion and hence a stronger dilution. 
This interplay can be appreciated more transparently by rewriting the mediator–mass and temperature dependence of the freeze–in abundance as
\begin{equation}
    Y_\chi(m_X)\;\propto\;\frac{\Gamma_{X\to \chi \chi}}{H}
    \;\propto\;
    \frac{\Gamma_{X\to \chi \chi}}{T^{2}}
    \;\sim\;
    \frac{(m_X^{2}-4m_{\chi}^{2})^{3/2}}{m_X^{4}}\,,
\end{equation}
which is precisely the scaling in eq. \eqref{eq:YXcc}.

As a consequence, increasing $m_X$ has two competing effects: on the one hand the decay rate grows, enhancing production; on the other hand, the earlier production time increases the expansion rate and suppresses the final abundance. 

\begin{figure}[t]
    \centering

    \begin{minipage}{0.48\linewidth}
        \centering
        \includegraphics[width=\linewidth]{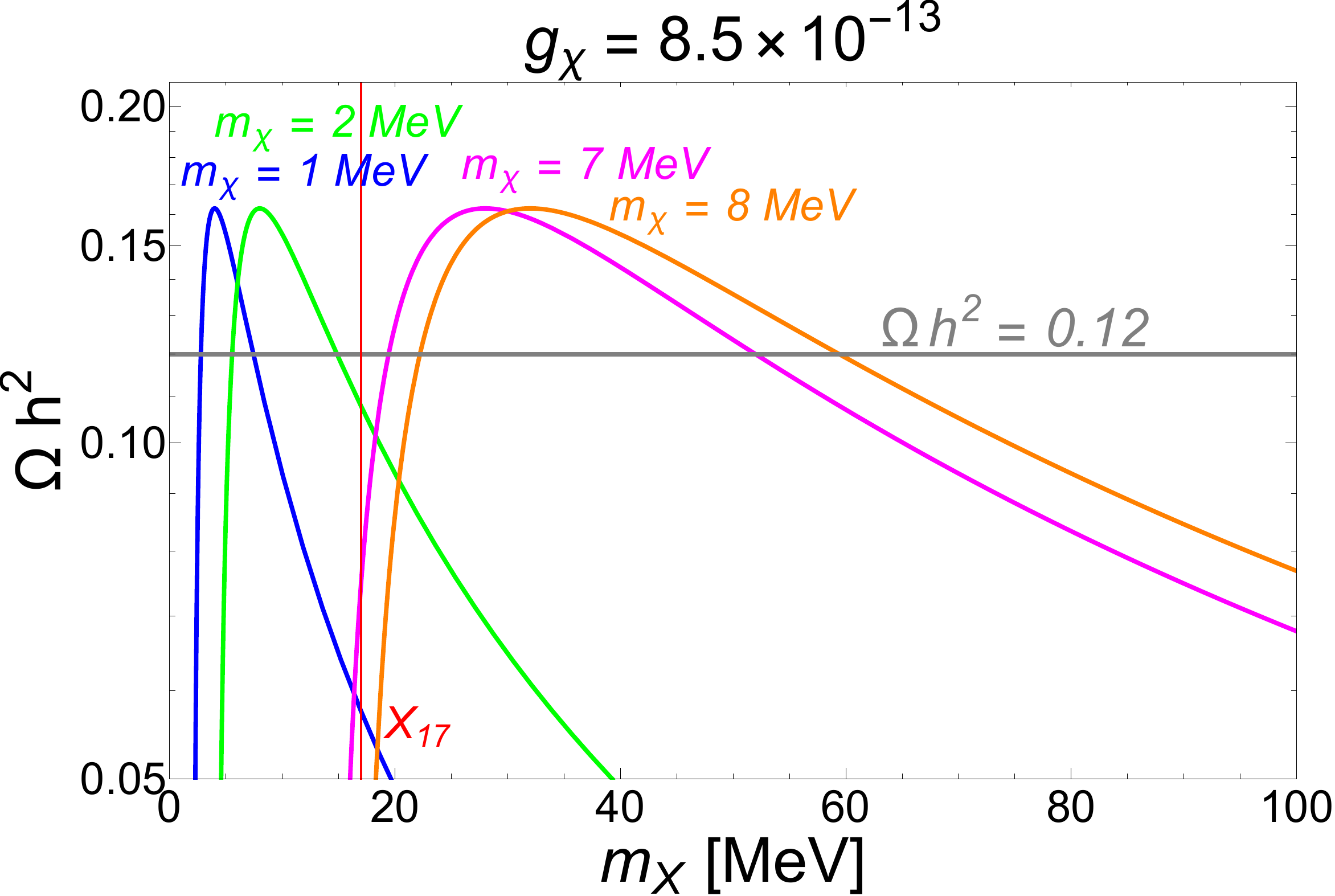}
    \end{minipage}
    \hfill
    \begin{minipage}{0.48\linewidth}
        \centering
        \includegraphics[width=\linewidth]{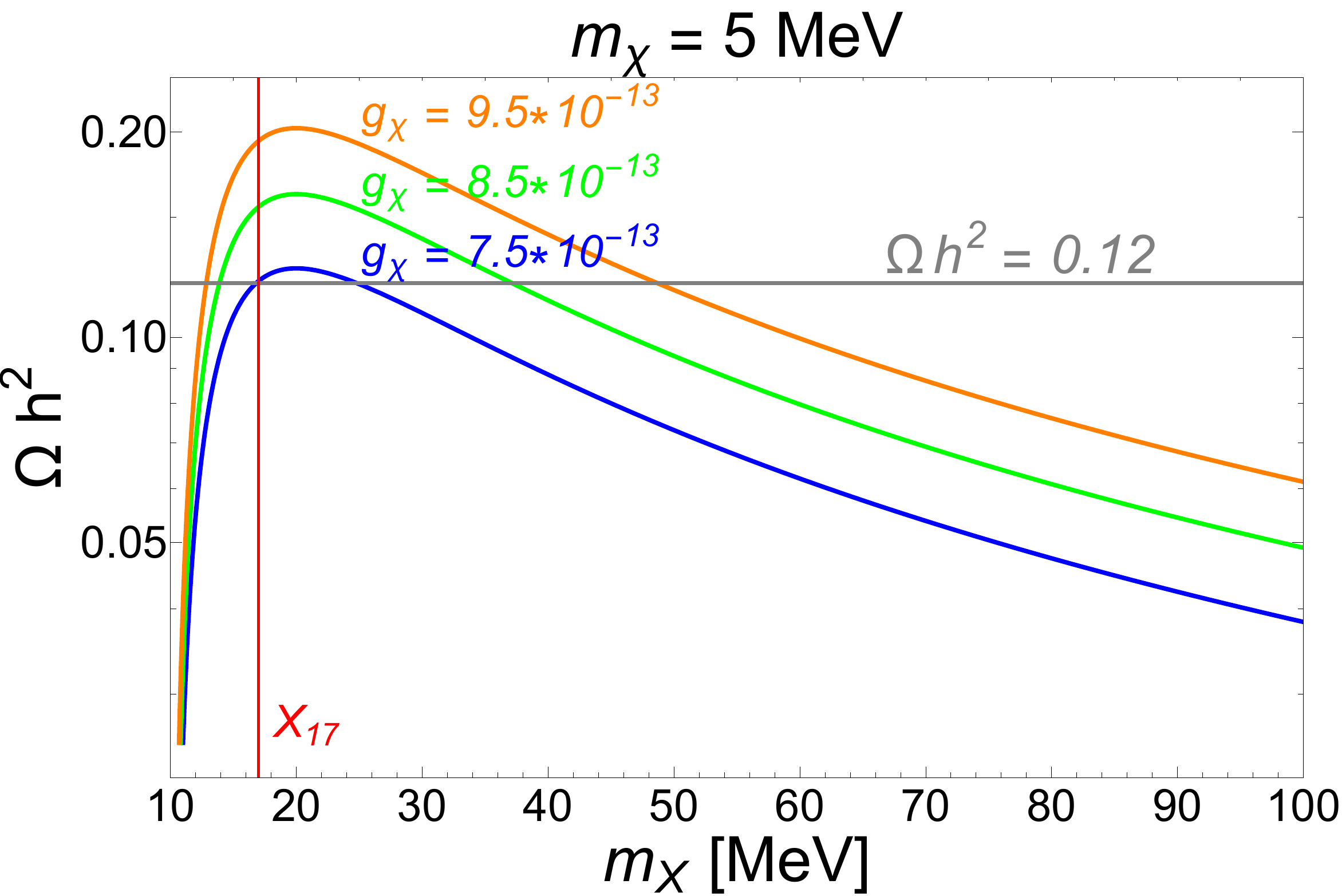}
    \end{minipage}
    \caption{Relic abundance in the resonant regime as a function of the mediator mass $m_X$. 
    The horizontal dashed line denotes the observed value $\Omega_{\rm DM}h^{2}=0.12$, while the vertical red line corresponds to $m_X = 17~\mathrm{MeV}$ associated with the $X_{17}$ hypothesis. 
    In the left panel we fix the coupling to $g_\chi = 8.5\times 10^{-13}$ and show the dependence on the mediator mass for four representative DM masses,
    $m_\chi = 1,\,2,\,7$ and $8~\mathrm{MeV}$. In the right panel instead we fix the DM mass to $m_\chi = 5~\mathrm{MeV}$ and plot the relic abundance for three benchmark values of the dark-sector coupling, namely $g_\chi = 7.5\times 10^{-13}$, $8.5\times 10^{-13}$ and $9.5\times 10^{-13}$. 
    }
    \label{fig:GLPlot}
\end{figure}

The interplay between these effects leads to a maximum of $\Omega_{\rm DM}h^{2}(m_X)$, so that the relic--abundance condition may be satisfied on both sides of this maximum, giving rise to two solutions for $m_X$.
The qualitative behavior described above is illustrated in fig. \ref{fig:GLPlot}. 
In both plots we show the relic abundance as a function of the mediator mass. 
The horizontal line corresponds to the observed value $\Omega_{\rm DM}h^{2}=0.12$, while the vertical red line marks the value $m_X = 17~\mathrm{MeV}$ associated with the $X_{17}$ hypothesis.

Unlike the non-resonant case, here both $g_\chi$ and $m_\chi$ are active parameters in the sense that variations in either of them significantly affect the resulting phenomenology. For this reason, we present two complementary plots: one in which the DM mass is kept fixed while the coupling is varied, and one in which the coupling is fixed and the DM mass is varied. This allows us to separately assess the phenomenological impact of each parameter.

In the left panel we fix the coupling to $g_\chi = 8.5\times 10^{-13}$ and show the dependence on the mediator mass for four representative DM masses, $m_\chi = 1,\,2,\,7$ and $8\,\mathrm{MeV}$.
In the right panel we instead fix the DM mass to $m_\chi = 5~\mathrm{MeV}$ and plot the relic abundance for three benchmark values of the dark-sector coupling, namely $g_\chi = 7.5\times 10^{-13}$, $8.5\times 10^{-13}$ and $9.5\times 10^{-13}$. 

\begin{figure}[t]
    \centering

    \begin{minipage}{0.48\linewidth}
        \centering
        \includegraphics[width=\linewidth]{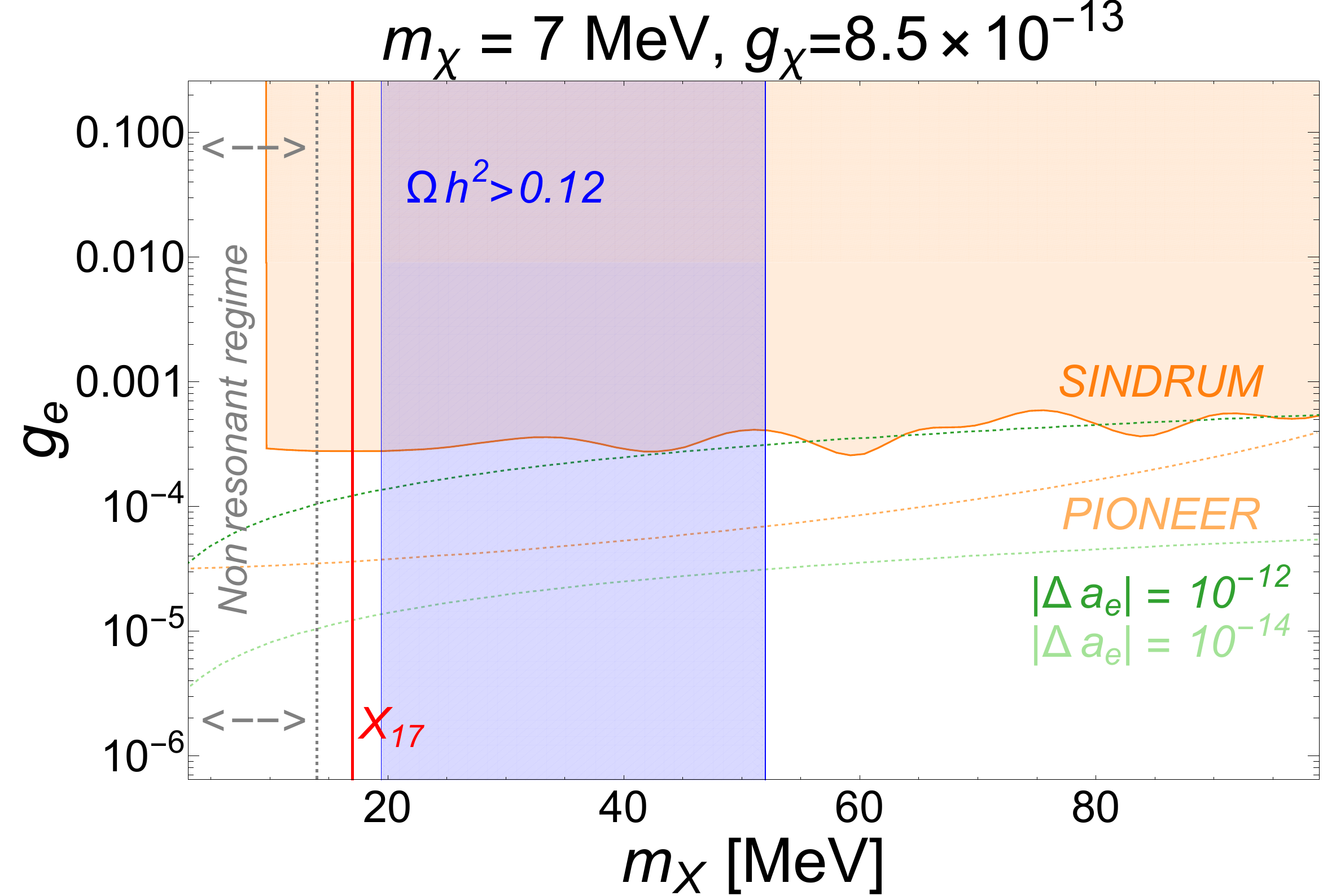}
    \end{minipage}
    \hfill
    \begin{minipage}{0.48\linewidth}
        \centering
        \includegraphics[width=\linewidth]{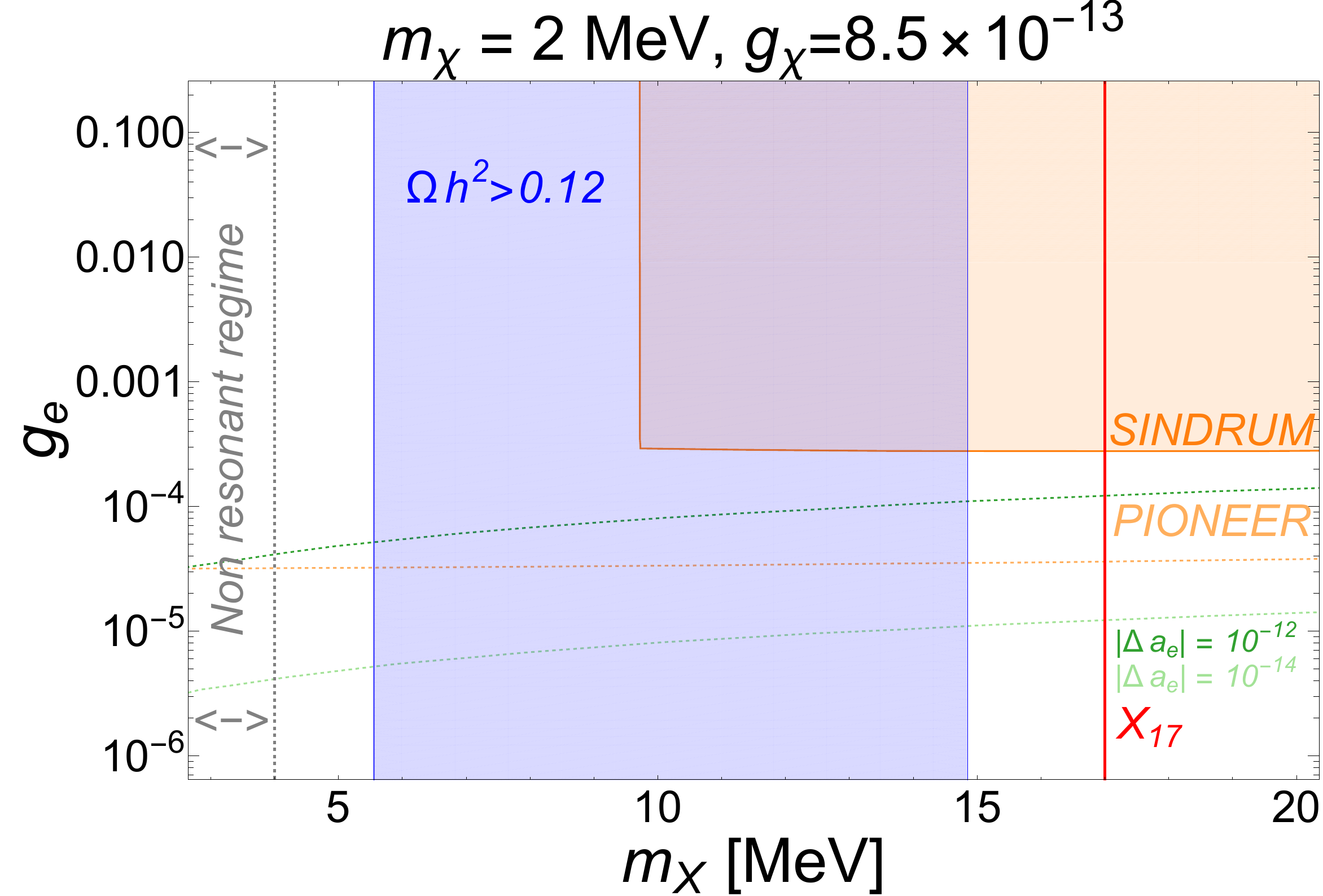}
    \end{minipage}
    \caption{Overlay of the regions excluded by DM overproduction on top of the low–energy constraints of ref.~\cite{DiLuzio:2025ojt} for the two benchmark scenarios described in the text.  In the left panel the relic abundance is reproduced in the narrow kinematic-increase region close to $m_X\simeq 2m_\chi$, for $m_\chi = 7~\mathrm{MeV}$ and $g_\chi = 8.5\times 10^{-13}$. In the right panel the correct relic density is obtained on the high–mass branch, for $m_\chi = 2~\mathrm{MeV}$ and the same coupling $g_\chi = 8.5\times 10^{-13}$. The vertical gray dashed line at $m_X = 2\,m_\chi$ marks the transition between the non--resonant
    and resonant production regimes.}
    \label{fig:OverProduction2}
\end{figure}

It is worth stressing that, in contrast with the non–resonant regime, varying the coupling $g_\chi$ in the resonant case does modify the qualitative behavior of the curves. The dependence is extremely sharp: the relic–abundance curves in fig. \ref{fig:GLPlot} are highly sensitive to small changes in $g_\chi$, which is in practice confined to a rather narrow interval, $g_\chi \sim (0.7\text{--}1)\times 10^{-12}$, as clear from Figure 3.4.
Within this range even modest variations of $g_\chi$ can shift the position of the maximum or remove one of the two solutions entirely.
When replacing the relic-density condition with the inequality in eq. \eqref{eq:RelAbIne}, the requirement becomes simply that freeze-in production through the $X$ mediator does not exceed the DM present in the Universe. As a consequence, instead of selecting isolated solutions, the allowed parameter space splits into two distinct regions.

A first, narrow allowed band appears close to threshold, where the relic abundance increases due to the kinematic increase of the decay rate, and is therefore confined around $m_X \gtrsim 2\,m_\chi\,$, hence this region is strongly controlled by the DM mass. 
A second, much broader allowed region emerges at larger mediator masses. 
In this domain the decrease driven by cosmological dilution opens up an extended parameter region that satisfies the non–overproduction condition.

To make this structure explicit, we overlay in fig.~\ref{fig:OverProduction2} the exclusion regions obtained from the non-overproduction condition on the low-energy plane $(m_X,g_e)$ already constrained in ref. \cite{DiLuzio:2025ojt}. For illustration, we consider two benchmark cases chosen such that the $X_{17}$ mass, $m_X = 17\,\mathrm{MeV}$, lies in different parts of the allowed region.

In the first case (left panel), we take $m_\chi = 7~\mathrm{MeV}$ and $g_\chi = 8.5\times 10^{-13}$. For these values, the point $m_X = 17~\mathrm{MeV}$ falls within the narrow region close to threshold, where the relic abundance is mainly controlled by the kinematic increase of the decay rate. The corresponding band allowed by the non-overproduction condition is plotted on top of the low-energy constraints of ref. \cite{DiLuzio:2025ojt}.

In the second case (right panel), we keep the coupling fixed at $g_\chi = 8.5\times 10^{-13}$ but choose a lighter DM mass, $m_\chi = 2~\mathrm{MeV}$. In this configuration the same value $m_X = 17~\mathrm{MeV}$ lies instead on the broader, high-mass side of the curve, where the relic density is dominated by cosmological dilution. The corresponding exclusion region from DM overproduction is again superimposed on the $(m_X,g_e)$ plane. In this way, both the threshold and dilution--dominated solutions associated with the $X_{17}$ mass are directly compared with the existing low--energy bounds.

\medskip
To summarize, in this section we have combined the relic–abundance requirement with the low–energy bounds in the $(m_X,g_e)$ plane. In the non–resonant regime, where DM is produced through off–shell scatterings, this comparison directly constrains the dark–sector coupling, leading to a robust lower bound $g_\chi \gtrsim 10^{-8}$. In the resonant regime, instead, DM is mainly produced from on–shell $X$ decays and the phenomenology is qualitatively different: the relic abundance becomes highly sensitive to the mediator mass and may admit two distinct solutions. Relaxing the assumption that $X$–mediated freeze–in is the only production mechanism, the non–overproduction condition selects two allowed regions, associated respectively with the kinematic increase near threshold and with the dilution–dominated branch at larger $m_X$.

\chapter{Conclusions}
\label{cap:Conclusions}

This thesis examined whether a light boson with a mass of order $17\,\mathrm{MeV}$, motivated by the recent $e^+e^-$ excess observed by PADME, can serve as a mediator between the SM and a minimal dark sector. 
DM remains one of the principal unresolved experimental puzzles of the Universe. Thus, although the experimental picture of  $17\,\mathrm{MeV}$ resonance is still evolving, it is valuable from a theoretical point of view to scrutinize the available data and relate it to the broader DM puzzle. The confirmation of such a particle would not only represent a major breakthrough but could also open a novel pathway toward addressing one of the central outstanding inconsistencies of the SM: the nature and origin of DM.

Firstly, in Chapter~\ref{cap:MotAndResObj} we outline the structure of the SM and review the experimental evidence supporting the existence of DM. We then introduce the main motivation for this thesis, summarizing the current status of the hints for a $17\,\mathrm{MeV}$ resonance reported by PADME and, earlier, by ATOMKI in nuclear transitions.

In Chapter~\ref{cap:TheFra}, we construct an EFT framework extending the SM with a real scalar mediator $X$, a Majorana fermion $\chi$, and a right-handed neutrino $\nu_R$. We classify all gauge-invariant operators up to dimension six that couple these new states to SM fields, providing a systematic basis for the phenomenological analysis developed in the subsequent chapters. To be more specific, we define a minimal benchmark scenario of interest for phenomenology, namely the one in which the $X$ mediator is a real scalar singlet and the DM particle $\chi$ coincides with the right–handed neutrino $\nu_R$.

In Chapter~\ref{cap:DMAbCalcu}, we apply this EFT to the early Universe, studying 
DM production through the freeze-in mechanism. Using analytic limits 
of the Boltzmann equation, we identify viable regions of parameter space in which 
thermal decays of DM reproduce the observed relic 
abundance. 

In Chapter~\ref{cap:DMExpSig}, we assess whether astrophysical ID probes can 
further constrain the model. Freeze-in generally implies the presence of weak couplings, which, indeed, are too tiny to be detected by existing ID experiments. However, our DM model remains testable with current and future low-energy observables and data, such as measurements of the anomalous magnetic moment of
the electron, as well as of rare muon and pion decays. 

The central outcomes of our analysis are presented in Figs. \ref{fig:RelicOverlay}-    \ref{fig:OverProduction2} of Sec. \ref{sec:CombinedConstraints}, where we combine the cosmological constraints derived in this work with the existing low-energy bounds—such as those from the electron anomalous magnetic moment and from rare muon and pion decays—reported in ref \cite{DiLuzio:2025ojt}.
This unified picture offers the most comprehensive assessment to date of the viable parameter space for the $X_{17}$-mediated dark-sector scenario explored here.

Taken together, our results demonstrate that a light MeV-scale mediator---such as
the hypothesized $X_{17}$---can be incorporated into a minimal and
gauge-invariant extension of the SM, can be produced through thermal decays in a
way that matches the observed relic abundance, and can remain compatible with a
broad set of experimental astrophysical bounds, such as indirect detection. 
Such a mediator can  offer a promising route toward unraveling the 
longstanding mystery of DM. The EFT framework ensures
that these conclusions are not tied to any specific ultraviolet completion,
while maintaining predictive power and internal consistency.

\medskip
The originality of the work presented in this thesis can be summarized as follows. 
In cap. \ref{cap:TheFra}, the operator tables and all amplitude calculations are original and work out for this specific project. 
In cap. \ref{cap:DMAbCalcu}, sec. \ref{sec:FICal} includes the original  computations (in particular sec. \ref{sec:Scateecc}); the approximations in sec. \ref{sec:regimes} and the numerical comparisons in the Appendix \ref{app:ConNum} are fully original. 
In cap. \ref{cap:DMExpSig}, sec. \ref{sec:LimTerCroSec} is original. All the figures \ref{fig:RelicOverlay}-\ref{fig:OverProduction2}, inspired by the bounds in Figure \ref{fig:PlotLuztot} (as computed in ref. \cite{DiLuzio:2025ojt}), would similarly contribute new original material.

\medskip
Several extensions can follow from this thesis and are necessary to ultimately 
prepare it for publication. A more complete EFT could include couplings to 
quarks, which would be essential for a unified interpretation of the ATOMKI 
anomalies. Future PADME runs will further clarify the spin of the mediator 
through improved sensitivity to the decay channel $X_{17}\to\gamma\gamma$. 
Moreover, while freeze-in scenarios typically imply extremely small interaction 
strengths, the presence of a light~($\sim$MeV-scale) mediator can enhance 
scattering at very low recoil energies, opening a window for DM direct 
detection. It will therefore be useful to investigate this type of measurements  in the near future to further constrain the parameter space identified in this thesis.
\newpage

\appendix

\chapter{Neutrino Coupling}
\label{app:NeuCou}

In this Appendix we will focus on the terms from the interaction Lagrangian (eq. \eqref{eq:LintBrutta}) that induce couplings of the dark fermion $\chi$ to active neutrinos, either directly or via mixing.

Our goal is to provide a qualitative but comprehensive discussion of their phenomenological implications. 
We show that astrophysical observations impose extremely stringent constraints on such interactions. 
These include, in particular, bounds from the stability of dark matter, limits on neutrino masses and mixing, and constraints from indirect detection.
As a consequence of these considerations, the couplings appearing in the neutrino sector must be strongly suppressed, which justifies neglecting them in the main body of the thesis.

In our model, the SM neutrino may interact with the dark fermion $\chi$ in two
distinct ways: 
\begin{itemize}
    \item firstly, through the Yukawa--like term
\begin{equation}
  y_{mix}\,\bar{\ell}\tilde{H}\chi
  \;\;\xrightarrow{\text{EWSB}}\;\;
  \theta\, m_\chi\,\bar{\nu}_L \chi \,
\label{eq:NeuMix}
\end{equation}
which induces a mixing between $\nu_L$ and $\chi$, recall that $\theta$ is the parameter introduced in eq.~(\ref{eq:theta}); 
\item secondly, through a trilinear interaction
\begin{equation}
  \frac{\hat{g}_{XDM\nu}}{\Lambda}\,X\,\bar{\ell}\tilde{H}\chi
  \;\;\xrightarrow{\text{EWSB}}\;\;g_\nu\,X\,\bar{\nu}_L \chi \, .
\label{eq:NeuYuk}
\end{equation}
\end{itemize}
As already discussed, these terms might allow DM to decay, undermining DM stability. 
However, a variety of experimental and observational inputs point towards
extremely suppressed $\theta$  and $g_\nu$, as we will show in this section. These results strongly disfavor neutrino interactions with our dark sector and, then, favor a portal-like interaction between the dark and the SM sectors. A theoretical explanation of the smallness of these couplings may rely on the introduction of an ad hoc discrete $Z_2$ symmetry that protects the dark sector, forbids the problematic operators, and at the same time stabilizes the DM candidate $\chi$. 

\section{Neutrino mixing term}

We are first considering the mixing term. Focusing our attention on eq.(\ref{eq:NeuMix}), there are three main experimental pieces of evidence to show why $\theta$ is ultra-suppressed: neutrino mass, DM lifetime and X-ray bounds.

The first handle comes from neutrino mass measurements. The most precise one comes from cosmology. It concerns the sum of the masses of three generations of neutrinos and it is taken from Ref. \cite{Elbers:2025vlz}
\begin{equation}
    \sum_{\nu}m_\nu<0.071 \,eV\,.
\end{equation}

In particular, the mixing between the SM neutrino and the Majorana fermion
$\chi$ induced by the dark–sector Lagrangian generates an additional contribution to the light neutrino mass.  
To make this explicit, it is useful to isolate from
$\mathcal{L}_{DS}$ the pieces that control the neutrino–$\chi$ mass
mixing:
\begin{equation}
    \mathcal{L}_{DS} \supset
    - \frac{1}{2}\,\hat m\,\bar{\chi}^c \chi
    + y_{mix}\,\bar{\ell}\tilde{H}\chi
    + \text{h.c.}
\end{equation}
After electroweak symmetry breaking, these terms generate a $2\times 2$
mass matrix in the $(\nu_L,\chi)$ basis:

\begin{equation}
    \mathcal{L}^{Broken}_{DS} \supset 
    -\frac{1}{2}
    \begin{pmatrix}
        \bar{\nu}_L & \bar{\chi}^c
    \end{pmatrix}
    \begin{pmatrix}
        0   & y_{mix}\,\frac{v}{\sqrt{2}}\\
       y_{mix}\,\frac{v}{\sqrt{2}} & \hat m
    \end{pmatrix}
    \begin{pmatrix}
        \nu_L \\ \chi
    \end{pmatrix}
    + \text{h.c.}
\label{eq:SSDiag}
\end{equation}

We assume the limit $\hat m \gg y_{mix}\,v/\sqrt{2}$, which is naturally
expected in a seesaw–like setup. Even in scenarios with relatively light
DM, this approximation remains consistent provided that $y_{mix}$ can be taken sufficiently small. In this sense, we can safely use the cosmological neutrino–mass bound to derive a conservative upper limit on $y_{mix}$. 
The mass matrix can be diagonalised perturbatively. To leading order one finds:
\begin{equation}
\begin{aligned}
        m_\chi&=\hat m\\
        \Delta m_\nu&=\frac{{y_{mix}^2\,v}^2}{2\,\hat m}
\end{aligned}
\label{eq:neumasscorr}
\end{equation}
corresponding to the physical mass of $\chi$ and a correction to the SM neutrino mass.

Using eq. (\ref{eq:neumasscorr}), it is convenient to express the correction
to the neutrino mass in terms of the active–sterile mixing angle:
\begin{equation}
  \Delta m_\nu \;\simeq\; \theta^2\,m_\chi\,.
\end{equation}
Requiring this contribution not to exceed the cosmological bound on the sum
of neutrino masses, $\Delta m_\nu \lesssim \sum_{\nu} m_\nu$, immediately gives
\begin{equation}
  \theta \;\lesssim\; \sqrt{\frac{\sum_{\nu} m_\nu}{m_\chi}}\,
  \qquad\Rightarrow\qquad
  \theta \;\lesssim\; 2.7\times 10^{-4}\,
  \sqrt{\frac{\text{MeV}}{m_\chi}}
\end{equation}

A much stronger constraint on $\theta$ may, instead, arise from the requirement of dark
matter stability, namely from the condition that the lifetime of the DM particle $\chi$ must be at least comparable to the age of the Universe, $T_U \simeq 10^{18}\,\rm s$, so that it does not decay away on cosmological time scales. The total decay width $\Gamma_\chi$ is related to the lifetime by
\begin{equation}
  \tau_\chi \;=\; \Gamma_\chi^{-1}\,,
\end{equation}
and the stability requirement can be written as
\begin{equation}
  \tau_\chi \;\gtrsim\; T_U
  \qquad\Longleftrightarrow\qquad
  \Gamma_\chi\,T_U \;\lesssim\; 1\,.
\label{eq:ReqStab}
\end{equation}
To illustrate how this translates into a bound on the active–sterile mixing angle $\theta$, let us consider a simple benchmark mass range. For $m_\chi < m_e$ the dominant decay channel is into SM neutrinos $(\chi \to 3\nu)$ with the decay width given by \cite{Cirelli:2024ssz}
\begin{equation}
\Gamma(\chi \to 3\nu)
= \frac{G_F^2 m_\chi^5}{96\pi^3}\,\theta^2
\;\approx\;
\frac{\theta^2}{33\,T_U}\left(\frac{m_\chi}{\text{keV}}\right)^5 \,,
\label{eq:GammaChi3nu}
\end{equation}
where $G_F$ is the Fermi constant, and $T_U$ denotes the age of the Universe.

Putting together the expression in eq. (\ref{eq:GammaChi3nu}) with the dark–matter stability requirement $\Gamma_\chi\,T_U \lesssim 1$ requirement in eq. (\ref{eq:ReqStab}), we obtain the following upper bound on the active–sterile mixing angle
\begin{equation}
\theta \;\lesssim\; 2\times 10^{-7}\,
\left(\frac{\text{MeV}}{m_\chi}\right)^{5/2}
\end{equation}
The stability bound above is stronger than the one derived from neutrino masses. 

We remark that the estimate above has been obtained assuming $m_\chi<m_e$, so that only invisible decays into neutrinos are kinematically accessible.
If the dark–matter particle were instead heavier than the electron, additional decay channels would open up, mediated by the same active–sterile mixing.
In this case, charged final states particles would unavoidably produce photons, either via final-state radiation or inverse Compton scattering on ambient radiation fields. 
Such electromagnetic signatures are much more tightly constrained observationally than purely neutrino final states, in particular by X-ray and gamma-ray telescopes.
As a consequence, once $m_\chi > m_e$ the corresponding bounds on the lifetime of $\chi$ become even stronger than those discussed above, implying an even more severe upper limit on the active–sterile mixing angle $\theta$.
Therefore, the constraint in Eq.~\eqref{eq:GammaChi3nu} should be regarded as conservative, and neutrino couplings must in general be extremely suppressed throughout the entire mass range of interest. 

Finally, the most significant bounds come from decay modes involving photons, since photons are easy to detect. 
The dominant such channel in this case is at one loop, namely $\chi \to \nu\,\gamma$, arising, for example, from the following diagram:

\begin{figure}[H]
    \centering
    \includegraphics[width=0.3\linewidth]{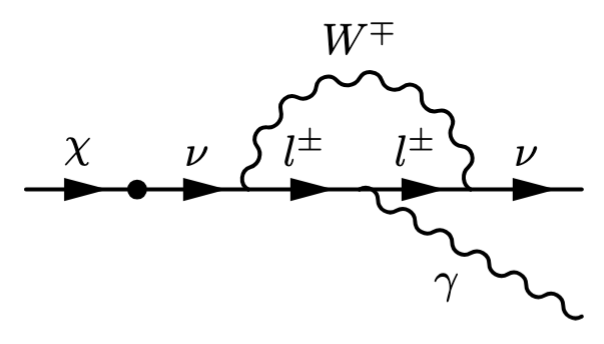}

\end{figure}

The corresponding decay width is (see Ref. \cite{Pal:1981rm})
\begin{equation}
\Gamma(\chi \to \nu \gamma)
= \frac{9 \alpha G_F^2 m_\chi^5}{256 \pi^4}\,\theta^2
\approx \frac{\theta^2}{4250\,T_U} \left( \frac{m_\chi}{\mathrm{keV}} \right)^5\,,
\end{equation}

where $\alpha$ is the usual electromagnetic fine-structure constant $\alpha \approx \frac{1}{137}$. Very roughly, current X–ray observations (see \cite{Essig:2013goa, Boudaud:2016mos, Liu:2020wqz}) require
\begin{equation}
  \tau_\chi\,\text{BR}(\chi \to \nu\gamma)
  \;=\;\frac{1}{\Gamma(\chi \to \nu\gamma)}\gtrsim\; 10^{8}\,T_U\,,
\end{equation}
i.e.\ the lifetime associated to the radiative decay channel must exceed the
age of the Universe by at least eight orders of magnitude. 

Imposing this X-ray lifetime bound on the decay rate
immediately translates into the following upper limit on the
active--sterile mixing angle:
\begin{equation}
  \theta \;\lesssim\; 2\times 10^{-10}\,
  \left(\frac{\mathrm{MeV}}{m_\chi}\right)^{5/2}.
\end{equation}

In summary, the active–sterile mixing angle $\theta$ is forced to be very small by three independent considerations: $i)$ correction to the light neutrino mass induced by a seesaw-like mechanism; $ii)$ requirement of DM stability; $iii)$ constraints by X--ray observations. Taken together, these arguments point to an ultra–suppressed mixing.   In particular, taking the most stringent constraint – the one from X–ray
observations – we find $\theta \lesssim 10^{-10}$ for a $\mathcal{O}(MeV)$ DM mass.

As a side remark, we note that such tiny mixing also implies that the
contribution of $\chi$ to active neutrino masses,
$m_\nu^{(\chi)} \simeq \theta^2 m_\chi$, is well below the largest measured
neutrino mass splittings. In other words, the observed neutrino masses must
predominantly arise from additional mechanisms beyond the simple seesaw induced
by the dark–matter state. 

\section{Neutrino pseudo-Yukawa term}

We now turn to the phenomenology of the pseudo–Yukawa interaction
\begin{equation}
  \frac{\hat{g}_{XDM\nu}}{\Lambda}\,X\,\bar{\ell}\tilde{H}\chi
  \;\;\xrightarrow{\text{EWSB}}\;\;g_\nu\,X\,\bar{\nu}_L \chi \,,
\end{equation}
and we will show, in close analogy with the mixing term discussion, that the coupling $g_\nu$ is also strongly constrained and effectively suppressed.

Firstly we will discuss the case in which the DM mass is larger than the mediator mass ($m_\chi > m_X$).
The opposite hierarchy ($m_\chi < m_X$) leads to an entirely similar phenomenological picture and will not be treated explicitly here.
In both situations existing experimental bounds force the coupling $g_\nu$ to be extremely small.

In the first regime, DM can decay via the two–body channel $\chi \;\to\; \nu\,X$, mediated by the coupling in eq. (\ref{eq:NeuYuk}).
If we assume this to be the dominant decay mode of $\chi$, the corresponding decay width is
\begin{equation}
  \Gamma(\chi \to \nu X)
  \;=\;
  \frac{g_\nu^2}{16\pi}\,m_\chi
  \left(1 - \frac{m_X^2}{m_\chi^2}\right)^{2},
  \label{eq:Gamma_chi_nuX}
\end{equation}
where we have neglected the SM neutrino mass.

As discussed above, DM stability on cosmological time–scales requires
its lifetime to exceed the age of the Universe,
\begin{equation}
  \tau_\chi \;\gtrsim\; T_U
  \qquad\Longleftrightarrow\qquad
  \Gamma_\chi\,T_U \;\lesssim\; 1\,.
\end{equation}
Identifying $\Gamma_\chi \simeq \Gamma(\chi \to \nu X)$
and inserting eq. (\ref{eq:Gamma_chi_nuX}), we obtain the bound
\begin{equation}
  g_\nu
  \;\lesssim\;
\sqrt{
    \frac{16\pi}
         {m_\chi\,T_U\left(1 - \dfrac{m_X^2}{m_\chi^2}\right)^{2}}
}\,.
\end{equation}
Thus, in the regime $m_\chi > m_X$, DM stability forces the
pseudo--Yukawa coupling $g_\nu$ to be very small, especially when the
mass splitting between $\chi$ and $X$ is not too close to threshold
($m_\chi \gg m_X$).

This bound is already extremely strong, roughly:
\begin{equation}
    g_\nu\lesssim10^{-20}\sqrt{\frac{\text{MeV}}{m_\chi}}\,.
\end{equation}

The discussion above shows that both sources of neutrino interactions with the dark sector the mixing term and the pseudo-Yukawa coupling are subject to extremely stringent phenomenological constraints.

In the case of mixing, the requirements coming from neutrino masses, DM stability and the absence of X-ray lines all force the mixing angle to be ultra-suppressed, $\theta \lesssim 10^{-10}$ for MeV-scale dark matter.  
Similarly, the pseudo-Yukawa coupling $g_\nu$ is constrained by the stability bound on $\chi$, leading to values as small as $g_\nu \lesssim 10^{-20}$ in the relevant mass range.

Such tiny couplings have no observable impact on the dark-sector phenomenology discussed in the main text.  
For this reason, throughout the thesis we consistently neglect neutrino interactions with the dark fermion $\chi$ and the mediator $X$, concentrating instead on the dominant couplings to electrons which control both the freeze-in production and the experimental signatures of the model.

\chapter{UV Completion}

In this appendix we introduce and discuss in a synthetic way a possible ultraviolet (UV) completion of the
effective theory discussed in Chapter \ref{cap:TheFra}.  

An immediate issue that emerges in this context concerns the couplings between the
dark sector and neutrinos.  
As discussed in Appendix \ref{app:NeuCou}, interactions of the dark fermion $\chi$ with SM neutrinos are subject to extremely stringent bounds coming from cosmology, indirect detection and neutrino–mass constraints.  
These considerations imply that all the operators in Lagrangian \eqref{eq:LintUV} inducing sizeable $\chi$–neutrino couplings must be highly suppressed, or absent altogether, in a realistic UV completion. A simple and economical way to account for this suppression is to invoke a discrete
$Z_2$ symmetry acting on the dark sector under which all SM particles and the $X_{17}$ boson are even while the DM particle $\chi$ is odd. 

The renormalizable terms involving the dark fermion $\chi$ at low energy (i.e. after EWSB) are
\begin{equation}
    \mathcal{L}_{\chi}
    =
    \bar{\chi}\,i\gamma_\mu{\partial}^\mu\chi
    - \frac{m_\chi}{2}\,\bar{\chi}^c\chi
    + g_\chi\,X\,\bar{\chi}^c\chi + \text{h.c.} \,.
\label{eq:LDM}
\end{equation}

The last term of \eqref{eq:LDM} is one of the two key ingredients that generate the coupling between the dark matter and the mediator $X$.  
The other ingredient, responsible for inducing the coupling of $X$ to electrons in the low–energy model discussed in section \ref{sec:DiagEva}, derives from the non–renormalizable operator 
\begin{equation}
    \mathcal{L}_{Xee}^{\rm eff}
    =
    \frac{g_{Xee}}{\Lambda}\,
    X_{17}\,\bar{\ell}\,H\,e_R
    + \text{h.c.}\,.
\end{equation}
This dimension–5 operator , that induce the low-energy coupling $g_e$ as in Eq.(2.11) after EWSB, must originate from an ultraviolet completion of the theory, which we now proceed to describe.

We consider a left-handed and a right-handed copies of the right-handed electron, $E_{L,R}$, which are color and $SU(2)_L$ singlets and which have weak hypercharge $Y=-1$. They are also even under the $Z_2$ symmetry. The renormalizable UV Lagrangian (omitting kinetic terms) reads

\begin{equation}
\begin{split}
     -\mathcal{L}_E=&\bar{L}Y_eHe_R+\bar{L}y_EHE_R+(\mu_E+\lambda_E\,X_{17})\bar{E}_Le_R\\
     &+(M_E+\lambda'_E\,X_{17})\bar{E}_LE_R+h.c.
\end{split}
\end{equation}
After electroweak symmetry breaking, $H$ acquires a vacuum expectation value and the terms above generate a $4\times 4$ mass matrix in the $(E_L,E_R,e_L,e_r)$ space.
In close analogy with the seesaw–like structure discussed in eq. \eqref{eq:SSDiag}, we diagonalize the mass spectrum within the assumption $M_E\gg  y_E \,v_H\gg \mu_E$. The new fermions $(E_L,E_R)$ pair in a new vector-like heavy state of mass $M_E+\mathcal{O}(y_E^2v_H^2/M_E)$.Notice that, parameterizing the coupling between the SM electron and the Higgs boson as $\kappa_e m_e\,h\bar{e}e$, we get $\kappa_e\sim 1+y_E^2v_H^2/m_e M_E$, well below the current limits $\kappa_e\lesssim \mathcal{O}(100)$~\gl{refs}. 

Finally, integrating the heavy fermion, we generate the interaction 

\begin{equation}
\frac{\lambda_Ey_Ev_H}{\sqrt{2}M_E}X_{17}\,\bar{e}e +\mathcal{O}(y_E^2v_H^2/M_E^2)\,.
\end{equation}
is generated. Thus, we can identify the scalar coupling $g_{eX}\sim \lambda_E\tilde{\mu}_E/M_E$. 

With a suitable choice of parameters, $(M_E,y_E\lambda_E)\sim \mathcal{O}(100\text{ TeV},0.01)$ provides $g_{eX}\sim 10^{-5}$ and a vector-like electron pair of mass $\sim 100$ TeV, which mixes with the electron with mixing angle $\theta_{eE}\sim y_Ev_H/M_E\sim 10^{-4}$.

The presence of heavy vector–like leptons offers a characteristic phenomenological handle at high–energy colliders, where they can be directly produced and searched for through their electroweak interactions.  
A detailed study of these signatures lies beyond the scope of the present thesis, and we refer the interested reader to ref. \cite{Poh:2017tfo}.

\chapter{Validity of the Narrow-Width and Non-Resonant Approximations}
\label{app:ConNum}

In section \ref{sec:regimes}, analytical expressions for the freeze-in abundance from the process $e^+e^- \to \chi\chi$ were obtained by simplifying the integral in eq. \eqref{eq:Yeecc} in different kinematic regimes of the DM mass with respect to the mediator mass. In this appendix we will assess the validity of those analytical approximations.

By performing a direct numerical integration over $\hat s$, we compare the approximate expressions derived in section \ref{sec:regimes} with the full result, thereby quantifying the accuracy of the corresponding analytic formulas.

\begin{figure}[t]
    \centering
    \includegraphics[width=0.8\linewidth]{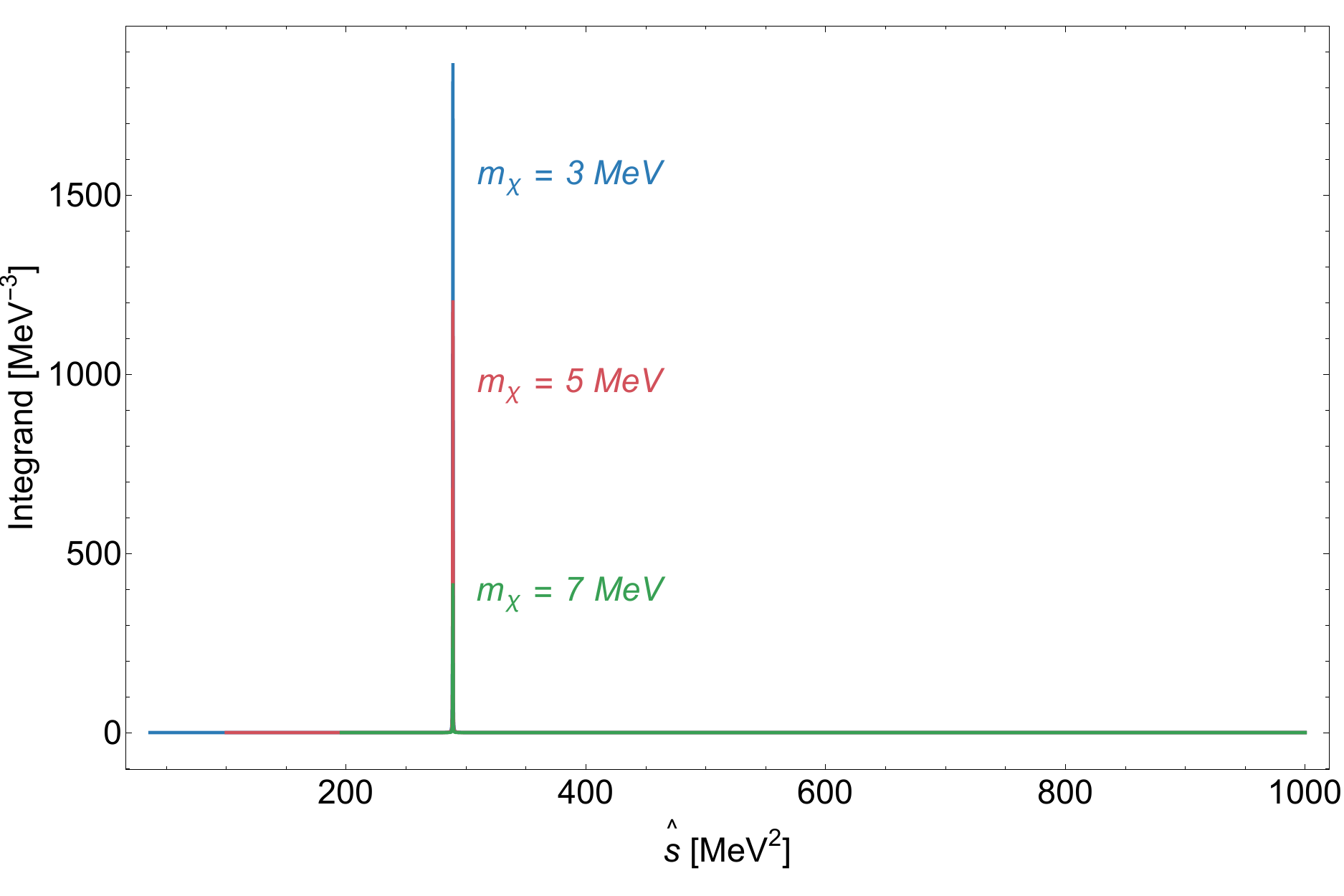}
    \caption{Integrand in the r.h.s. of Eq.(\ref{eq:defI}) as a function of the center–of–mass energy $\hat s$ for the benchmark $m_X=17 \mathrm{MeV}$. The three curves correspond to $m_\chi=\{3,5,7\}\,\mathrm{MeV}$ (as indicated). A narrow Breit–Wigner peak is clearly visible at $\hat s=m_X^2$; away from the pole the integrand decreases rapidly. Axes are linear.}
    \label{fig:BWPea}
\end{figure}

In the expression for the abundance in eq. \eqref{eq:Yeecc}, the purely kinematical
integration over the Mandelstam variable $\hat s$ can be isolated as
\begin{equation}
I(m_\chi,m_X,\Gamma_X,m_e)
\equiv
\int_{\hat s_{\min}}^{\infty} 
d\hat s\;
\frac{\left(\hat s-4m_\chi^{2}\right)^{3/2}
      \left(\hat s-4m_e^{2}\right)^{3/2}}
     {\hat s^{5/2}\left[\Gamma_X^{2}m_X^{2}
     +\left(m_X^{2}-\hat s\right)^{2}\right]}\,.
\label{eq:defI}
\end{equation}
We begin by considering the resonant regime, in which the center-of-mass energy approaches the pole of the mediator propagator, $\hat s \simeq m_X^2$. The presence of a sharp peak at $\hat s = m_X^2$ is clearly visible in fig. \ref{fig:BWPea}, where, for simplicity, we have chosen the benchmark value $m_X = 17~\mathrm{MeV}$ and plotted the integrand for various values of $m_\chi$.

In this situation, the numerical evaluation of the integral in eq. \ref{eq:Yeecc} requires particular care, due to the Breit-Wigner enhancement near the resonance. In order to properly resolve the peak and ensure numerical stability, we split the integration domain into two separate intervals, namely
from $0$ to $m_X^2$ and from $m_X^2$ to $+\infty$. The integral is then computed numerically in each interval and the two contributions are summed.
This procedure avoids instabilities associated with sampling across the resonance and guarantees the convergence of the numerical integration in the resonant regime.

We now compare the numerical value of the integral, $I_{\rm Num}$, obtained with the procedure described above, with the analytic expression derived in the narrow-width approximation, $I_{\rm NWA}$, see eq.~\ref{eq:BW}. In order to quantify the accuracy of the NWA, we define the relative percentage error as
\begin{equation}
    {\rm Err}
    =
    \left|
    \frac{I_{\rm NWA}-I_{\rm Num}}{I_{\rm NWA}}
    \right|
    \times 100\,.
\label{eq:ErrNWA}
\end{equation}
This quantity provides a direct measure of the deviation between the exact numerical result and the analytic approximation obtained in the resonant limit.

We evaluate ${\rm Err}$ for a range of DM masses $m_\chi$ in the resonant regime and display the resulting percentage error as a function of $m_\chi$. We find that the difference between the numerical evaluation and the NWA result is extremely small over the whole range of dark–matter masses considered.  
Scanning over $\mathcal{O}(10^3)$ values of $m_\chi$ in the interval ($1\,MeV-8.5\,MeV$), and computing the corresponding ${\rm Err}$ for each point, we evaluate the mean percentage deviation and its standard deviation.  
The resulting averaged error is

\begin{equation}
    {\rm Err}_{\rm mean}\;\approx\;3\times 10^{-5}\,\% ,
\end{equation}

with a standard deviation

\begin{equation}
    \sigma_{\rm Err}\;\approx\;3\times 10^{-4}\,\% \,
\end{equation}

Both the tiny value of the mean error and the small spread of the distribution confirm that the narrow–width approximation reproduces the full numerical integration with excellent accuracy throughout the resonant regime.
Such a tiny value confirms that the narrow–width approximation reproduces the full numerical integration with excellent accuracy throughout the resonant regime, and validates the use of the analytic expression in our analysis.

In addition to the explicit numerical check above, the validity of the NWA in a thermal bath can be assessed analytically following the criterion proposed in ref. \cite{Heeck:2014zfa}.  
That work is devoted to a different physical scenario from the one considered here.  
However, in section F of their third chapter, the authors explicitly study the validity of the NWA for resonant production processes and derive the range of temperatures where the approximation provides an accurate description.  

In our case, the role of the mediator discussed in Ref. \cite{Heeck:2014zfa} is played by the scalar $X$. Translating their results to our notation, and defining the dimensionless width parameter as $\varepsilon \;\equiv\; \frac{\Gamma_X}{m_X}$ the corresponding validity range can be expressed in terms of the variable $z$
\begin{equation}
    \sqrt{\varepsilon} \;\lesssim\; z \;\lesssim\; 14\,\sqrt{\ln \varepsilon^{-1}}\,.
\end{equation}
Using the parameter of our setup, $m_X \approx 17\,\mathrm{MeV}$ and a benchmark value for the electron coupling $g_e \approx 5\times 10^{-4}$ , the validity range of the narrow-width approximation translates into the following temperature window:
\begin{equation}
    1.0\times 10^{-4} \;\lesssim\; z \;\lesssim\; 4.3\,.
\end{equation}
Since the freeze-in production in our model is dominated around $z \approx 1$, which lies well inside this interval, we conclude that the NWA provides an accurate description of the resonant dynamics in our case.

The analysis above focuses on the resonant regime.  
For completeness, we have also performed an analogous test in the non–resonant regime, $m_\chi > m_X/2$.  
In this case the integral $I(m_\chi,m_X,\Gamma_X,m_e)$ in eq. \eqref{eq:defI} is compared with the non–resonant analytic approximation derived in sec. \ref{sec:regimes}, and we define a relative error ${\rm Err}_{\rm n.r.}$ in complete analogy with eq. \eqref{eq:ErrNWA}.  

Scanning over $\mathcal{O}(10^5)$ values of $m_\chi$ in the non–resonant domain ($8.5~{\rm MeV}\;\lesssim\; m_\chi \;\lesssim\; 1~{\rm GeV}$), and computing ${\rm Err}_{\rm n.r.}$ for each point, we find that the typical deviation between the analytic estimate and the full numerical result remains at the sub–percent level.  
In particular, the mean value of the relative error is

\begin{equation}
    {\rm Err}_{{\rm mean},\,{\rm n.r.}}\;\approx\;0.01\,\%\,,
\end{equation}

with a standard deviation

\begin{equation}
    \sigma_{{\rm Err},\,{\rm n.r.}}\;\approx\;0.07\%\,.
\end{equation}

We therefore conclude that, also in the non–resonant regime, the analytic approximation used in sec.~\ref{sec:regimes} provides an accurate description of the exact numerical result for the purposes of our phenomenological analysis.

\backmatter

\newpage
\phantomsection
\addcontentsline{toc}{chapter}{\bibname}
\bibliography{Extra/bibliografia}


\chapter{Acknowledgments}

We would first like to express our deepest gratitude to Federico Mescia for his constant guidance, support, and encouragement throughout the entire development of this thesis.
We would also like to express our sincere gratitude to Claudio Toni for his continuous support and availability. His experience and expertise on the $X_{17}$ topic have significantly contributed to clarifying many theoretical and phenomenological aspects addressed in this thesis.

\end{document}